\documentclass[showpacs,amssymb,prd,superscriptaddress,twocolumn,aps,nofootinbib]{revtex4-1}
\usepackage{graphicx, epsfig, amssymb} %include figure files
\usepackage{amsmath, amsfonts}
\usepackage{bm} %include bold math: \bm{} creates bold letters in math mode
\usepackage[breaklinks]{hyperref}
\usepackage{color}

\def\nn{\nonumber}
\def\be{\begin{equation}}
\def\ee{\end{equation}}
\def\beq{\begin{eqnarray}}
\def\eeq{\end{eqnarray}}
\def\lp{{\ell+1}}
\def\lm{{\ell-1}}
\def\lpp{{\ell+2}}
\def\lmm{{\ell-2}}
\def\pa{\partial}

\def\th{\vartheta}
\def\ii{{{\rm i}\,}}

\def\cQ{{\cal Q}}

\begin{document}

\title{Perturbations of slowly rotating black holes:\\ massive vector fields in the Kerr metric}

\author{Paolo Pani}\email{paolo.pani@ist.utl.pt}
\affiliation{CENTRA, Departamento de F\'{\i}sica, Instituto Superior T\'ecnico, Universidade T\'ecnica de Lisboa - UTL,
Av.~Rovisco Pais 1, 1049 Lisboa, Portugal.}

\author{Vitor Cardoso}\email{vitor.cardoso@ist.utl.pt}
\affiliation{CENTRA, Departamento de F\'{\i}sica, Instituto Superior T\'ecnico, Universidade T\'ecnica de Lisboa - UTL,
Av.~Rovisco Pais 1, 1049 Lisboa, Portugal.}
\affiliation{Department of Physics and Astronomy, The University of Mississippi, University, MS 38677, USA.}

\author{Leonardo Gualtieri}\email{Leonardo.Gualtieri@roma1.infn.it}
\affiliation{Dipartimento di Fisica, Universit\`a di Roma ``La Sapienza'' \& Sezione INFN Roma1, P.A. Moro 5, 00185, Roma, Italy.}

\author{Emanuele Berti}\email{berti@phy.olemiss.edu}
\affiliation{Department of Physics and Astronomy, The University of Mississippi, University, MS 38677, USA.}
\affiliation{California Institute of Technology, Pasadena, CA 91109, USA}

\author{Akihiro Ishibashi}\email{akihiro.ishibashi@kek.jp}
\affiliation{Theory Center, Institute of Particle and Nuclear Studies,
High Energy Accelerator Research Organization (KEK), Tsukuba, 305-0801, Japan}
\affiliation{Department of Physics, Kinki University, Higashi-Osaka 577-8502, Japan}

\begin{abstract} 
  We discuss a general method to study linear perturbations of slowly
  rotating black holes which is valid for any perturbation field, and
  particularly advantageous when the field equations are not
  separable. As an illustration of the method we investigate massive
  vector (Proca) perturbations in the Kerr metric, which do not appear
  to be separable in the standard Teukolsky formalism.
  Working in a perturbative scheme, we discuss two important effects
  induced by rotation: a Zeeman-like shift of nonaxisymmetric
  quasinormal modes and bound states with different azimuthal
  number $m$, and the coupling between axial and polar modes with
  different multipolar index $\ell$.
  We explicitly compute the perturbation equations up to second order
  in rotation, but in principle the method can be extended to any
  order. Working at first order in rotation we show that polar and
  axial Proca modes can be computed by solving two decoupled sets of
  equations, and we derive a single master equation describing axial
  perturbations of spin $s=0$ and $s=\pm 1$.
  By extending the calculation to second order we can study the
  superradiant regime of Proca perturbations in a self-consistent
  way. For the first time we show that Proca fields around Kerr black
  holes exhibit a superradiant instability, which is significantly
  stronger than for massive scalar fields. Because of this
  instability, astrophysical observations of spinning black holes
  provide the tightest upper limit on the mass of the photon:
  $m_\gamma \lesssim 4\times 10^{-20}$~eV under our most conservative
  assumptions.  Spin measurements for the largest black holes could
  reduce this bound to $m_\gamma \lesssim 10^{-22}$~eV or lower.
\end{abstract}


\maketitle

%\tableofcontents

%%%%%%%%%%%%%%%%%%%%%%%%%
\section{Introduction}
%%%%%%%%%%%%%%%%%%%%%%%%%
Linear perturbations of black holes (BHs) play a major role in
physics.  Many astrophysical processes can be modeled as small
deviations from analytically known BH backgrounds: for instance,
perturbative calculations of quasinormal modes (QNMs) are useful to
describe the late stages of compact binary mergers or gravitational
collapse~\cite{Nollert:1999ji,Kokkotas:1999bd,Ferrari:2007dd,
  Berti:2009kk,Konoplya:2011qq}, and BH perturbation theory provides
an accurate description of the general relativistic dynamics of
extreme mass-ratio
inspirals~\cite{AmaroSeoane:2007aw,Barack:2009ux,Poisson:2011nh}. Even
when the understanding of complex astrophysical phenomena requires
numerical relativity, the set up of the numerical simulations and the
interpretation of the results are easier when we can rely on prior
perturbative knowledge of the problem (see
e.g.~\cite{Berti:2007fi,Berti:2010ce}).

Besides their interest in modeling gravitational-wave sources for
present and future detectors in Einstein's theory and in various
proposed extensions of general relativity, perturbative studies of BH
dynamics are also relevant in the context of high-energy
physics~\cite{Cardoso:2012qm}.  Perturbation theory can shed light on
several open issues, such as the stability properties of BH spacetimes
in higher dimensions and in asymptotically anti-de Sitter
spacetimes. Within the gauge-gravity duality, some of the correlation
functions and transport coefficients are related to the lowest order
BH QNMs. In a semiclassical treatment of BH evaporation, the
calculation of greybody factors (which may be of direct interest for
ongoing experiments) relies heavily on our ability to understand wave
scattering in rotating BH spacetimes.

BH perturbation theory is a useful tool to investigate issues in
astrophysics and high-energy physics as long as the radial and angular
parts of the perturbation equations are separable. This usually
happens when the background spacetime has special symmetries.  If
separable, the perturbation equations in the frequency domain reduce
to a system of ordinary differential equations
(ODEs)~\cite{1992mtbh.book.....C}. Separability is the norm if the
background spacetime is spherically
symmetric. Teukolsky~\cite{Teukolsky:1973ha} discovered that a large
class of perturbation equations is exceptionally separable in the Kerr
metric, the underlying reason being that the Kerr spacetime is of type
D in the Petrov classification. Separability is much more difficult to
achieve in the Kerr-Newman metric~\cite{1992mtbh.book.....C} and in
higher spacetime dimensions~\cite{Durkee:2010qu}. In the Kerr
background, perturbations induced by massless fields with integer and
half-integer spins (including Dirac and Rarita-Schwinger fields
\cite{Teukolsky:1973ha,TorresdelCastillo:1990aw,1992mtbh.book.....C})
are all separable, but this does not seem to be possible for some
classes of perturbations, such as massive vector (Proca)
perturbations~\cite{Rosa:2011my}.

Here we discuss a general method to study linear perturbations of slowly
rotating BH backgrounds that is particularly useful when the perturbation
variables are not separable. The method is an extension of Kojima's work on
perturbations of slowly rotating neutron stars
\cite{Kojima:1992ie,1993ApJ...414..247K,1993PThPh..90..977K}. Slowly rotating
backgrounds are ``close enough'' to spherical symmetry that an approximate
separation of the perturbation equations in radial and angular parts becomes
possible. The field equations can be Fourier transformed in time and expanded in
spherical harmonics and they reduce, in general, to a coupled system of ODEs.
This approach can be seen as a two-parameter perturbative expansion
\cite{Bruni:2002sma}, where the small parameters are the amplitude of the
perturbation and the angular velocity of the background.
The method we present can in principle be extended to any order in the rotation
parameter; here we derive the perturbation equations explicitly up to second
order.

To first order in rotation, the final system of coupled ODEs can be
simplified by dropping terms that couple perturbations with different
values of the harmonic indices. This is a surprisingly good
approximation: extending previous arguments by Kojima
\cite{1993PThPh..90..977K}, we show that the neglected coupling terms
can affect the frequencies and damping times of the QNMs only at
second or higher order in the rotation rate.  We support the
analytical argument by numerical results, showing that the modes
computed with and without the coupling terms coincide to first order
in the rotation rate.  At second order in rotation, the coupling of
perturbations with different harmonic indices cannot be
neglected. However, a notion of ``conserved quantum number'' $\ell$ is
preserved: perturbations with given parity and harmonic index $\ell$
are coupled with perturbations with opposite parity and harmonic
indices $\ell\pm1$, and with perturbations with the same parity and
harmonic indices $\ell\pm2$~\cite{ChandraFerrari91}.

The main limitation of the present approach is the slow-rotation
approximation. Currently this is not a restriction in extensions of
general relativity including quadratic curvature corrections, where
rotating BH solutions are only known in the slow-rotation
limit~\cite{Yunes:2009hc,Konno:2009kg,
  Pani:2009wy,Pani:2011gy,Yagi:2012ya}.  Furthermore, as we show in
this paper, a slow-rotation approximation is sufficient to describe
important effects, like the superradiant instability of massive fields
around rotating BHs. While a second-order expansion is needed to
describe superradiance \emph{consistently}, even the first-order
approximation provides accurate results well beyond the nominal region
of validity of the approximation. Similar extrapolations have been
used in the past to predict the existence and timescale of $r$-mode
instabilities in the relativistic theory of stellar perturbations (see
e.g.~\cite{Andersson:1997xt,Kojima:1997vv,Lockitch:1998nq,Lockitch:2000aa}),
and it is not unreasonable to expect that some quantities computed in
the small-rotation regime (e.g. reflection coefficients) can be safely
extrapolated to higher spin values. For all these reasons we are
confident that perturbative studies of slowly rotating BHs will
further our understanding of several interesting open problems in
astrophysics and high-energy physics.

As an interesting testing ground of the slow-rotation approximation,
here we focus on massive vector (Proca) perturbations of slowly
rotating Kerr BHs. It is well known that massive bosonic fields in
rotating BH spacetimes can trigger superradiant
instabilities~\cite{Press:1972zz,Damour:1976kh,Cardoso:2004nk,
  Cardoso:2005vk,Dolan:2007mj,Rosa:2009ei,Cardoso:2011xi}.  For scalar
fields the instability is well studied. It is regulated by the
dimensionless parameter $M\mu$ (in units $G=c=1$), where $M$ is the BH
mass and $m_s=\mu\hbar$ is the bosonic field mass\footnote{In this
  paper, with a slight abuse of notation, we will use $\mu$ for both
  the scalar field mass ($\mu=m_s/\hbar$) and the vector field mass
  ($\mu=m_v/\hbar$). The meaning should be clear from the context.},
and it is strongest for maximally spinning BHs, when $M\mu\sim 1$. For
a solar mass BH and a field of mass $m_s\sim 1$~eV the parameter
$M\mu\sim10^{10}$.  In this case the instability is exponentially
suppressed~\cite{Zouros:1979iw} and in many cases of astrophysical
interest the instability timescale would be larger than the age of the
universe. However, strong superradiant instabilities ($M\mu\sim1$) can
occur either for light primordial BHs which may have been produced in
the early
universe~\cite{1971MNRAS.152...75H,1966AZh....43..758Z,1974MNRAS.168..399C}
or for ultralight exotic particles found in some extensions of the
standard model, such as the ``string axiverse'' scenario
\cite{Arvanitaki:2009fg,Arvanitaki:2010sy}. In this scenario, massive
scalar fields with $10^{-33}~{\rm eV}<m_s<10^{-18}~{\rm eV}$ could
play a key role in cosmological models. Superradiant instabilities may
allow us to probe the existence of such ultralight bosonic fields by
producing gaps in the mass-spin BH Regge
spectrum~\cite{Arvanitaki:2009fg,Arvanitaki:2010sy}, by modifying the
inspiral dynamics of compact
binaries~\cite{Cardoso:2011xi,Yunes:2011aa, Alsing:2011er} or by
inducing a ``bosenova'', i.e. collapse of the axion cloud (see
e.g. ~\cite{Kodama:2011zc,Yoshino:2012kn,Mocanu:2012fd}).

Similar instabilities are expected to occur for massive hidden $U(1)$
vector fields, which are also a generic feature of extensions of the
standard model
\cite{Goodsell:2009xc,Jaeckel:2010ni,Camara:2011jg,Goldhaber:2008xy}.
Massive vector perturbations of rotating BHs are expected to induce a
superradiant instability, but an explicit demonstration of this effect
has been lacking. While this problem has been widely studied for
massive scalar fields~\cite{Press:1972zz,Damour:1976kh,Cardoso:2004nk,
  Cardoso:2005vk,Dolan:2007mj,Rosa:2009ei,Cardoso:2011xi}, the case of
massive vector fields is still uncharted territory; incursions in the
topic seem to be restricted to nonrotating
backgrounds~\cite{Gal'tsov:1984nb,Herdeiro:2011uu,Rosa:2011my,Konoplya:2005hr}.  The main reason is
that the Proca equation, which describes massive vector fields, does
not seem to be separable in the Kerr background. In the slow-rotation
approximation we can reduce the problem of finding QNMs to a tractable
system of coupled ODEs, where polar perturbations with angular index
$\ell$ are generically coupled to axial perturbations with index
$\ell\pm 1$ (and viceversa).

%%%
\begin{figure}[thb]
\begin{center}
\epsfig{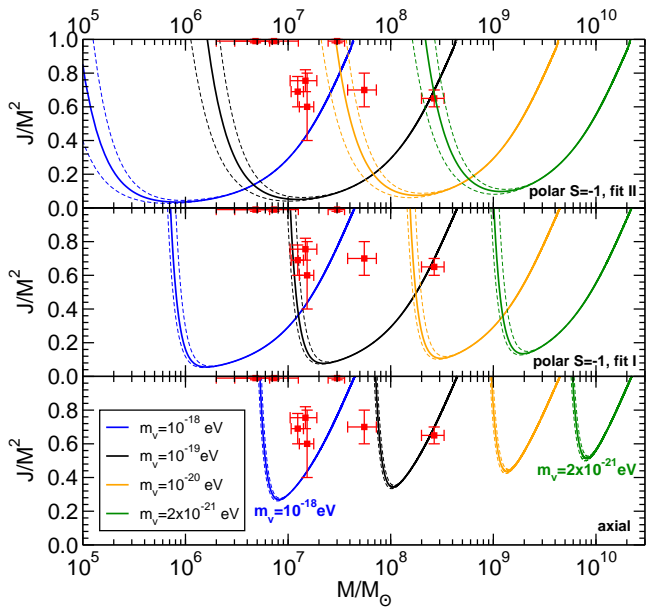}
\caption{(color online) Contour plots in the BH Regge plane~\cite{Arvanitaki:2010sy}
  corresponding to an instability timescale shorter than a typical
  accretion timescale, $\tau_{\rm Salpeter}=4.5\times 10^7$~yr, for
  different values of the vector field mass $m_v={{\mu}}\hbar$ (from left to right: $m_v=10^{-18}{\rm eV}$, $10^{-19}{\rm eV}$, $10^{-20}{\rm eV}$, $2\times10^{-21}{\rm eV}$). For
  polar modes we consider the $S=-1$ polarization, which provides the strongest instability, and we use both Eq.~\eqref{fitII} (fit II, top panel) and
  Eq.~\eqref{fit} (fit I, middle panel), and for axial modes we use
  Eq.~\eqref{fit} (bottom panel).  Dashed lines bracket our estimated
  numerical errors.  The experimental points (with error bars) refer
  to the mass and spin estimates of supermassive BHs listed in Table~2
  of~\cite{Brenneman:2011wz}; the rightmost point corresponds to the
  supermassive BH in Fairall~9~\cite{Schmoll:2009gq}. Supermassive BHs
  lying above each of these curves would be unstable on an observable
  timescale, and therefore they exclude a whole range of Proca field
  masses.\label{fig:ReggePlane} }
\end{center}
\end{figure}
In this paper we derive the Proca perturbation equations up to second
order in rotation, and for the first time we show that rotating BHs
are indeed unstable to massive vector perturbations in the superradiant regime. 
At first order in
rotation, perturbations with a given parity and angular index $\ell$
are coupled to perturbations with $\ell\pm1$ and opposite parity
(cf.~Eqs.~\eqref{pol1RDb}, \eqref{pol2RDb} and \eqref{axialRDb}
below).  However we show both analytically and numerically that this
coupling does not affect the eigenfrequencies of the BH. In addition,
the axial sector is described by a single master equation
[Eq.~(\ref{master}) below], which is valid for both massive scalar and
axial Proca perturbations. At second order in rotation the structure
of the equations remain the same, but the $\ell\pm1$ couplings do
affect the eigenfrequency spectrum.

Quasinormal frequencies, damping times of stable modes and instability
timescales of unstable modes in the slow-rotation limit can be
computed using standard methods
\cite{Leaver:1985ax,Berti:2009kk,Dolan:2007mj,Rosa:2011my}. As a test
of our approximation scheme we revisit two problems for which
``exact'' solutions are known in the Kerr background. We study QNMs of
massless vector perturbations to first order in rotation and we find
that our slow-rotation method reproduces known results
\cite{Berti:2009kk} with accuracy better than $1\%$ when the
dimensionless Kerr parameter $\tilde a=J/M^2 \lesssim 0.3$.  In
addition we compute stable and unstable \emph{bound-state} modes
(which are spatially localized within the vicinity of the BH and decay
exponentially at spatial infinity~\cite{Dolan:2007mj}) of massive
scalar perturbations up to second order in rotation, and we compare
our results to the numerical calculations by Dolan
\cite{Dolan:2007mj}. We find good quantitative agreement with
numerical results for nonsuperradiant frequencies, and we reproduce
the imaginary part of superradiantly unstable modes within a factor 3
for $\tilde a\lesssim0.7$, i.e. even for moderately large spin.

Extrapolations of our perturbative results indicate that, because of
the strong superradiant instability of Proca perturbations,
astrophysical BH spin measurements may set the most stringent
bounds on the masses of vector fields~\cite{paperPRL}.  This is shown in
Fig.~\ref{fig:ReggePlane}, where we summarize the astrophysical
implications of our results (cf. Sect.~\ref{sec:astro} for details).
To be more specific, in Fig.~\ref{fig:ReggePlane} we show exclusion
regions in the ``BH Regge plane'' (cf. Fig.~3
of~\cite{Arvanitaki:2010sy}), i.e. we plot contours corresponding to
an instability timescale of the order of a typical accretion timescale
(the Salpeter time $\tau_{\rm Salpeter}=4.5\times 10^7$~yr) for four
different masses of the Proca field ($m_v=10^{-18}$~eV, $10^{-19}$~eV,
$10^{-20}$~eV and $2\times10^{-21}$~eV) and for axial modes (bottom
panel) and polar modes (top and middle panels). 
For polar modes we consider only the $S=-1$ polarization, which provides the strongest instability (cf. Sec.~\eqref{res_num} for details).
While our numerical
results for the axial modes are supported by an analytical formula, in
the polar case we have used two different functions to fit the
numerical data at second order in the BH spin.  The plot shows that
essentially \emph{any} spin measurement for supermassive BHs with
$10^6M_\odot\lesssim M\lesssim 10^9M_\odot$ would exclude a wide range
of vector field masses.

The rest of the paper is organized as follows. In
Sec.~\ref{sec:general} we outline the general method, which is valid
for any stationary and axisymmetric background and for any kind of
perturbation.  In Sec.~\ref{sec:scalar} we study massive Klein-Gordon
perturbations of slowly rotating Kerr BHs up to second order in
rotation and outline how the method can be extended, at least in
principle, to higher orders.  In Sec.~\ref{sec:proca} we specialize to
Proca perturbations of slowly rotating Kerr BHs. In
Sec.~\ref{sec:num_methods} we set up the eigenvalue problem for
quasinormal modes and bound states, and in Sec.~\ref{sec:result} we
present our numerical results. Section~\ref{sec:astro} deals with the
implications of the superradiant instability for astrophysics and
fundamental physics. In Sec.~\ref{sec:conclusion} we draw some
conclusions and we discuss possible future applications of the method.
To improve readability we relegate some technical material to the
appendices. Appendix~\ref{app:coeffs} lists the coefficients appearing
in the scalar and Proca perturbation equations. Appendix
\ref{app:full} shows explicitly the perturbation equations for a Proca
field on a slowly rotating Kerr background to first order in rotation,
and appendix~\ref{res_ana} generalizes Detweiler's analytical
calculation of the unstable massive scalar modes of a Kerr
BH~\cite{Detweiler:1980uk} to axial perturbations of a massive vector
field up to linear order in rotation. 

To reduce the risk of typographical errors and facilitate comparison
with our results, we made many of our calculations available online as
{\em Mathematica} notebooks~\cite{url}.

%%%%%%%%%%%%%%%%%%%%%%%%%%%%%%%%%%%%%%%%%%%%%%%%%%%%%%
\section{Perturbations of a slowly rotating black hole: general method}
\label{sec:general}
%%%%%%%%%%%%%%%%%%%%%%%%%%%%%%%%%%%%%%%%%%%%%%%%%%%%%%
In this section we outline the strategy to study generic (scalar,
vector, tensor, etcetera) perturbations of \emph{any} stationary,
axisymmetric background up to second order in the rotation parameter,
but in principle the formalism can be extended to \emph{any} order.
We consider the most general stationary axisymmetric spacetime
%%%%
\begin{equation}
 ds^2_0=-H^2dt^2+Q^2dr^2+r^2K^2\left[d\vartheta^2+\sin^2\vartheta(d\varphi-Ldt)^2\right]\,,\nn
\end{equation}
where $H$, $Q$, $K$ and $L$ are functions of $r$ and $\vartheta$
only. To second order, the metric above can be expanded
as~\cite{Hartle:1967he}
%%%
\begin{align}
ds^2_0=&-F(r)\left[1+F_2\right]dt^2 +B(r)^{-1}\left[1+\frac{2B_2}{r-2M}\right]dr^2\nn\\
&+r^2(1+k_2)\left[d\th^2+\sin^2\th(d\varphi-\varpi dt)^2\right]	\,,
\label{metric2}
\end{align}
%%%%
where $M$ is the mass of the spacetime, $\varpi$ is a function of $r$
linear in the rotation parameter, and $F_2$, $B_2$ and $k_2$ are
functions of $r$ and $\th$ quadratic in the rotation parameter.
Because $H$, $K$ and $Q$ all transform like scalars under rotation,
they can be expanded in scalar spherical harmonics which, due to
axisymmetry, reduce to the Legendre polynomials $P_\ell(\th)$. As
shown in Ref.~\cite{Hartle:1967he}, only $\ell=0$ and $\ell=2$
polynomials contribute at second order in rotation; therefore $F_2$
can be expressed as $F_2(r,\th)=F_{r}(r)+F_{\th}(r)P_2(\th)$, and the
same applies to $B_2(r,\th)$ and $k_2(r,\th)$.
%%%%%

At first order in rotation the background~\eqref{metric2} can be
written in the simpler form
\begin{align}
ds^2_0=&-F(r)dt^2 +B(r)^{-1}dr^2+r^2d^2\Omega \nn\\
&-2\varpi(r)\sin^2\th d\varphi dt\,.
\label{metric1}
\end{align}
%%%%
Note that, given a nonrotating metric, the gyromagnetic function
$\varpi(r)$ can be computed using the approach originally developed by
Hartle~\cite{Hartle:1967he}.  For example, slowly rotating Kerr BHs
[cf. Eq.~\eqref{kerrmetric} below] correspond to $F(r)=B(r)=1-2M/r$
and $\varpi=2M^2\tilde a/r$, where $M$ and $J=M^2\tilde a$ are the
mass and the angular momentum of the BH. Furthermore, the general
metrics~\eqref{metric2} and ~\eqref{metric1} encompass (among others)
slowly rotating Kerr-Newman BHs and BH solutions in modified gravity
theories; the latter have been derived analytically only up to first
order~\cite{Yunes:2009hc,Pani:2011gy} (and more recently to second
order~\cite{Yagi:2012ya}) in the BH spin.

Scalar, vector and tensor field equations in the background
metric~\eqref{metric2} can be linearized in the field perturbations
(that we shall schematically denote by $\delta X$) as well as in the
BH angular momentum. We neglect terms of second order in the
perturbation amplitude $\delta X$ and terms of third order in the
rotation parameter $\tilde a$, Fourier transform the perturbations,
and expand them in tensor spherical harmonics:
\begin{equation}
\delta X_{\mu_1\dots}(t,r,\vartheta,\varphi)=
\delta X^{(i)}_{\ell m}(r){\cal Y}_{\mu_1\dots}^{\ell m\,(i)}e^{-\ii\omega t}\,,
\label{expa}
\end{equation}
where ${\cal Y}_{\mu_1\dots}^{\ell m\,(i)}$ is a basis of scalar,
vector or tensor harmonics (depending on the tensorial nature of the
perturbation $\delta X$) and the frequency $\omega$ is, in general,
complex.  The perturbation variables $\delta X^{(i)}_{\ell m}(r)$ can
be classified as ``polar'' or ``axial'' depending on their behavior
under parity transformations ($\vartheta\rightarrow\pi-\vartheta$,
$\varphi\rightarrow\varphi+\pi$): polar and axial perturbations are
multiplied by $(-1)^\ell$ and $(-1)^{\ell+1}$, respectively.

With the introduction of the expansion (\ref{expa}), the linear
response of the system is fully characterized by the quantities
$\delta X^{(i)}_{\ell m}(r)$. The perturbation equations, expanded in
spherical harmonics and Fourier transformed in time, yield a coupled
system of ODEs in the perturbation functions $\delta X^{(i)}_{\ell
  m}(r)$. In the case of a spherically symmetric background,
perturbations with different values of $(\ell,\,m)$, as well as
perturbations with opposite parity, are decoupled. In a rotating,
axially symmetric background, perturbations with different values of
$m$ are decoupled\footnote{From now on we will append the relevant
  multipolar index $\ell$ to any perturbation variable but we will
  omit the index $m$, because in an axisymmetric background 
it is possible to decouple the perturbation equations so that all quantities
have the same value of $m$.
} but perturbations with different values of
$\ell$ are not.
However, in the limit of slow rotation there is a Laporte-like
``selection rule''~\cite{ChandraFerrari91}: at first order in
$\tilde{a}$, perturbations with a given value of $\ell$ are only
coupled to those with $\ell\pm1$ and \emph{opposite} parity, similarly
to the case of rotating stars. At second order, perturbations with a
given value of $\ell$ are also coupled to those with $\ell\pm2$ and
\emph{same} parity, and so on.

In general, the perturbation equations can be written in the following
form:
\begin{eqnarray}
0&=&{\cal A}_{\ell}+\tilde a m \bar{\cal A}_{{\ell}}+\tilde{a}^2 \hat{{\cal A}}_\ell\nn\\
&+&\tilde a ({\cal Q}_{{\ell}}\tilde{\cal P}_{\ell-1}+{\cal Q}_{\ell+1}\tilde{\cal P}_{\ell+1})\nn\\
&+&\tilde{a}^2 \left[\cQ_\lm \cQ_\ell \breve{{\cal A}}_\lmm + \cQ_\lpp \cQ_\lp 
\breve{{\cal A}}_\lpp \right]+{\cal O}(\tilde{a}^3)\,,\nn\\\label{epF1c}\\
%%%%%
0&=&{\cal P}_{\ell}+\tilde a m \bar{\cal P}_{{\ell}}+\tilde{a}^2 \hat{{\cal P}}_\ell\nn\\
&+&\tilde a ({\cal Q}_{{\ell}}\tilde{\cal A}_{\ell-1}+{\cal Q}_{\ell+1}\tilde{\cal A}_{\ell+1})\nn\\
&+&\tilde{a}^2 \left[\cQ_\lm \cQ_\ell \breve{{\cal P}}_\lmm + \cQ_\lpp \cQ_\lp 
\breve{{\cal P}}_\lpp \right]+{\cal O}(\tilde{a}^3)\,.\nn\\\label{epF2c}
\end{eqnarray}
%%%%%
Here 
\begin{equation}
{\cal Q}_\ell=\sqrt{\frac{\ell^2-m^2}{4\ell^2-1}}
\,;\label{Qpm}
\end{equation}
%%%%
${\cal A}_{\ell}$, $\bar {\cal A}_{\ell}$, $\tilde {\cal A}_{\ell}$,
$\hat {\cal A}_{\ell}$, $\breve {\cal A}_{\ell}$ are \emph{linear}
combinations of the axial perturbations and of their derivatives, with
multipolar index $\ell$; similarly, ${\cal P}_{\ell}$, $\bar {\cal
  P}_{\ell}$, $\tilde {\cal P}_{\ell}$, $\hat {\cal P}_{\ell}$,
$\breve {\cal P}_{\ell}$ are linear combinations of the polar
perturbations and of their derivatives, with index $\ell$. The
dimensionless parameter $\tilde a$ keeps track of the order of the
various terms in the slow-rotation expansion.

The structure of Eqs.~\eqref{epF1c}--\eqref{epF2c} is the following.
Perturbations with a given parity and index $\ell$ are coupled to: (i)
perturbations with \emph{opposite} parity and index $\ell\pm1$ at
order $\tilde{a}$; (ii) perturbations with \emph{same} parity and
\emph{same} index $\ell$ up to order $\tilde{a}^2$; (iii)
perturbations with \emph{same} parity and index $\ell\pm2$ at order
$\tilde{a}^2$.

From Eq.~\eqref{Qpm} it follows that ${\cal Q}_{\pm m}=0$, and
therefore if $|m|=\ell$ the coupling of perturbations with index
$\ell$ to perturbations with indices $\ell-1$ and $\ell-2$ is
suppressed. Since the contribution $|m|=\ell$ dominates the linear
response of the system, this general property of
Eqs.~\eqref{epF1c}--\eqref{epF2c} is reminiscent of a ``propensity
rule'' in atomic theory, which states that transitions $\ell\to\ell+1$
are strongly favored over transitions $\ell\to\ell-1$
(cf.~\cite{ChandraFerrari91} and references therein).

Due to the coupling between different multipolar indices, the spectrum
of the solutions of Eqs.~\eqref{epF1c}--\eqref{epF2c} is extremely
rich. However, if we are interested in the characteristic modes of the
slowly rotating background to first or to second order in $\tilde{a}$,
the perturbation equations can be considerably simplified.  We discuss
the truncation to first and to second order in the next two sections,
respectively.
%%%%%%%%%%%%%%%%%%%%%%%%%
\subsection{Eigenvalue spectrum to first order}
%%%%%%%%%%%%%%%%%%%%%%%%%
Let us consider the first order expansion of Eqs.~\eqref{epF1c} and
\eqref{epF2c}. We expand all quantities to first order and we ignore
the terms $\hat {\cal A}_{\ell}$, $\breve {\cal A}_{\ell}$, $\hat
{\cal P}_{\ell}$ and $\breve {\cal P}_{\ell}$, which are multiplied by
$\tilde{a}^2$.  Furthermore the terms ($\tilde {\cal
  P}_{\ell},\,\tilde {\cal A}_{\ell}$) in Eqs.~\eqref{epF1c} and
\eqref{epF2c} do not contribute to the eigenfrequencies at first order
in $\tilde{a}$.  This was first shown by
Kojima~\cite{1993PThPh..90..977K} using symmetry arguments for the
axisymmetric case $m=0$. Here we extend his argument to any value of
$m$, for the generic set of equations~\eqref{epF1c} and
\eqref{epF2c}. Let us start by noting that, at first order,
Eqs.~\eqref{epF1c} and \eqref{epF2c} are invariant under the
simultaneous transformations
\begin{subequations}
\begin{eqnarray}
& a_{\ell m}\to\mp a_{\ell -m}\,,\qquad  p_{\ell m}\to \pm p_{\ell -m}\,,\\
& \tilde{a}\to-\tilde{a}\,,\qquad \hspace{1cm} m\to -m\,,
\end{eqnarray}
\end{subequations}
where $a_{\ell m}$ ($p_{\ell m}$) schematically denotes all of the
axial (polar) perturbation variables with indices $(\ell,m)$. The
invariance follows from the linearity of the terms in
Eqs.~\eqref{epF1c} and \eqref{epF2c} and from the fact that the ${\cal
  Q}_{\ell}$'s are \emph{even} functions of $m$. The boundary
conditions that define the characteristic modes of the BH are also
invariant under the transformation above (cf. Eqs.~\eqref{BChor} and
~\eqref{BCinf} below). Therefore in the slow-rotation limit the
eigenfrequencies can be expanded as
%
%%%
\begin{equation}
 \omega=\omega_0+m\,\omega_1 \tilde{a}+\omega_2 
\tilde{a}^2+{\cal O}(\tilde{a}^3)\,, \label{exp_omega}
\end{equation}
%%%
where $\omega_0$ is the eigenfrequency of the nonrotating spacetime
and $\omega_n$ is the $n$-th order correction (note that $\omega_1$
and $\omega_2$ are generically polynomials in $m$ but, due to the
above symmetry, $\omega_1$ is an \emph{even} polynomial).
Crucially, only the terms ($\bar {\cal P}_{\ell},\,\bar {\cal
  A}_{\ell}$) in Eqs.~\eqref{epF1c} and \eqref{epF2c} can contribute
to $\omega_1$. Indeed, due to the factor $\tilde{a}$ in front of all
terms ($\bar {\cal P}_{\ell},\,\bar {\cal A}_{\ell}$, $\tilde {\cal
  P}_{\ell},\,\tilde {\cal A}_{\ell}$) and to their linearity, at
first order in $\tilde{a}$ we can simply take the zeroth order (in
rotation) expansion of these terms. That is, to our level of
approximation the terms ($\bar {\cal P}_{\ell},\,\bar {\cal
  A}_{\ell}$, $\tilde {\cal P}_{\ell},\,\tilde {\cal A}_{\ell}$) in
Eqs.~\eqref{epF1c} and \eqref{epF2c} only contain the perturbations of
the \emph{nonrotating}, spherically symmetric background. Since the
latter do not explicitly depend on $m$, the $m$ dependence in
Eq.~\eqref{exp_omega} can only arise from the terms ($\bar {\cal
  P}_{\ell},\,\bar {\cal A}_{\ell}$) to zeroth order (recall that the
${\cal Q}_{\ell}$'s are even functions of $m$, so they do not
contribute).

In the case of slowly rotating stars, this argument is reinforced by
numerical simulations for the $m=0$ axisymmetric
modes~\cite{1993PThPh..90..977K} and in the general nonaxisymmetric
case~\cite{Ferrari:2007rc}, which show that the coupling terms do not
affect the first order corrections to the QNMs.  In this work we shall
verify that the same result holds also for massless vector
(e.g. electromagnetic) modes and for Proca modes of a Kerr BH.
If we are interested in the modes of the rotating background to ${\cal
  O}(\tilde{a})$ the coupling terms can therefore be neglected, and
the eigenvalue problem can be written in the form
\begin{eqnarray}
{\cal A}_{\ell}+\tilde a m \bar{\cal A}_{{\ell}}&=&0\,,\label{epF1}\\
{\cal P}_{\ell}+\tilde a m \bar{\cal P}_{{\ell}}&=&0\,.\label{epF2}
\end{eqnarray}
In these equations the polar and axial perturbations (as well as
perturbations with different values of the harmonic indices) are
decoupled from each other, and can be studied independently.

%%%%%%%%%%%%%%%%%%%%%%%%%
\subsection{Eigenvalue spectrum to second order}
%%%%%%%%%%%%%%%%%%%%%%%%%
For what concerns the calculation of the eigenfrequencies to second
order in $\tilde{a}$, the equations above can be simplified as
follows. To begin with, we remark that all of the coefficients in
Eqs.~\eqref{epF1c} and \eqref{epF2c} are \emph{linear} combinations of
axial or polar perturbation functions, $a_{\ell m}$, $p_{\ell m}$,
respectively. These functions can be expanded in the rotation
parameter as
\begin{eqnarray}
a_{\ell m}&=&a^{(0)}_{\ell m}+\tilde a\,a^{(1)}_{\ell m}+\tilde a^2a^{(2)}_{\ell m}\nn\\
p_{\ell m}&=&p^{(0)}_{\ell m}+\tilde a\,p^{(1)}_{\ell m}+\tilde a^2p^{(2)}_{\ell m}\,.
\end{eqnarray}
The terms $\breve{{\cal A}}_{\ell\pm2}$ and $\breve{{\cal
    P}}_{\ell\pm2}$ are multiplied by factors $\tilde{a}^2$, so they
only depend on the zeroth-order perturbation functions, $a^{(0)}_{\ell
  m}$, $p^{(0)}_{\ell m}$. The terms $\tilde{{\cal A}}_{\ell\pm1}$ and
$\tilde{{\cal P}}_{\ell\pm1}$ are multiplied by factors $\tilde{a}$,
so they only depend on zeroth- and first-order perturbation functions
$a^{(0)}_{\ell m}$, $p^{(0)}_{\ell m}$, $a^{(1)}_{\ell m}$,
$p^{(1)}_{\ell m}$.

Since in the nonrotating limit axial and polar perturbations are
decoupled, a possible consistent set of solutions of the system
(\ref{epF1c})--(\ref{epF2c}) has $a^{(0)}_{\ell\pm2 m}\equiv0$;
another consistent set of solutions of the same system has
$p^{(0)}_{\ell\pm2 m}\equiv0$. Such solutions, which we can call,
following Refs.~\cite{Lockitch:1998nq,Lockitch:2000aa}, ``axial-led''
and ``polar-led'' perturbations respectively, can be found by solving
the following subsets of the equations (\ref{epF1c})--(\ref{epF2c}):
\begin{eqnarray}
&&{\cal A}_{\ell}+\tilde a m \bar{\cal A}_{{\ell}}+\tilde{a}^2 \hat{{\cal  A}}_\ell+
\tilde a ({\cal Q}_{{\ell}}\tilde{\cal P}_{\ell-1}+
{\cal Q}_{\ell+1}\tilde{\cal P}_{\ell+1})=0\,,\nn\\\label{epF1bis1}\\
%%%%
&&{\cal P}_{\ell+1}+\tilde a m \bar{\cal P}_{{\ell+1}}+
\tilde a {\cal Q}_{{\ell+1}}\tilde{\cal A}_{\ell}=0\,,\label{epF1bis2}\\
&&{\cal P}_{\ell-1}+\tilde a m \bar{\cal P}_{{\ell-1}}+
\tilde a {\cal Q}_{{\ell}}\tilde{\cal A}_{\ell}=0\,,\label{epF1bis3}
\end{eqnarray}
%%%%%
and
\begin{eqnarray}
&&{\cal P}_{\ell}+\tilde a m \bar{\cal P}_{{\ell}}+\tilde{a}^2 \hat{{\cal P}}_\ell+
\tilde a ({\cal Q}_{{\ell}}\tilde{\cal A}_{\ell-1}+{\cal Q}_{\ell+1}
\tilde{\cal A}_{\ell+1})=0\,,\nn\\\label{epF2bis1}\\
%%%%
&&{\cal A}_{\ell+1}+\tilde a m \bar{\cal A}_{{\ell+1}}
+\tilde a {\cal Q}_{{\ell+1}}\tilde{\cal P}_{\ell}=0\,,\label{epF2bis2}\\
&&{\cal A}_{\ell-1}+\tilde a m \bar{\cal A}_{{\ell-1}}+
\tilde a {\cal Q}_{{\ell}}\tilde{\cal P}_{\ell}=0\,.\label{epF2bis3}
\end{eqnarray}
%%%%%
In the second and third equations of the two systems above we have
dropped the $\tilde{{\cal A}}_{\ell\pm2}$ and $\tilde{{\cal
    P}}_{\ell\pm2}$ terms, because these only enter at zeroth order,
and we have set $a^{(0)}_{\ell\pm2 m}\equiv0$ and $p^{(0)}_{\ell\pm2
  m}\equiv0$.

Interestingly, within this perturbative scheme a notion of ``conserved
quantum number'' $\ell$ is still meaningful, even though, for any
given $\ell$, rotation couples terms with opposite parity and
different multipolar index. In the most relevant case from the point
of view of superradiant instabilities, i.e. the case $\ell=m$, the
last equations~\eqref{epF1bis3} and~\eqref{epF2bis3} are automatically
satisfied, because ${\cal Q}_m=0$ and perturbations with $\ell<m$
automatically vanish.

It is worth stressing that, even though in principle there may be
modes which do not belong to the classes of ``axial-led'' or
``polar-led'' perturbations, all solutions belonging to one of these
classes which fulfill the appropriate boundary conditions defining
QNMs or bound states are also solutions of the full system
\eqref{epF1c}--\eqref{epF2c} and belong to the eigenspectrum (up to
second order in the rotation rate). This is particularly relevant
since, as we shall show, these classes contain also superradiantly
unstable modes.

In the case of massive scalar perturbations, as we show in the next
section, the coupling to $\ell\pm2$ can be eliminated by defining a
suitable linear combination of the eigenfunctions with different
multipolar indices [cf. Eq.~\eqref{Zell}], so that our procedure
allows us to compute the entire BH spectrum up to second order. 
By a similar, suitable ``rotation'' in eigenfunction space it may be possible to
recast Eqs.~\eqref{epF1c}--\eqref{epF2c} in the
forms~\eqref{epF1bis1}--\eqref{epF1bis3} and
~\eqref{epF2bis1}--\eqref{epF2bis3}, but we did not find a general
proof of this conjecture. If the conjecture is correct, studying the
``axial-led'' and ``polar-led'' systems may be sufficient to describe
the {\it entire} BH spectrum at second order in rotation.

To summarize, the eigenfrequencies (or at least a subset of the
eigenfrequencies) of the general system~\eqref{epF1c}--\eqref{epF2c}
can be found, at first order in $\tilde{a}$, by solving the two
decoupled sets~\eqref{epF1} and ~\eqref{epF2} for axial and polar
perturbations, respectively. At second order in $\tilde{a}$ we must
solve either the set~\eqref{epF1bis1}--\eqref{epF1bis3} or the set
~\eqref{epF2bis1}--\eqref{epF2bis3} for ``axial-led'' and
``polar-led'' modes, respectively.  We conclude this section by noting
that this procedure can be applied to any slowly rotating spacetime
and to any kind of perturbation. For concreteness, in the next section
we shall specialize the method to the case of massive scalar and
vector perturbations of the Kerr metric.
%%%
\section{Massive scalar perturbations of slowly rotating Kerr black holes}\label{sec:scalar}
%%%
In order to illustrate how the slow-rotation expansion works, in this
section we begin by working out the simplest case. We consider the
massive Klein-Gordon equation for a scalar field perturbation $\phi$
around a rotating BH, and we work out the perturbation equations up to
second order in rotation. The formalism applies to a generic
stationary and axisymmetric background, but it is natural to focus on
the case of interest in general relativity, i.e. the Kerr metric in
Boyer-Lindquist coordinates:
%%%%
%
\begin{eqnarray}
 &&ds_{\rm Kerr}^2=-\left(1-\frac{2Mr}{\Sigma}\right)dt^2
+\frac{\Sigma}{\Delta}dr^2-\frac{4rM^2}{\Sigma}\tilde{a}\sin^2\vartheta d\varphi dt   \nonumber\\
&&+{\Sigma}d\vartheta^2+
\left[(r^2+M^2\tilde{a}^2)\sin^2\vartheta +
\frac{2rM^3}{\Sigma}\tilde{a}^2\sin^4\vartheta \right]d\varphi^2\,,\nn\\
\label{kerrmetric}
\end{eqnarray}
where $\Sigma=r^2+M^2\tilde{a}^2\cos^2\vartheta$,
$\Delta=(r-r_+)(r-r_-)$ and $r_\pm=M(1\pm\sqrt{1-\tilde a^2})$.
In what follows we shall expand the metric and all other quantities of
interest to second order in $\tilde{a}$. At this order, the event
horizon $r_+$, the Cauchy horizon $r_-$ and the outer ergosphere
$r_{S^+}$ can be written in the form
%%%
\begin{eqnarray}
&& r_+=2M\left(1-\frac{\tilde{a}^2}{4}\right)\,,\quad r_-=\frac{M\tilde{a}^2}{2}\,, \nn\\
&&r_{S^+}=2M\left(1-\cos^2\th\frac{\tilde{a}^2}{4}\right)\,.
\label{cauchyhor}
\end{eqnarray}
%%%
In particular, note that second-order corrections are necessary for the
ergoregion to be located outside the event horizon.

The massive Klein-Gordon equation reads
\begin{equation}
\square\phi=\mu^2\phi\,,\label{KG}
\end{equation}
%%%
where $m_s=\mu\hbar$ is the mass of the scalar field. We decompose the
field in spherical harmonics:
%%%
\begin{equation}
 \phi=\sum_{\ell m}\frac{\Psi_\ell(r)}{\sqrt{r^2
+\tilde{a}^2 M^2}}e^{-\ii\omega t}Y^\ell(\vartheta,\varphi)\,,
\end{equation}
%%%
and expand the square root above to second order in $\tilde{a}$.
Schematically, we obtain the following equation:
\begin{equation}
 A_{\ell} Y^\ell+ D_\ell \cos^2\vartheta Y^\ell=0\,,\label{eq_expY}
\end{equation}
%%%
where a sum over $(\ell,m)$ is implicit, and the explicit form of
$A_\ell$ and $D_\ell$ is given in Appendix~\ref{app:coeff_scalar}.
The crucial point is that $D_\ell$ is proportional to $\tilde{a}^2$,
so the second term in the equation above is zero to first order in
rotation. If we consider a first-order expansion in rotation the
scalar equation is already decoupled, and it can be cast in the form
%%%
\begin{equation}
 \hat{\cal D}_2\Psi_\ell-\left[\frac{4m M^2
\tilde{a}\omega}{r^3}+F \frac{2M}{r^3}\right]\Psi_\ell=0\,,\label{scalar1st}
\end{equation}
where $F=1-2M/r$, $dr/dr_*=F$ and we defined the operator~\cite{Rosa:2011my}
\begin{equation}
 \hat{\cal D}_2=\frac{d^2}{d r_*^2}+\omega^2
-F\left[\frac{\ell(\ell+1)}{r^2}+\mu^2\right]\,.\label{D2}
\end{equation}
Equation~\eqref{scalar1st} coincides with Teukolsky's master
equation~\cite{Teukolsky:1973ha} for spin $s=0$ perturbations expanded
to first order in $\tilde{a}$.  The coupling to perturbations with
indices $\ell\pm1$ vanishes for a simple reason: Klein-Gordon
perturbations are polar quantities, and at first order the
Laporte-like selection rule implies that polar perturbations with
index $\ell$ should couple to axial perturbations with $\ell\pm1$, but
the latter are absent in the spin-0 case. At second order,
perturbations with harmonic index $\ell$ are coupled to perturbations
with the \emph{same} parity and $\ell\pm2$, but this coupling does not
contribute to the eigenfrequencies for the reasons discussed in the
previous section. Therefore we must solve a single scalar equation for
given values of $\ell$ and $m$, that we write schematically as
\begin{equation}
{\cal P}_{\ell}+\tilde a m \bar{\cal P}_{{\ell}}+\tilde{a}^2 \hat{{\cal P}_\ell}=0\,.\label{eqSCAL2}
\end{equation}
%%%

In order to confirm this general result, let us separate the angular part of
Eq.~\eqref{eq_expY}. This can be achieved by using the
identities~\cite{Kojima:1992ie}
\begin{eqnarray}
\cos\th Y^{\ell}&=&\cQ_{\ell+1}Y^{\ell+1}+\cQ_{\ell}Y^{\ell-1}\,,\label{ident1}\\
\sin\th \partial_\vartheta Y^{\ell}&=&
{\cal Q}_{\ell+1}\ell Y^{\ell+1}-{\cal Q}_{\ell}(\ell+1)Y^{\ell-1}\,,\label{ident2}\\
%%%%
\cos^2\th Y^{\ell}&=&\left(\cQ_\lp^2 + \cQ_\ell^2\right)Y^\ell\nn\\
&&+\cQ_\lp \cQ_\lpp Y^\lpp + \cQ_\ell \cQ_\lm Y^\lmm\,,\nn\\\label{ident3}\\
%%%%
\cos\th\sin\th \pa_\th Y^{\ell}&=&\left(\ell\cQ_\lp^2 -(\ell+1)\cQ_\ell^2\right)Y^\ell\nn\\
&&+\cQ_\lp \cQ_\lpp \ell Y^\lpp \nn\\
&&- \cQ_\ell \cQ_\lm (\ell+1) Y^\lmm\,,\label{ident4}
\end{eqnarray}
%%%
as well as the orthogonality property of scalar spherical harmonics:
%%%
\begin{equation}
 \int Y^{\ell} Y^{*\,\ell'}d\Omega=\delta^{\ell\ell'}\,.\label{orthoscalar}
\end{equation}
%%%
The result reads, schematically,
%%%%
\begin{eqnarray}
&& A_\ell+(\cQ_{\ell+1}^2+\cQ_{\ell}^2) D_\ell\nn\\
&&+\cQ_{\ell-1}\cQ_{\ell} D_{\ell-2}+\cQ_{\ell+2}
\cQ_{\ell+1} D_{\ell+2 }=0\label{final_schematic}\,.
\end{eqnarray}
%%%%

By repeated use of the identity~\eqref{ident1} we can separate the
perturbation equations at \emph{any order} in $\tilde{a}$. Indeed,
because of the expansion in $\tilde{a}$, only combinations of the form
$(\cos\vartheta)^n Y^\ell$ will appear. As we discuss in the next
section this is also true for spin-1 or spin-2 perturbations, except
that now the perturbation equations will contain combinations of
vector and tensor spherical harmonics, and this introduces terms such
as $(\sin\vartheta)^n \partial_\vartheta Y^\ell$, which can be
decoupled in a similar fashion by repeated application of the
identities listed above. This procedure is well known in quantum
mechanics, and the coefficients ${\cal Q}_{\ell}$ are related to the
usual Clebsch-Gordan coefficients.

Using the explicit form of the coefficients given in
Appendix~\ref{app:coeff_scalar}, the field
equations~\eqref{final_schematic} schematically read
%%%%
\begin{eqnarray}
\frac{d^2\Psi_\ell}{dr_*^2}+V_\ell \Psi_\ell+\tilde{a}^2&&
\Bigg[U_{\ell+2} \Psi_{\ell+2}+U_{\ell-2} \Psi_{\ell-2} \nn\\
&&+W_{\ell+2} \frac{d^2\Psi_{\ell+2}}{dr_*^2}+
W_{\ell-2} \frac{d^2\Psi_{\ell-2}}{dr_*^2}\Bigg]=0\,,\nn\\\label{final_tortoise}
\end{eqnarray}
%%%%
where we have defined the tortoise coordinate via $dr/dr_*\equiv
f=\Delta/(r^2+a^2)$ (expanded at second order) and $V$, $U$ and $W$
are some potentials, whose explicit form is not needed here.

Note that the coupling to the $\ell\pm2$ terms is proportional to
$\tilde{a}^2$. For a calculation accurate to second order in
$\tilde{a}$ the terms in parenthesis can be evaluated at zeroth order,
and therefore the functions $\Psi_{\ell\pm2}^{(0)}$ must be solutions
of
%%%
\begin{equation}
 \frac{d^2\Psi_{\ell\pm2}^{(0)}}{dr_*^2}+V_{\ell\pm2}^{(0)} \Psi_{\ell\pm2}^{(0)}=0\,.
\end{equation}
%%%
By substituting these relations in Eq.~\eqref{final_tortoise} we get
\begin{eqnarray}
 \frac{d^2\Psi_\ell}{dr_*^2}+V_\ell \Psi_\ell&&+\tilde{a}^2
\left(U_{\ell+2}^{(0)}-V_{\ell+2}^{(0)}W_{\ell+2}^{(0)}\right) \Psi_{\ell+2}^{(0)}\nn\\
&&+\tilde{a}^2\left(U_{\ell-2}^{(0)}-
V_{\ell-2}^{(0)}W_{\ell-2}^{(0)}\right)  \Psi_{\ell-2}^{(0)}=0\,.
\end{eqnarray}
Finally, making use of the expressions for $V$, $U$ and $W$, the field
equations can be reduced to
%%%%
\begin{eqnarray}
 &&\frac{d^2\Psi_\ell}{dr_*^2}+V_\ell \Psi_\ell=\frac{\tilde{a}^2 
M^2(r-2M)\left(\mu ^2-\omega ^2\right)}{r^3}\nn\\
&&\times \left[\cQ_{\ell+1} \cQ_{\ell+2} \Psi_{\ell+2}^{(0)}
+\cQ_{\ell-1} \cQ_{\ell} \Psi_{\ell-2}^{(0)}\right]\,,\label{final}
\end{eqnarray}
%%%
where the potential is given by
%%%
\begin{eqnarray}
 V_\ell&&=\omega^2-\left(1-\frac{2M}{r}\right)\left[
\frac{\ell(\ell+1)}{r^2}+\frac{2M}{r^3}+\mu^2\right]\nn\\
&&-\frac{4\tilde{a}m\omega M^2}{r^3}\nn\\
&&+\frac{\tilde{a}^2 M^2}{r^6}\left[-24 M^2-4 M r 
\left(\ell(\ell+1)-3+r^2 \mu ^2\right)\right.\nn\\
&&\left.+2 M r^3 \omega ^2+r^2 \left(\ell(\ell+1)+m^2
+r^2 (\mu^2 -\omega^2 ) -1\right)\right.\nn\\
&&\left.-r^3 (r-2M) (\mu^2 -\omega^2 ) \left(\cQ_{\ell}^2
+\cQ_{\ell+1}^2\right)\right]\,.\label{scalpot}
\end{eqnarray}
%%%%
As we previously discussed, the couplings to terms with indices
$\ell\pm2$ can be neglected in the calculation of the modes. In the
scalar case this can be shown explicitly as follows. If we define
%%%
\begin{equation}
 Z_\ell=\psi_\ell-\tilde{a}^2\left[c_{\ell+2}\psi_{\ell+2}-c_{\ell}\psi_{\ell-2}\right]\,, \label{Zell}
\end{equation}
%%%
where
\begin{equation}
c_{\ell}=\frac{M^2\left(\mu ^2-\omega ^2\right) \cQ_\lm \cQ_\ell}{2 (2\ell-1) }\,,
\end{equation}
%%%
then, at second order in rotation, Eq.~\eqref{final} can be written as
a single equation for $Z_\ell$:
%%%
\begin{equation}
 \frac{d^2 Z_\ell}{dr_*^2}+V_\ell Z_\ell=0\,,\label{final2}
\end{equation}
%%%
which can be solved by standard methods. This equation
coincides with Teukolsky's master equation~\cite{Teukolsky:1973ha} for
spin $s=0$ perturbations expanded at second order in $\tilde{a}$. This
is a nontrivial consistency check for the slow-rotation expansion. In
particular, the coefficients ${\cal Q}_{\ell}$ in Eq.~\eqref{scalpot}
agree with an expansion of Teukolsky's spheroidal eigenvalues to
second order in $\tilde{a}$~\cite{Berti:2005gp}. In fact, by extending
our procedure to arbitrary order in $\tilde{a}$ we can reconstruct the
Teukolsky scalar potential order by order. This can be viewed as an
independent check of the standard procedure, which consists of
expanding the angular equation to obtain the angular eigenfrequencies
(see~\cite{Berti:2005gp} and references therein).

In addition, as discussed in the previous section, the fact that
neglecting the $\ell\pm2$ couplings is equivalent to a field
redefinition implies that the entire BH spectrum of (massive scalar)
QNMs and bound states can be found by neglecting those couplings.

Finally, the near-horizon behavior of Eq.~\eqref{final2} reads
%%%
\begin{equation}
Z_\ell\sim e^{-\ii k_H r_*}\,,
\end{equation}
%%%%
where $k_H=\omega-m\Omega_H$ and $\Omega_H\sim \tilde{a}/(4M)+{\cal
  O}(\tilde{a}^3)$. Note that, by virtue of the second order
expansion, we get precisely $V_\ell\sim k_H^2$ close to the horizon.

The possibility to obtain a {\em single} equation for any given $\ell$
and $m$ is a special feature of scalar perturbations. The underlying
reason is that scalar perturbations have definite parity, so the
mixing between perturbations of different parity cannot occur. As we
show in the next section, this property does not necessarily hold for
perturbations of higher spin.
%%%%%%%%%%%%%%%%%%%%%%%%%%%%%%%%%%%%%%%%%%%%%%%%%%%%%%
\section{Massive vector perturbations of slowly rotating Kerr black holes}
\label{sec:proca}
%%%%%%%%%%%%%%%%%%%%%%%%%%%%%%%%%%%%%%%%%%%%%%%%%%%%%%  

The field equations of a massive vector field (also known as Proca's equation)
read
%%%
\begin{equation}
\Pi^\nu\equiv \nabla_\sigma F^{\sigma\nu}-\mu^2 A^\nu=0\,,\label{proca}
\end{equation}
%%%
where $F_{\mu\nu}=\partial_\mu A_\nu-\partial_\nu A_\mu$ and $A_\mu$
is the vector potential. Maxwell's equations are recovered when
$\mu=m_v/\hbar=0$, where $m_v$ is the mass of the vector field.  Note
that, as a consequence of Eq.~\eqref{proca}, the Lorenz condition
$\nabla_\mu A^\mu=0$ is automatically satisfied, i.e. in the massive
case there is no gauge freedom and the field $A_\mu$ propagates
$2s+1=3$ degrees of freedom~\cite{Rosa:2011my}.

In the $\mu=0$ case, Teukolsky showed that the equations for spin-$1$
perturbations around a Kerr BH are separable~\cite{Teukolsky:1973ha},
the angular part being described by spin-$1$ spheroidal harmonics. In
the case of a massive field the separation does not appear to be
possible, and one is left with a set of coupled partial differential
equations.  To avoid these difficulties we shall consider the
slow-rotation limit of the Kerr metric and apply the approach
described in Sec.~\ref{sec:general}.
The procedure is equivalent in spirit to that described for scalar
perturbations, but it is more involved due to the spin-1 nature of the
vector field.

\subsection{Harmonic expansion of the Proca equation on a slowly rotating background}

Following the notation of~\cite{Gerlach:1980tx,Ferrari:2007rc} we set
$x^{\mu}=(t,r,x^b)$ with $x^b=(\vartheta,\varphi)$. We also introduce
the metric of the two-sphere $\gamma_{ab}={\rm
  diag}(1,\sin^2\vartheta)$.

Any vector field can be decomposed in a set of vector spherical
harmonics~\cite{Gerlach:1980tx}
\begin{eqnarray}
\mathbf{Y}_b^{\ell}&=&\left(\partial_\vartheta Y^{\ell},\partial_\varphi Y^{\ell}\right)\,,\nonumber\\
\mathbf{S}_b^{\ell}&=&\left(\frac{1}{\sin\vartheta}\partial_\varphi Y^{\ell},
-\sin\vartheta\partial_\vartheta Y^{\ell}\right)\,,
\end{eqnarray}
where %$b\equiv(\vartheta,\varphi)$ and 
$Y^{\ell}(\vartheta,\varphi)$
are the scalar spherical harmonics. We expand the electromagnetic
potential as follows~\cite{Rosa:2011my}:
%%%%%%%%
\begin{equation}
\delta A_{\mu}(t,r,\vartheta,\varphi)=\sum_{\ell,m}\left[
 \begin{array}{c} 0 \\ 0\\
u_{(4)}^\ell \mathbf{S}_b^{\ell}/\Lambda\\
 \end{array}\right]+\sum_{\ell,m}\left[ \begin{array}{c}u_{(1)}^\ell Y^{\ell}/r\\u_{(2)}^\ell Y^{\ell}/(r f) \\
 u_{(3)}^\ell \mathbf{Y}_b^{\ell}/\Lambda\\ \end{array}\right],
\label{expansion_maxwell}
\end{equation}
%%%%%%%%%
where $\Lambda=\ell(\ell+1)$.
Because of their transformation properties under parity, the functions
$u_{(i)}^\ell$ belong to the \emph{polar} sector when $i=1,2,3$, and to the
\emph{axial} sector when $i=4$. In the nonrotating case the two sectors are
decoupled~\cite{Rosa:2011my}.
The Proca equation~\eqref{proca}, linearized in the perturbations $u_{(i)}^\ell$
($i=1,2,3,4$), can be written in the following form:
%%%
\begin{eqnarray}
 \delta\Pi_{I}&\equiv&  \left(A^{(I)}_{\ell}+{\tilde A}^{(I)}_{\ell}\cos\th
+D_\ell^{(I)}\cos^2\th\right)Y^\ell \nn\\
&&+\left(B^{(I)}_{\ell}+\tilde{B}^{(I)}_{\ell}\cos\th\right)
\sin\th\pa_{\th}Y^\ell=0\,,\label{eq1i}\\ 
%%%
 \delta\Pi_{\vartheta}&\equiv&   \left(\alpha_{\ell}+\rho_\ell\sin^2\th\right)
\pa_{\th}Y^\ell-\ii m\beta_\ell\frac{Y^\ell}{\sin\th}\nn\\
&&+\left(\eta_\ell+\sigma_\ell\cos\th\right)\sin\th Y^\ell=0\,, \label{eq2}\\
 \frac{\delta\Pi_{\varphi}}{\sin\th}&\equiv&  \left(\beta_{\ell}+
\gamma_\ell\sin^2\th\right)\pa_{\th}Y^\ell+\ii m\alpha_\ell\frac{Y^\ell}{\sin\th}\nn\\
&&+\left(\zeta_\ell+\lambda_\ell\cos\th\right)\sin\th Y^\ell=0\,, \label{eq3}
% && \beta_{\ell}\pa_{\th}Y^\ell+\ii m\alpha_{\ell}\frac{Y^\ell}{\sin\th}+\zeta_{\ell}\sin\th Y^\ell=0\,,\label{eq4}\nonumber\\
\end{eqnarray}
%%%%
where a sum over $(\ell,m)$ is implicit and $I$ denotes either the $t$
component or the $r$ component.  The various radial coefficients in
the equations above are given in terms of the perturbation functions
$u_{(i)}^\ell$ in Appendix~\ref{app:coeff}.

The perturbation equations~\eqref{eq1i}--\eqref{eq3} can be simplified
by using the Lorenz identity $\nabla_\mu A^\mu=0$. To second order in
rotation, and for the background metric~\eqref{kerrmetric}, this
condition reads
%%%
\begin{eqnarray}
\delta\Pi_{L}&\equiv&\left(A^{(2)}_{\ell}+{\tilde A}^{(2)}_{\ell}\cos\th+D_\ell^{(2)}\cos^2\th\right)Y^\ell \nn\\
&&+\left(B^{(2)}_{\ell}+\tilde{B}^{(2)}_{\ell}\cos\th\right)\sin\th\pa_{\th}Y^\ell=0\,,\label{eqL}
\end{eqnarray}
%%%
where the various coefficients are again listed in
Appendix~\ref{app:coeff}.

Each of the coefficients in Eqs.~\eqref{eq1i}--\eqref{eqL} is a linear
combination of perturbation functions with either polar or axial
parity (cf.~Appendix~\ref{app:coeff}). Therefore we can divide them
into two sets:
\begin{eqnarray}
 &&\text{Polar:}\qquad A^{(j)}_{\ell}\,,\quad \alpha_{\ell}\,,\quad \zeta_{\ell}\,,
\quad \tilde{B}_{\ell}\,,\quad D_{\ell}\,,\quad \rho_{\ell}\,,\quad \sigma_{\ell}\,,\nn\\
%%%
 &&\text{Axial:}\qquad \tilde A^{(j)}_{\ell}\,,\quad B^{(j)}_{\ell}\,,\quad 
\beta_{\ell}\,,\quad \eta_{\ell}\,,\quad \lambda_{\ell}\,,\quad \gamma_{\ell}\,,\nn
\end{eqnarray}
%%%
where $j=0,1,2$.  In order to separate the angular variables in
Eqs.~\eqref{eq1i}--\eqref{eqL} we compute the following integrals:
\begin{subequations}
\begin{align}
&\int\delta\Pi_I Y^{*\,{\ell}}d\Omega\,,\quad (I=t,r,L)\,;\\
&\int\delta\Pi_a \mathbf{Y}_b^{*\,{\ell}}\gamma^{ab}d\Omega\,,\quad (a\,,b=\vartheta,\varphi)\,;\\
&\int\delta\Pi_a \mathbf{S}_b^{*\,{\ell}}\gamma^{ab}d\Omega\,,\quad (a\,,b=\vartheta,\varphi)\,.
\end{align}
\end{subequations}
Using the orthogonality properties of scalar and vector harmonics,
Eq.~\eqref{orthoscalar}, the relations
\begin{align}
%&\int Y^{\ell} Y^{*\,\ell'm'}d\Omega=\delta^{\ell\ell'}\delta^{mm'}\,,\nn\\
&\int \mathbf{Y}_b^{\ell} \mathbf{Y}_b^{*\,\ell'}\gamma^{ab} d\Omega=\int \mathbf{S}_b^{\ell} 
\mathbf{S}_b^{*\,\ell'}\gamma^{ab} d\Omega=\Lambda\delta^{\ell\ell'}\,,\nn\\
&\int \mathbf{Y}_b^{\ell} \mathbf{S}_b^{*\,\ell'}\gamma^{ab} d\Omega=0\,,
\end{align}
as well as the identities~\eqref{ident1}--\eqref{ident4},
% %%
%
%%
we find the following \emph{radial} equations:
\begin{eqnarray}
&& A_\ell^{(I)}+\cQ_{\ell+1}^2\left[D_\ell^{(I)}+\ell \tilde{B}_\ell^{(I)}
\right]+\cQ_{\ell}^2\left[D_\ell^{(I)}-(\ell+1) \tilde{B}_\ell^{(I)}\right]+\nn\\
&& \cQ_\ell\left[\tilde{A}_{\ell-1}^{(I)}+(\ell-1)B_{\ell-1}^{(I)}\right]
+\cQ_{\ell+1}\left[\tilde{A}_{\ell+1}^{(I)}-(\ell+2)B_{\ell+1}^{(I)}\right]+\nn\\
&& \cQ_{\ell-1}\cQ_\ell\left[D_{\ell-2}^{(I)}+(\ell-2)\tilde{B}_{\ell-2}^{(I)}\right]\nn\\
&&+\cQ_{\ell+2}\cQ_{\ell+1}\left[D_{\ell+2}^{(I)}-(\ell+3)\tilde{B}_{\ell+2}^{(I)}
\right]=0\,,\label{eqset1sec}\\
%%%%%%
&& \Lambda \alpha_\ell-\ii m\zeta_\ell+\cQ_\lp^2\ell\left[\ell\rho_\ell
+\sigma_\ell\right]+\cQ_\ell^2(\ell+1)\left[(\ell+1)\rho_\ell-\sigma_\ell\right]\nn\\
&&+\cQ_\ell\left[-(\ell+1)\eta_\lm-\ii m((\ell-1)\gamma_\lm+\lambda_\lm)\right]\nn\\
&&+\cQ_\lp\left[\ell\eta_\lp+\ii m((\ell+2)\gamma_\lp-\lambda_\lp)\right]\nn\\
&&-\cQ_\lm \cQ_{\ell}(\ell+1)\left[(\ell-2)\rho_\lmm+\sigma_\lmm\right]\nn\\
&&+\cQ_\lpp \cQ_\lp \ell\left[-(\ell+3)\rho_\lpp+\sigma_\lpp\right]=0\,,\label{eqset2sec} \\
%%%%%%
&& \Lambda \beta_\ell+\ii m\eta_\ell+\cQ_\lp^2\ell\left[\ell\gamma_\ell+
\lambda_\ell\right]+\cQ_\ell^2(\ell+1)\left[(\ell+1)\gamma_\ell-\lambda_\ell\right]\nn\\
&&+\cQ_\ell\left[-(\ell+1)\zeta_\lm+\ii m((\ell-1)\rho_\lm+\sigma_\lm)\right]\nn\\
&&+\cQ_\lp\left[\ell\zeta_\lp-\ii m((\ell+2)\rho_\lp-\sigma_\lp)\right]\nn\\
&&-\cQ_\lm \cQ_{\ell}(\ell+1)\left[(\ell-2)\gamma_\lmm+\lambda_\lmm\right]\nn\\
&&+\cQ_\lpp \cQ_\lp \ell\left[-(\ell+3)\gamma_\lpp+\lambda_\lpp\right]=0\,.\label{eqset3sec}
\end{eqnarray}
%%%%%
Note that Eqs.~\eqref{eqset1sec}--\eqref{eqset3sec} have exactly the
same structure as Eqs.~\eqref{epF1c}--\eqref{epF2c}.

%%%%%%%%%%%%%%
\subsection{Proca perturbation equations at first order}
%%%%%%%%%%%%%
%
In order to make the equations more tractable, in this section we
focus on the first-order corrections only. The second-order analysis
is presented in Sec.~\ref{sec:Proca2nd} below. At first order,
Eqs.~\eqref{eqset1sec}--\eqref{eqset3sec} simplify to
\begin{align}
&A^{(I)}_{{\ell}}+\cQ_{{\ell}}\left[{\tilde A}^{(I)}_{\ell-1}+(\ell-1){B}^{(I)}_{\ell-1}\right]\nn\\
&+\cQ_{\ell+1}\left[{\tilde A}^{(I)}_{\ell+1}
-(\ell+2){ B}^{(I)}_{\ell+1}\right]=0\,,\label{eqset1}\\
&\Lambda\alpha_{\ell}-\ii m\zeta_{\ell}\nn\\
&-{\cal Q}_{\ell}(\ell+1)\eta_{\ell-1}+{\cal Q}_{\ell+1}\ell \eta_{\ell+1}=0\,,\label{eqset2}\\
&\Lambda\beta_{\ell}+\ii m\eta_{\ell}\nn\\
&-{\cal Q}_{\ell}(\ell+1)\zeta_{\ell-1}+{\cal Q}_{\ell+1}\ell \zeta_{\ell+1}=0\,,\label{eqset3}
\end{align}
where now the coefficients are linear in $\tilde a$.

To begin with, let us focus on the equations for monopole perturbations,
$\ell=m=0$.  The longitudinal mode of a massive vector field (unlike the
massless case) is dynamical. Since $m=0$, the monopole only excites axisymmetric
modes.  In the nonrotating case these are described by a single equation
belonging to the polar sector~\cite{Konoplya:2005hr} (see
also~\cite{Rosa:2011my}); for $\ell=0$, only the first two components
$u_{(1)}^{0}$ and $u_{(2)}^{0}$ are defined. However, in the slowly rotating
case these components are coupled to the $\ell=1$ axial component $u_{(4)}^{1}$
through Eq.~\eqref{eqset1}.

When $\ell=m=0$, $\cQ_{0}=0$ and Eq.~\eqref{eqset1} reduces to
\begin{equation}
A^{(I)}_{0}+\cQ_{1}\left[{\tilde A}^{(I)}_{1}-2{ B}^{(I)}_{1}\right]=0\,,\label{epmonopole} 
\end{equation}
where $I=0,1,2$. This is an extreme example of the ``propensity rule'' discussed
in the general case: at first order the monopole is only coupled with axial
perturbations with $\ell=1$.
Using the explicit form of the coefficients $A^{(I)},{\tilde A}^{(I)},B^{(I)}$
given in Appendix~\ref{app:coeff} (truncated at first order), the equations
above can be written as a single equation for $u_{(2)}^{0}$:
\begin{align}
&
\left[\frac{d^2}{dr_*^2}+\omega^2-F\left(\frac{2(r-3M)}{r^3}+\mu^2\right)\right]u_{(2)}^{0}\nn\\
&=\frac{2\ii\sqrt{3} \tilde aM^2\omega F}{r^3}u_{(4)}^{1}\,,
\label{monopole1}
\end{align}
%%%
where at first order the tortoise coordinate $r_*$ is the same as in the
Schwarzschild case, and it is defined by $dr/dr_*=F$. The source term
$u_{(4)}^{1}$ is the solution of Eq.~\eqref{eqset1} with $\ell=1$. To first
order in $\tilde a$ we can approximate $u_{(4)}^{1}$ (that is multiplied by
$\tilde a$ in the source term) by its zeroth-order expansion in $\tilde a$,
which is a solution of
%%%
\begin{equation}
 \left\{\frac{d^2}{d r_*^2}+\omega^2-F\left[\frac{2}{r^2}
+\mu^2\right]\right\} u_{(4)}^{1}=0\,.\label{monopole0}
\end{equation}
%%%%
Note that Eq.~\eqref{monopole0} is precisely the axial perturbation equation in
the nonrotating case for $\ell=1$~\cite{Rosa:2011my}. Eqs.~\eqref{monopole1}
and \eqref{monopole0} fully describe the dynamics of the polar $\ell=0$
perturbations in the slow-rotation approximation.  As we proved in
Sec.~\ref{sec:general}, the coupling on the right-hand side of
Eq.~\eqref{monopole1} does not affect the QNMs to first order in rotation. In
addition, since for the monopole $m=0$, to this order the frequency is the same
as in the Schwarzschild case, which is extensively discussed in
Ref.~\cite{Rosa:2011my}.

Let us now turn to modes with $\ell>0$. The equations for $\ell>0$ at first
order in $\tilde{a}$ are derived in Appendix~\ref{app:full} by using the Lorenz
condition~\eqref{lorenz_full} in order to eliminate $u_{(1)}^\ell$.  The polar
sector is fully described by the system
\begin{align}
& \hat{\cal D}_2
 u_{(2)}^\ell-\frac{2F}{r^2}\left(1-\frac{3M}{r}\right)\left[u_{(2)}^\ell-u_{(3)}^\ell
\right]\nn\\
&= \frac{2\tilde aM^2 m }{\Lambda r^5 \omega } \left[\Lambda  
\left(2 r^2 \omega ^2+3 F^2\right) u_{(2)}^\ell\right.\nn\\
&\left.+3 F \left(r \Lambda  F {u'}_{(2)}^\ell-
\left(r^2 \omega ^2+\Lambda  F\right) u_{(3)}^\ell\right)\right]\nn\\
&-\frac{6\ii\tilde{a} M^2F\omega}{\Lambda r^3}\left[(\ell+1){\cal Q}_{\ell}
 u_{(4)}^{\ell-1}-\ell {\cal Q}_{\ell+1 }u_{(4)}^{\ell+1}\right]\,,\label{pol1RDb}\\
& \hat{\cal D}_2 u_{(3)}^\ell+\frac{2F\Lambda}{r^2}u_{(2)}^\ell=\nn\\
&\frac{2 \tilde aM^2 m }{r^5 \omega }\left[2 r^2 
\omega ^2 u_{(3)}^\ell+3 r F^2 {u'}_{(3)}^\ell-3 
\left(\Lambda+r^2 \mu ^2\right) F u_{(2)}^\ell\right]\,,\label{pol2RDb}
\end{align}
%%%
where here and in the following a prime denotes derivation with respect to $r$ and the operator $\hat{\cal D}_2$ is defined in Eq.~\eqref{D2}.
Note that while Eq.~\eqref{pol2RDb} only involves polar perturbations, in
Eq.~\eqref{pol1RDb} we also have a coupling to $u_{(4)}^{\ell\pm1}$.

On the other hand, the axial sector leads to
%%%
\begin{eqnarray}
&& \hat{\cal D}_2 u_{(4)}^\ell-\frac{4 \tilde aM^2 m\omega}{r^3}u_{(4)}^\ell\nn\\
&&=-\frac{6 \ii \tilde a M^2 F}{r^5 \omega }\left[(\ell+1) {\cal Q}_{\ell m} 
\psi^{\ell-1} -\ell {\cal Q}_{\ell+1m} \psi^{\ell+1}\right] \label{axialRDb}
\end{eqnarray}
%%%
(see Appendix~\ref{app:full} for details), where we have defined the polar
function
%%%%%
\begin{equation}
 \psi^\ell=\left(\Lambda+r^2\mu ^2\right)u_{(2)}^\ell-(r-2M){u'}_{(3)}^\ell\,.
\end{equation}
%%%%%
Note that this term is similar to Eq.(25) in Ref.~\cite{Rosa:2011my}.
As expected, the axial perturbation $u_{(4)}^\ell$ is coupled to the polar
functions with $\ell\pm1$.  Equations~\eqref{pol1RDb}, \eqref{pol2RDb}
and~\eqref{axialRDb} describe the massive vector perturbations of a Kerr
BH to first order in $\tilde a$ for $\ell>0$. These equations reduce to those in
Ref.~\cite{Rosa:2011my} when $\tilde a=0$.

Since the right-hand side of Eq.~\eqref{axialRDb} is proportional to ${\tilde
  a}$, in our perturbative framework we can \emph{first} solve for
$u_{(2)}^{\ell\pm1}$ and $u_{(3)}^{\ell\pm1}$ to zeroth order in the rotation
rate, and then use these solutions as a source term in Eq.~\eqref{axialRDb}.  As
we proved in Sec.~\ref{sec:general}, the coupling on the right-hand side of
Eq.~\eqref{axialRDb} does not affect the QNMs to first order in rotation. In
Sec.~\ref{sec:result} we shall verify this property numerically. Therefore, for the
purpose of computing eigenfrequencies to first order in $\tilde a$ we can simply
consider the following polar equations:
\begin{align}
& \hat{\cal D}_2
 u_{(2)}^\ell-\frac{2F}{r^2}\left(1-\frac{3M}{r}\right)\left[u_{(2)}^\ell-u_{(3)}^\ell
\right]\nn\\
&= \frac{2\tilde aM^2 m }{\Lambda r^5 \omega } \left[\Lambda  
\left(2 r^2 \omega ^2+3 F^2\right) u_{(2)}^\ell\right.\nn\\
&\left.+3 F \left(r \Lambda  F {u'}_{(2)}^\ell-
\left(r^2 \omega ^2+\Lambda  F\right) u_{(3)}^\ell\right)\right]\,,\label{pol1RDbb}\\
& \hat{\cal D}_2 u_{(3)}^\ell+\frac{2F\Lambda}{r^2}u_{(2)}^\ell=\nn\\
&\frac{2 \tilde aM^2 m }{r^5 \omega }\left[2 r^2 \omega ^2 u_{(3)}^\ell+3 r F^2 {u'}_{(3)}^\ell-3 
\left(\Lambda+r^2 \mu ^2\right) F u_{(2)}^\ell\right]\,,\label{pol2RDbb}
\end{align}
as well as the decoupled axial equation
%%%
\begin{equation}
 \hat{\cal D}_2 u_{(4)}^\ell-\frac{4 \tilde aM^2 m\omega}{r^3}u_{(4)}^\ell=0\,.\label{axialRD}
\end{equation}
%%%
Within our perturbative scheme, any eigenfrequency of Eq.~\eqref{axialRD} is
also a solution of the coupled Eq.~\eqref{axialRDb} as long as
$u_{(2)}^{\ell\pm1}=u_{(3)}^{\ell\pm1}=0$, which is a trivial solution of
Eqs.~\eqref{pol1RDb} and~\eqref{pol2RDb} for $\ell\pm1$ at zeroth
order. Furthermore, consistently with our argument in Sec.~\ref{sec:general}, we
have checked that there are no other modes to order ${\cal O}(\tilde a)$
(cf. Sec.~\ref{sec:result}).

Note the similarity between Eq.~\eqref{axialRD}, describing axial Proca modes,
and Eq.~\eqref{scalar1st} for massive scalar perturbations.  Indeed, one can
show that the generalization of Eq.~\eqref{axialRD} to the background
metric~\eqref{metric1} (i.e. without specializing to the slowly rotating Kerr
metric) can be written in a form that includes also massive \emph{scalar}
perturbations. This ``master equation'' reads
\begin{eqnarray}
&& F B\Psi_{\ell}''+\frac{1}{2}\left[B'F+F'B\right]\Psi_{\ell}'
+\left[\omega^2-\frac{2 m \varpi(r) \omega}{r^2}\right.\nn\\
&&\left.-F\left(\frac{\Lambda}{r^2}+\mu^2+(1-s^2)
\left\{\frac{B'}{2r}+\frac{B F'}{2r F}\right\}\right)\right]\Psi_{\ell}=0\,,\nn\\\label{master}
\end{eqnarray}
and it can be simplified by introducing a generalized tortoise coordinate $y(r)$
such that $dr/dy=\sqrt{FB}$. In the equation above $s$ is the spin of the
perturbation ($s=0$ for scalar perturbations and $s=\pm1$ for vector
perturbations with axial parity). In the nonrotating case Eq.~\eqref{master} is
exact; it also includes gravitational perturbations of a Schwarzschild BH if
$s=\pm 2$ and $F=B=1-2M/r$. In the slowly rotating case Eq.~\eqref{master} is a
complete description of massive scalar perturbations for $s=0$, whereas for
$s=\pm1$ it describes the axial sector without the $\ell\to\ell\pm1$ couplings,
i.e. it is a generalization of Eq.~\eqref{axialRD}.

%%%%%%%%%%%%%%
\subsection{Proca perturbation equations at second order}\label{sec:Proca2nd}
%%%%%%%%%%%%%
In this section we briefly present the derivation of the Proca eigenvalue
problem to second order in $\tilde{a}$.

By using the Lorenz condition to eliminate the spurious $u_{(1)}^\ell$ mode,
Eqs.~\eqref{eqset1sec}--\eqref{eqset3sec} can be written as
%%%%
\begin{eqnarray}
 \mathbf{{\cal D}_A}\mathbf{\Psi_A}^\ell+\mathbf{V_A}\mathbf{\Psi_A}^\ell&=&0\,,\label{systA0}\\
 \mathbf{{\cal D}_P}\mathbf{\Psi_P}^\ell+\mathbf{V_P}\mathbf{\Psi_P}^\ell&=&0\,,\label{systP0}
\end{eqnarray}
%%%
where $\mathbf{{\cal D}_{A,P}}$ are second order differential
operators, $\mathbf{V_{A,P}}$ are matrices,
$\mathbf{\Psi_A}^\ell=(u_{(4)}^\ell,u_{(2)}^{\ell\pm1},u_{(3)}^{\ell\pm
  1},u_{(4)}^{\ell\pm2})$ and
$\mathbf{\Psi_P}^\ell=(u_{(2)}^\ell,u_{(3)}^\ell,u_{(4)}^{\ell\pm
  1},u_{(2)}^{\ell\pm 2},u_{(3)}^{\ell\pm 2})$. The explicit form of
the equations above is 
quite lengthy; therefore we do not show it in this article, but we make it
available online~\cite{url}. It can be obtained
using the procedure explained above and the coefficients listed in
Appendix~\ref{app:coeff}.  The function $u_{(1)}^{\ell}$ can be
obtained from the Lorenz condition once the three dynamical degrees of
freedom are known. Note that Eqs.~\eqref{systA}-\eqref{systP} are
particular cases of Eqs.~\eqref{epF1bis1}-\eqref{epF1bis3} and
Eqs.~\eqref{epF2bis1}-\eqref{epF2bis3}, respectively.

If we are interested in the eigenfrequencies up to second order in $\tilde{a}$
we can drop the couplings to $\ell\pm2$ perturbations, for reasons discussed in
Sec.~\ref{sec:general}. Therefore, a consistent subset of
Eqs.~\eqref{eqset1sec}--\eqref{eqset3sec} reads
%%%
%%%%
\begin{eqnarray}
  0&=& A_\ell^{(i)}+\cQ_{\ell+1}^2\left[D_\ell^{(i)}+\ell \tilde{B}_\ell^{(i)}\right]\nn\\
  &&+\cQ_{\ell}^2\left[D_\ell^{(i)}-(\ell+1) \tilde{B}_\ell^{(i)}\right]
  +\cQ_\ell\left[\tilde{A}_{\ell-1}^{(i)}+(\ell-1)B_{\ell-1}^{(i)}\right]\nn\\
  &&+\cQ_{\ell+1}\left[\tilde{A}_{\ell+1}^{(i)}-(\ell+2)B_{\ell+1}^{(i)}\right]\,,\nn\\
%%%%%
0&=&\Lambda \alpha_\ell-\ii m\zeta_\ell+\cQ_\lp^2\ell\left[\ell\rho_\ell+\sigma_\ell\right]\nn\\
&&+\cQ_\ell^2(\ell+1)\left[(\ell+1)\rho_\ell+\sigma_\ell\right]\nn\\
&&+\cQ_\ell\left[-(\ell+1)\eta_\lm-\ii m((\ell-1)\gamma_\lm+\lambda_\lm)\right]\nn\\
&&+\cQ_\lp\left[\ell\eta_\lp+\ii m((\ell+2)\gamma_\lp-\lambda_\lp)\right]\,,\nn\\
%%%%%%
0&=& \Lambda \beta_\ell+\ii m\eta_\ell+\cQ_\lp^2\ell\left[\ell\gamma_\ell+\lambda_\ell\right]\nn\\
&&+\cQ_\ell^2(\ell+1)\left[(\ell+1)\gamma_\ell+\lambda_\ell\right]\nn\\
&&+\cQ_\ell\left[-(\ell+1)\zeta_\lm+\ii m((\ell-1)\rho_\lm+\sigma_\lm)\right]\nn\\
&&+\cQ_\lp\left[\ell\zeta_\lp-\ii m((\ell+2)\rho_\lp-\sigma_\lp)\right]\,.\label{eqsec}
\end{eqnarray}
%%%
%
As in the scalar case, the second-order coefficients generally contain second
derivatives of the perturbation functions, i.e.  ${u''}_{(i)}^\ell$,
${u''}_{(i)}^{\ell\pm1}$ and ${u''}_{(i)}^{\ell\pm2}$. Since the coefficients
are already of second order, we can use the perturbation equations {\em in the
  nonrotating limit} in order to eliminate these second derivatives.  After
some manipulation we get the final two sets of equations, that can be
schematically written as
%%%%
\begin{eqnarray}
 \mathbf{{{\cal D}}'_A}\mathbf{\Upsilon_A}^\ell+\mathbf{{V'}_A}\mathbf{\Upsilon_A}^\ell&=&0\,,\label{systA}\\
 \mathbf{{{\cal D}'}_P}\mathbf{\Upsilon_P}^\ell+\mathbf{{V'}_P}\mathbf{\Upsilon_P}^\ell&=&0\,,\label{systP}
\end{eqnarray}
%%%
where $\mathbf{{{\cal D}'}_{A,P}}$ are second order differential operators,
$\mathbf{\Upsilon_A}^\ell=(u_{(4)}^\ell,u_{(2)}^{\ell\pm1},u_{(3)}^{\ell\pm 1})$,
$\mathbf{\Upsilon_P}^\ell=(u_{(2)}^\ell,u_{(3)}^\ell,u_{(4)}^{\ell\pm 1})$, and $\mathbf{{V'}_{A,P}}$ are
matrices.

%In general, 
We remark that
for a given value of $\ell$ and $m$, $\mathbf{\Upsilon_A}$
and $\mathbf{\Upsilon_P}$ are five- and four-dimensional vectors,
respectively,
while
%.  This result must be compared with 
$\mathbf{\Psi_A}$ and $\mathbf{\Psi_P}$ in Eqs.~\eqref{systA0} and \eqref{systP0}
%which 
are seven- and eight-dimensional vectors, respectively. This is a very
convenient simplification, because large systems of equations are
computationally more demanding.  Furthermore the couplings to
perturbations with index $\ell-1$ vanish if $\ell=m$, because as
usual the factors ${\cal Q}_m=0$, and in this case we are left with
two subsystems of dimension three.

%

%%%%%%%%%%%%%%%%%%%%%%%%%%%%%%%%%%%%%%%%%%%%%%%%%%%%%%%
\section{The eigenvalue problem for quasinormal modes and bound states}
\label{sec:num_methods}
%%%%%%%%%%%%%%%%%%%%%%%%%%%%%%%%%%%%%%%%%%%%%%%%%%%%%%
One of the key advantages of the slow-rotation approximation is that the
perturbation equations can be solved using well-known numerical
approaches. The integration proceeds exactly in the same way
as in the nonrotating case. For example, for Proca perturbations of a slowly
rotating Kerr BH we can compute the characteristic modes by following a
procedure similar to the Schwarzschild case discussed in
Ref.~\cite{Rosa:2011my}, as follows.

The perturbation equations~\eqref{systA}--\eqref{systP}, together with
appropriate boundary conditions at the horizon and at infinity, form an
eigenvalue problem for the frequency spectrum.  In the near-horizon limit,
$\mathbf{{\cal D}_{A,P}}\to d^2/d r_*^2$ and $\mathbf{V_{A,P}}\to(\omega-m\Omega_H)^2$, so that at
the horizon we impose purely ingoing-wave boundary conditions:
%%%
\begin{equation}
 u_{(i)}^\ell\sim u_{(i)H}^\ell e^{-\ii k_H r_*}\,,\label{BChor}
\end{equation}
%%%
where $u_{(i)H}^\ell$ is a constant and
\begin{equation}
 k_H=\omega-m\Omega_H= \omega-\frac{m \tilde{a}}{4M}+{\cal O}(\tilde{a}^3)\,.
\end{equation}
%%%
We have introduced the horizon angular frequency $\Omega_H=a/(2M r_+)$ and
expanded it to second order.

In the massless case, due to the boundary condition~\eqref{BChor},
superradiant scattering for scalar, electromagnetic and gravitational
perturbations is possible when
$\omega_R<m\Omega_H$~\cite{Teukolsky:1974yv}, i.e. (to second order in
rotation) when
\begin{equation}
\tilde{a}>\frac{4 M\omega_R}{m}\,,\label{superradiance_cond}
\end{equation}
where $\omega_R$ is the real part of the mode frequency, i.e.
$\omega=\omega_R+\ii\omega_I$. Superradiance is possible because the
energy flux at the horizon
%%%
\begin{equation}
 \dot{E}_{r_+}\equiv\lim_{r\to r_+}\int d\vartheta d\varphi \sqrt{-g}T^r_t \label{Edot}
\end{equation}
%%%
is \emph{negative} when $k_H<0$. In the equation above $T_{\mu\nu}$ is
the stress-energy tensor of the perturbation.
% ; in our case $\sqrt{-g}T_{\mu\nu}=-2\delta {\cal L}_{\rm Proca}/\delta g^{\mu\nu}$, where ${\cal L}_{\rm Proca}$ is the Proca Lagrangian.
The same argument can be applied to massive scalar perturbations.

The case of massive vector perturbations is more involved. For purely
axial perturbations close to the horizon, by using Eq.~\eqref{Edot}
and the stress-energy tensor of a Proca field we get $\dot
E_{r_+}=\sum_{\ell m}\dot{E}^\ell_{r_+}$ with
%%%
\begin{equation}
 \dot{E}_{r_+}^\ell=\frac{\omega k_H}{4\ell(\ell+1) M^6}|u_{(4)H}^{\ell}|^2+{\cal O}(\tilde a^3)\,,
\end{equation}
which shows that the energy flux across the horizon is negative when
$k_H<0$. The general case involves terms proportional to each
component $u_{(i)H}^{\ell}$, as well as terms proportional to $\mu^2$.

As we discuss in Sec.~\ref{sec:result} below, our results turn out to
be very accurate for moderate values of $\tilde a$ and they are
reliable even in the superradiant regime, defined by
Eq.~\eqref{superradiance_cond}, as long as $\omega_R M\ll1$.
Superradiant scattering leads to instabilities for massive scalar
perturbations~\cite{Detweiler:1980uk,Dolan:2007mj}.
In the Proca case, the numerical results discussed in
Sec.~\ref{sec:result} below show that, when the superradiant
condition~\eqref{superradiance_cond} is met, the imaginary part of the
modes crosses zero. Thus our numerical data show hard evidence, for
the first time, that massive vector fields trigger (as expected) a
superradiant instability.
% %
%

The asymptotic behavior of the solution at infinity reads
%%%
\begin{equation}
 u_{(i)}^\ell\sim B_{(i)} e^{-k_\infty r} r^{-\frac{M(\mu^2-2\omega^2)}{k_\infty}}
+C_{(i)} e^{k_\infty r} r^{\frac{M(\mu^2-2\omega^2)}{k_\infty}}\,,\label{BCinf}
\end{equation}
where $k_\infty=\sqrt{\mu^2-\omega^2}$, so that $\rm{Re}[k_\infty]>0$.
%%%
The boundary conditions $B_{(i)}=0$ yield purely outgoing waves at
infinity, i.e. QNMs~\cite{Berti:2009kk}. If instead we impose
$C_{(i)}=0$ we get states that are spatially localized within the
vicinity of the BH and decay exponentially at spatial infinity,
i.e. bound states (see e.g.~\cite{Dolan:2007mj,Rosa:2011my}). Stable
QNMs are more challenging to compute than bound states. In the former
case the boundary conditions above imply that purely outgoing waves
blow up as $r_*\to\infty$, whereas purely ingoing waves are
exponentially suppressed at infinity. For bound states, direct
integration of the equations combined with a shooting method is
sufficient, but QNMs are more efficiently computed by other means, for
example via continued fraction methods~\cite{Berti:2009kk}.

%%%%%%%%%%%%%%%%%%%%%%%%%%%%%%%%%%%%%%%%%%%%%%%%%%%%%%%%%%%%%%%%%%%%%%
\subsection{On the superradiant regime and the second order expansion}
%%%%%%%%%%%%%%%%%%%%%%%%%%%%%%%%%%%%%%%%%%%%%%%%%%%%%%%%%%%%%%%%%%%%%%
%
Here we comment on some important features that would be missed in a
first order treatment. First, at second order the structure of the
background metric is remarkably different, because all metric
coefficients acquire ${\cal O}(\tilde{a}^2)$ corrections. This affects
the location of the event horizon, of the ergosphere and of the inner
Cauchy horizon [cf. Eq.~\eqref{cauchyhor}]. Second-order corrections
are known to approximate the Kerr solution much better than a
first-order expansion (see e.g.~\cite{Berti:2004ny}).

From a dynamical point of view, axisymmetric modes acquire
second-order corrections that break the $m=0$ degeneracy of the
first-order case.  Most importantly, at second order the superradiant
regime of vector and scalar fields can be described within our
perturbative approach in a \emph{self-consistent} way. The
superradiant condition~\eqref{superradiance_cond} may look like a
first-order effect.  However, at the onset of superradiance $\omega
M\sim{\tilde a}$ (at least in the most relevant cases, i.e. when $m$
is of order unity). In this case, terms like $\omega^2$ in the field
equations (which are crucial, in particular in the study of the mode
spectrum) are of the same order of magnitude as second order
quantities.

The field equations are of second differential order, so the linearized field
equations (at \emph{first} order in $\tilde{a}$) contain at most terms of order
%%%%
\begin{equation}
 \omega M\,,\quad (\omega M)^2\,,\quad \tilde{a}\,,\quad \tilde{a}\omega
 M\,. \label{listterms}
\end{equation}
%%%%
In principle, terms of order $\tilde{a}(\omega M)^2$ would also be
allowed, but those do not appear in the linearized Proca
equations~\eqref{eqset1sec}--\eqref{eqset3sec}. In the massless case
this is consistent with a first-order expansion of the Teukolsky
equation, which does not contain any $\tilde{a}(\omega M)^2$ term for
scalar, vector and tensor perturbations.

Thus, if $\omega M\sim{\tilde a}$ the second and
fourth terms listed above would be as large as second-order terms in
$\tilde{a}$, which are neglected in a first-order expansion.
In a second-order expansion, instead, one also keeps terms
proportional to $\tilde{a}^2$. This would be enough to consistently
describe the superradiant regime up to first order in $\tilde{a}$,
i.e. up to terms $\sim\tilde{a}\omega$ and $\sim\omega^2$. For a
related discussion in the case of neutron star r-modes, we refer the
reader to Ref.~\cite{Lockitch:2001hq}.

%%%%%%%%%%%%%%%%%%%%%%%%%%%%%%%%%%%%%%%%%%%%%%%%%%%%%%%%%%%%%%%%%%%%%%%%%%%%%
\subsection{Continued fraction method for quasinormal modes and bound states}
%%%%%%%%%%%%%%%%%%%%%%%%%%%%%%%%%%%%%%%%%%%%%%%%%%%%%%%%%%%%%%%%%%%%%%%%%%%%%
From a conceptual point of view, numerical calculations in the
slow-rotation approximation are not more complicated than in the
nonrotating case.  For example, in order to apply the continued
fraction method we can write down a recurrence relation similar to
Eqs.~(31) and (32) of Ref.~\cite{Rosa:2011my}, starting from
Eqs.~\eqref{pol1RDb}, \eqref{pol2RDb} and \eqref{axialRD}, by imposing
the ansatz
\begin{equation}
 u_{(i)}^\ell=(r-r_+)^{-2 i k_H}r^{\nu }e^{-q r}\sum_n a_n^{(i)}(r-r_+)^n\,,
\end{equation}
where $\nu=-q+\omega^2/q+2 i k_H$ and $q=\pm k_\infty$ for bound
states and for QNMs, respectively.
For the axial equation this ansatz leads to a three-term recurrence
relation of the form
%%%%
\begin{eqnarray}
 \alpha_0 a_1^{(4)}+\beta_0 a_0^{(4)}&=&0\,,\nn\\
 \alpha_n a_{n+1}^{(4)}+\beta_n a_n^{(4)}+\gamma_{n} a_{n-1}^{(4)}&=&0\,,\quad n>0\nn\,,
\end{eqnarray}
%%%%%
whose coefficients (to first order in $\tilde{a}$, and
setting $M=1$ in these equations only) read
\begin{eqnarray}
 \alpha_n&=& -4 (1+n) q^2 (1+n-4 \ii \omega)-4 \ii \tilde a m (1+n) q^2\,,\label{alphan}\\
 \beta_n&=& 4 q \left(q (1+\ell(\ell+1)+2 n^2+q (3+4 q)\right.\nn\\
 &&\left.+n (2+6 q)-s^2\right)-4 \ii q (1+2 n+3 q) \omega \nn\\
&&-(1+2 n+12 q) \omega ^2+4 \ii \omega ^3)\nn\\
&&+4 \ii \tilde a m q \left(q+2 n q+3 q^2-2 \ii q \omega -\omega ^2\right)\,,\label{betan}\\
 \gamma_n&=& -4 \left[(q (n + q) - 2 \ii q \omega - \omega^2)^2 - s^2 q^2\right]\nn\\
&&-4 \ii \tilde a m q \left(q (n+q)-2 \ii q \omega -\omega ^2\right)\,,\nn\\\label{gamman}
\end{eqnarray}
for $s=\pm1$. If we set $s=0$, the equations above are also valid for massive
\emph{scalar} perturbations of a Kerr BH in the slowly rotating limit because, as we have shown above, there exists a
single master equation describing both massive scalar
and axial vector modes [cf. Eq.~\eqref{master}]. We have investigated the massive scalar case to test the robustness of our
results, because in this case the perturbation equations on a generic Kerr
background are separable~\cite{Detweiler:1980uk} and the eigenvalues can be
computed by a direct solution of the Teukolsky equation~\cite{Berti:2009kk,Dolan:2007mj}.

In the slowly rotating case the polar sector leads to a six-term, matrix-valued~\cite{Rosa:2011my} recurrence relation
%%%
\begin{eqnarray}
&&\boldsymbol{\alpha}_0 \mathbf{U}_{1} + \boldsymbol{\beta}_0 \mathbf{U}_{0} = 0\,, \nn\\
&&\boldsymbol{\alpha}_1 \mathbf{U}_{2} + \boldsymbol{\beta}_1 \mathbf{U}_{1} + \boldsymbol{\gamma}_1 \mathbf{U}_{0} = 0\,, \quad  \hspace{2.5cm} n > 0\,,\nn \\
&&\boldsymbol{\alpha}_2 \mathbf{U}_{3} + \boldsymbol{\beta}_2 \mathbf{U}_{2} + \boldsymbol{\gamma}_2 \mathbf{U}_{1} + \boldsymbol{\delta}_2 \mathbf{U}_{0} = 0\,, \quad \hspace{1.28cm} n > 1\,,\nn \\
&&\boldsymbol{\alpha}_3 \mathbf{U}_{4} + \boldsymbol{\beta}_3 \mathbf{U}_{3} + \boldsymbol{\gamma}_3 \mathbf{U}_{2} + \boldsymbol{\delta}_3 \mathbf{U}_{1} + \boldsymbol{\rho}_3 \mathbf{U}_{0} = 0\,, \quad n > 2\,,\nn \\
&&\boldsymbol{\alpha}_n \mathbf{U}_{n+1} + \boldsymbol{\beta}_n \mathbf{U}_{n} + \boldsymbol{\gamma}_n \mathbf{U}_{n-1} + \boldsymbol{\delta}_n \mathbf{U}_{n-2} \nn\\
&&+ \boldsymbol{\rho}_n \mathbf{U}_{n-3} + \boldsymbol{\sigma}_n \mathbf{U}_{n-4} = 0\,, \quad \hspace{2.7cm} n > 3\,,\nn 
\end{eqnarray}
%%%%
where $\mathbf{U}_n=(a_n^{(2)},a_n^{(3)})$ is a two-dimensional
vectorial coefficient and $\boldsymbol{\alpha}_n$,
$\boldsymbol{\beta}_n$, $\boldsymbol{\gamma}_n$,
$\boldsymbol{\delta}_n$, $\boldsymbol{\rho}_n$ and
$\boldsymbol{\sigma}_n$ are $2\times2$ matrices, whose explicit form
we do not present here for brevity but it is available online~\cite{url}.  By using a matrix-valued Gaussian
elimination~\cite{Leaver:1990zz,Berti:2009kk,1999PhRvE..59.5344S} the
system above can be reduced to a three-term matrix-valued recurrence
relation, which can be solved with the method discussed in
Ref.~\cite{Rosa:2011my}.

The continued fraction method works very well for both QNMs and
bound-state modes. We computed QNMs and checked that they yield the
correct limit in the massless case (see Sec.~\ref{check} below), but
in the following we will focus mainly on bound states, that can become
unstable in the superradiant regime.

%%%%%%%%%%%%%%%%%%%%%%%%%%%%%%%%%%%%%%%%%%%%%%%%%%%%%%%%%%%%%%%%%%%%%%%%%%%%%%%%%%%%%
\subsection{Direct integration and Breit-Wigner resonance method for bound states}
\label{sec:BW}
%%%%%%%%%%%%%%%%%%%%%%%%%%%%%%%%%%%%%%%%%%%%%%%%%%%%%%%%%%%%%%%%%%%%%%%%%%%%%%%%%%%%%%

To compute bound state frequencies it is possible to use either a direct
integration method or the Breit-Wigner resonance method. 
We start with a series expansion of the solution close to the horizon:
\begin{equation}
  u_{(i)}^\ell\sim e^{-\ii k_H r_*}\sum_n b_n^{(i)}(r-r_+)^n\,,\label{series_hor}
\end{equation}
where $r_+$ is expanded to second order and the coefficients $b_n^{(i)}$ $(n>0$)
can be computed in terms of $b_0^{(i)}$ by solving the near-horizon equations
order by order. In the direct integration method, the field equations are
integrated outwards up to infinity, where the condition $C_{(i)}=0$ in
Eq.~\eqref{BCinf} is imposed (see Ref.~\cite{Rosa:2011my} for details).

The Breit-Wigner resonance method, also known as the standing-wave
approach~\cite{1969ApJ...158....1T,Chandrasekhar:1992ey,Ferrari:2007rc,Berti:2009wx},
is well suited to computing QNMs and bound states of the system of
equations \eqref{pol1RDb}--(\ref{axialRDb}) in the case of slowly
damped modes, i.e. those with $\omega_I\ll\omega_R$. In this case the
eigenvalue problem can be solved by looking for minima of a
real-valued function of a real
variable~\cite{Chandrasekhar:1992ey}. We briefly explain the procedure
below, where we extend it to deal with a system of coupled
equations. For clarity we only consider first-order corrections, but
our argument applies also to the second-order case described by
Eqs.~\eqref{systA}--\eqref{systP} and, in principle, to any order.

Since $\ell=0,1,2,..$, the full system~\eqref{pol1RDb},~\eqref{pol2RDb}
and~\eqref{axialRDb} formally contains an \emph{infinite} number of equations.
In practice, we can truncate it at some given value of $\ell$, compute the modes
as explained below, and finally check convergence by increasing the truncation
order. Let us suppose we truncate the axial sector at $\ell=L$ and the polar
sector at $\ell=L+1$, i.e. for a given $m$ we assume
\begin{equation}
 u_{(4)}^{\ell}\equiv0\,,\qquad u_{(j)}^{\ell+1}\equiv0\qquad 
{\rm when}\qquad \ell\geq L\,,
\end{equation}
%%%
with $j=1,2,3$ denoting the polar perturbations. All perturbations vanish
identically for $\ell<|m|$.

When $m=0$, the truncation above reduces the system to $N=3L$ coupled
second-order ODEs for $L-1$ axial functions and $2L+1$ polar
functions, including the monopole, described by
Eq.~\eqref{monopole1}. When $|m|>0$ the truncated system contains
$N=3L-3|m|+2$ second-order ODEs (for $L-|m|$ axial functions and
$2L-2|m|+2$ polar functions). In all cases we are left with a system
of $N$ second-order ODEs for $N$ perturbation functions, which we
collectively denote by $y_{(p)}$ ($p=1,...,N$).

At the horizon each function is described by ingoing and outgoing waves.  We
impose a purely ingoing wave boundary condition analogous to
Eq.~\eqref{series_hor},
\begin{equation}
  y_{(p)}\sim e^{-\ii k_H r_*}\sum_n c_n^{(p)}(r-r_+)^n\,,\label{series_hor2}
\end{equation}
%%%
where again the coefficients $c_n^{(p)}$ $(n>0$) can be computed in terms of
$c_0^{(p)}$. A family of solutions at infinity is then characterized by $N$
parameters, corresponding to the $N$-dimensional vector of the near-horizon
coefficients, $\mathbf{c_0}=\{c_0^{(p)}\}$. At infinity we look for
exponentially decaying solutions, which correspond to bound, slowly damped modes. 
The spectrum of these
modes can be obtained as follows. We first choose a suitable orthogonal basis
for the $N$-dimensional space of the initial coefficients $c_0^{(p)}$ (see also
Ref.~\cite{Rosa:2011my}). We perform $N$ integrations from the horizon to
infinity and construct the $N\times N$ matrix
\begin{eqnarray}
 \label{matrix_coupled}
\mathbf{S}_m(\omega)=\lim_{r\to\infty}
\begin{pmatrix}
y_{(1)}^{(1)} & y_{(1)}^{(2)}   & ... & y_{(1)}^{(N)}\\ 
y_{(2)}^{(1)} & y_{(2)}^{(2)}   & ... & ... \\
... & ...   & ... & ... \\
... & ...   & ... & ... \\
y_{(N)}^{(1)} & ...   & ... & y_{(N)}^{(N)} \\
\end{pmatrix}\,,
\end{eqnarray}
where the superscripts denote a particular vector of the basis,
i.e. $y_{(p)}^{(1)}$ corresponds to $\mathbf{c_0}=\{1,0,0,...,0\}$,
$y_{(p)}^{(2)}$ corresponds to $\mathbf{c_0}=\{0,1,0,...,0\}$ and
$y_{(p)}^{(N)}$ corresponds to $\mathbf{c_0}=\{0,0,0,...,1\}$.
Finally, the bound-state frequency
$\omega_0=\omega_R+\ii\omega_I$ is obtained by imposing
\begin{equation}
 \det{\mathbf{S}_m(\omega_0)}=0\,.\label{det_modes}
\end{equation}
So far we have not imposed the Breit-Wigner assumption $\omega_I\ll
\omega_R$. In fact, the procedure discussed above can be used to perform a
general direct integration in the case of coupled systems.

By expanding Eq.~\eqref{det_modes} about $\omega_R$ and assuming $\omega_I\ll
\omega_R$ we get~\cite{Ferrari:2007rc}
\begin{equation}
\det{\mathbf{S}_m(\omega_0)}\simeq\det{\mathbf{S}_m(\omega_R)}
+\ii\omega_I\left.\frac{d\left[\det{\mathbf{S}_m(\omega)}\right]}{d\omega}\right|_{\omega_R}=0\,,
\end{equation}
which gives a relation between ${d\left[ \det{\mathbf{S}_m(\omega)}
    \right]}/{d\omega}$ and $\det\mathbf{S}_m$ at
$\omega=\omega_R$. We consider the function $\det{\mathbf{S}_m}$
restricted to real values of $\omega$. A Taylor expansion for (real)
$\omega$ close to $\omega_R$ yields (using the relation above):
\begin{equation}
\det{\mathbf{S}_m(\omega)}\simeq\det{\mathbf{S}_m
(\omega_R)}\left[1-\frac{\omega-\omega_R}{\ii\omega_I}
\right]\propto\omega-\omega_R-\ii\omega_I\,.
\end{equation}
Therefore, on the real--$\omega$ axis, close to the real part of the mode, 
\begin{equation}
 |\det{\mathbf{S}_m\left(\omega\right)}|^2\propto
\left(\omega-\omega_R\right)^2+\omega_I^2\,.\label{parabola}
\end{equation}
To summarize, to find the slowly damped modes it is sufficient to integrate the
truncated system $N$ times for real values of the frequency $\omega$, construct
the matrix ${\mathbf{S}_m\left(\omega\right)}$ and find the minima of the
function $|\det{\mathbf{S}_m}|^2$, which represent the real part of the
modes. Then the imaginary part (in modulus) of the mode can be extracted through
a quadratic fit, as in Eq.~\eqref{parabola}.
We note that the same method can be straightforwardly extended to compute the
slowly damped modes of the general systems~\eqref{epF1bis1}--\eqref{epF1bis3}
and \eqref{epF2bis1}--\eqref{epF2bis3}.
\begin{table}[htb]
 \begin{tabular}{c|cc}
		&	$M\omega_R$ & $M\omega_I$  \\
\hline
No coupling, $\ell=1$ (DI)		&	$0.099484532$ & $1.1\cdot 10^{-7}$ 	\\
Full system, $L=1,2,3,4$ (BW)		&	$0.099484563$ & $1.1\cdot 10^{-7}$  	 \\
\hline
No coupling, $\ell=1$ (DI)		&	$0.09987320753$ & $4\cdot 10^{-11}$ 	 \\
Full system, $L=2,3,4$ (BW)		&	$0.09987183250$ & $5\cdot 10^{-11}$  	 
\end{tabular}
\caption{Examples of polar (top rows) and axial (bottom rows)
  bound-state modes for $m=1$, $M\mu=0.1$ and $\tilde a=0.1$ computed at first order. Modes
  were computed via direct integration (DI) of the
  system~\eqref{pol1RDb}--\eqref{pol2RDb} \emph{without} couplings
  $\ell\to\ell\pm1$, as well as with the Breit-Wigner (BW) method
  applied to the \emph{full} system~\eqref{pol1RDb}--\eqref{pol2RDb}
  and~\eqref{axialRDb}, for different truncation orders $L$.  The
  modes are insensitive to the truncation order within the quoted
  numerical accuracy. 
  \label{tab:coupling}}
\end{table} 

By applying the Breit-Wigner method we have numerically verified the
argument given in Sec.~\ref{sec:general}, i.e. that the
$\ell\to\ell\pm1$ couplings do not affect the eigenfrequencies in the
slow-rotation limit. As an example, in Table~\ref{tab:coupling} we
compare two modes ($m=1$, $\tilde a=0.1$, $M\mu=0.1$) of the full
system computed with the Breit-Wigner resonance method for several
truncation orders $L$ and the same modes computed with a direct
integration of the system of equations \emph{without} the
$\ell\to\ell\pm1$ couplings. Up to numerical accuracy the mode is
insensitive to the truncation order and, most importantly, it agrees
very well with that computed for the system without couplings. Note
that $L=1$ corresponds to the uncoupled polar system with
$\ell=1$. Therefore the small discrepancy is not due to the coupling
terms, but to some inherent numerical error of the resonance method,
which becomes less accurate when the imaginary part of the mode is
tiny.

%%%%%%%%%%%%%%%%%%%%%%%%%%%%%%%%%%%%%%%%%%%%%%%%%%%%%%%%%%
\section{Results}
\label{sec:result}
%%%%%%%%%%%%%%%%%%%%%%%%%%%%%%%%%%%%%%%%%%%%%%%%%%%%%%%%%%
%%%%%%%%%%%%%%%%%%%%%%%%%%%%%%%%%%%%%%%%%%%%%%%%%%%%%%%%%%%%%%%%%
\subsection{Numerical procedure}
%%%%%%%%%%%%%%%%%%%%%%%%%%%%%%%%%%%%%%%%%%%%%%%%%%%%%%%%%%%%%%%%%
In principle, by integrating the full systems~\eqref{systA} and
~\eqref{systP} for a given truncation order $L$ and a given value of
$m$ one can obtain the full spectrum of quasinormal modes and bound modes for
\emph{both} the axial and polar sectors and for any $\ell<L$.  We have
computed bound states and QNMs via the Breit-Wigner
procedure of Sec.~\ref{sec:BW}. We double-checked the results using
two additional, independent techniques, also described above: the
continued fraction method and direct integration of the reduced
equations~\eqref{systA} and ~\eqref{systP}. The results agree within
numerical accuracy. The Breit-Wigner procedure and the continued
fraction method have been implemented up to first order in the
rotation parameter; the direct integration, instead, has been extended
up to second order in $\tilde a$, in order to validate the results of
the first-order integration, as we shall discuss below. Furthermore,
when $\omega_I\ll\omega_R$ we found very good agreement with the
Breit-Wigner method (cf. Table~\ref{tab:coupling}).

Exploring the full parameter space of the eigenfrequencies at second
order in $\tilde{a}$ is numerically demanding.  For this reason, in
Sec.~\ref{res_num} below we will present a more extensive survey of
results at first order, and compare them to the second-order
calculation in Sec.~\ref{sec:res_num2} in selected cases.
%
%%%%%%%%%%%%%%%%%%%%%%%%%%%%%%%%%%%%%%%%%%%%%%%%%%%%%%%%%%%%%%%%%
\subsection{A consistency check: massless vector perturbations}
\label{check}
%%%%%%%%%%%%%%%%%%%%%%%%%%%%%%%%%%%%%%%%%%%%%%%%%%%%%%%%%%%%%%%%%
As a preliminary test of our method, we have computed the QNMs of
massless vector (i.e., electromagnetic) perturbations of a Kerr BH to
first order in $\tilde{a}$. These results can be compared with those
obtained by solving the Teukolsky equation without imposing the
slow-rotation approximation (see e.g.~\cite{Berti:2009kk}). In the
massless limit the axial equation~\eqref{axialRD} reduces to
\begin{equation}
\frac{d^2u_{(4)}}{dr_*^2}+\left(\omega^2-\frac{4 \tilde aM^2 m
\omega}{r^3}-\frac{\ell(\ell+1)}{r^2}F\right)u_{(4)}=0\,.\label{axialRD2}
\end{equation}
For the polar sector in the massless limit, we can define exactly the
same master variable as in the nonrotating case~\cite{Ruffini}.
The polar sector in the massless limit is described by a fourth-order
equation.  As in the nonrotating case~\cite{Rosa:2011my}, one solution
of this equation is a pure-gauge mode and can be eliminated.
In the slow-rotation approximation we can recast the other solution in
terms of a master function, which also satisfies
Eq.~\eqref{axialRD2}. Therefore axial and polar perturbations have the
same spectra.  This is consistent with the fact that electromagnetic
perturbations of the full Kerr geometry are described by a single
master equation in the Teukolsky formalism~\cite{Berti:2009kk}.

Due to isospectrality, in the massless limit we only need to solve
Eq.~\eqref{axialRD2}.  The modes can be computed via the continued fraction
method introduced above, where the coefficients of the recurrence relation can
be obtained by setting $\mu=0$ and $s=\pm1$ in
Eqs.~\eqref{alphan}--\eqref{gamman}.

\begin{figure}[htb]
\begin{center}
\epsfig{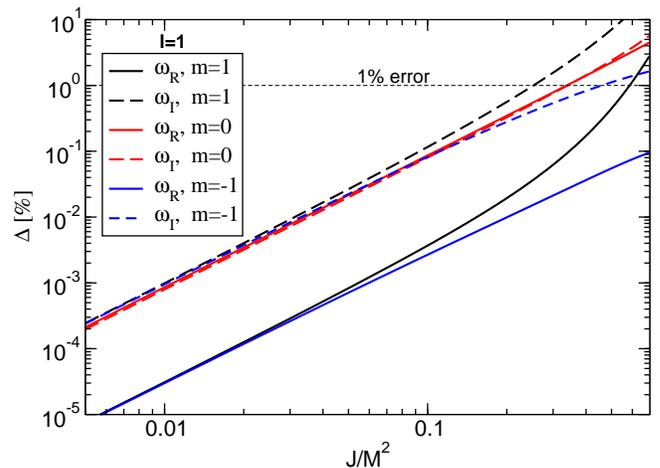}
\caption{(color online) Percentage difference between QNMs of massless vector
  perturbations in the slow-rotation limit at first order and the
  ``full'' numerical solution of the Teukolsky equation in the Kerr
  metric~\cite{Berti:2009kk}. The deviation scales like $\tilde{a}^2$
  as $\tilde{a}\to0$. \label{fig:comparison}}
\end{center}
\end{figure}
We compared our results with the exact massless vector modes of a Kerr
BH, computed by solving the Teukolsky equation, for $\ell=1$ and
different values of $m$ (see Fig.~\ref{fig:comparison}).  The first
order approximation performs very well, even for relatively large
values of the BH rotation rate.  Remarkably, the real and imaginary
parts of the QNMs computed with our approach deviate by less than
$1\%$ from their exact values if $\tilde a\lesssim0.3$.

\begin{figure}[htb]
\begin{center}
\begin{tabular}{c}
 \epsfig{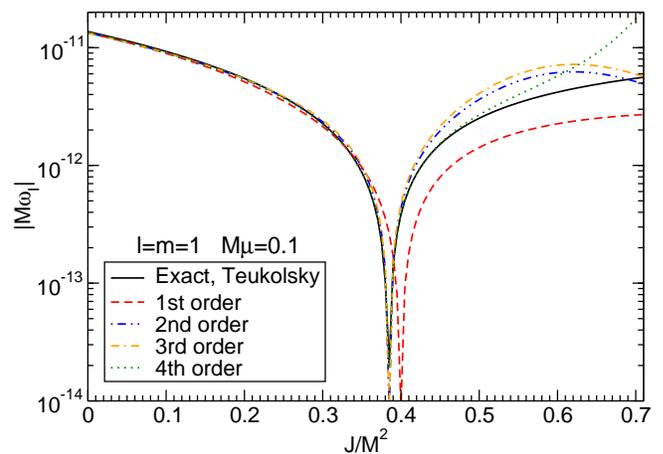}
\end{tabular}
\caption{(color online) Comparison between the exact, Teukolsky-based result and the
  results obtained by our slowly rotating approximation at first and
  second order for the imaginary part of the scalar fundamental
  bound-state mode with $\ell=m=1$ and $\mu M=0.1$. For comparison, we
  have also computed the same mode as obtained by expanding the
  Teukolsky equation at third and fourth order.
  \label{fig:scalar}}
\end{center}
\end{figure}
%

%%%%%%%%%%%%%
\subsection{A second test: bound state modes for scalar perturbations}
%%%%%%%%%%%%%
We can also investigate the accuracy of the slow-rotation
approximation for massive scalar perturbations, another case in which
the Teukolsky equation can be solved exactly (see
e.g.~\cite{Dolan:2007mj}).  In Fig.~\ref{fig:scalar} we show the
imaginary part of the bound state modes with $\ell=m=1$ and
$M\mu=0.1$, computed by solving Eq.~\eqref{final2} at first and second
order via direct integration, against the exact Teukolsky-based result
obtained with the continued fraction
method~\cite{Dolan:2007mj,Berti:2009kk}. For comparison, we also show
the results obtained by solving Teukolsky's equation at third and
fourth order.  As shown in Fig.~\ref{fig:scalar}, the imaginary part
crosses the axis when the superradiant condition is satisfied. In the
stable branch ($\tilde{a}\lesssim0.4$) even first-order results are in
good quantitative agreement with the ``exact'' calculation. In the
unstable branch ($\tilde{a}\gtrsim0.4$) the first-order approximation
is in only qualitative agreement with the exact result, but the
second-order approximation is in quantitative agreement with the
numerics at the onset of the instability. It is still quite remarkable
that even first-order results can correctly reproduce the onset of the
instability. We shall return to this consideration below, when we will
deal with the massive vector case.

\begin{figure}[htb]
\begin{center}
\epsfig{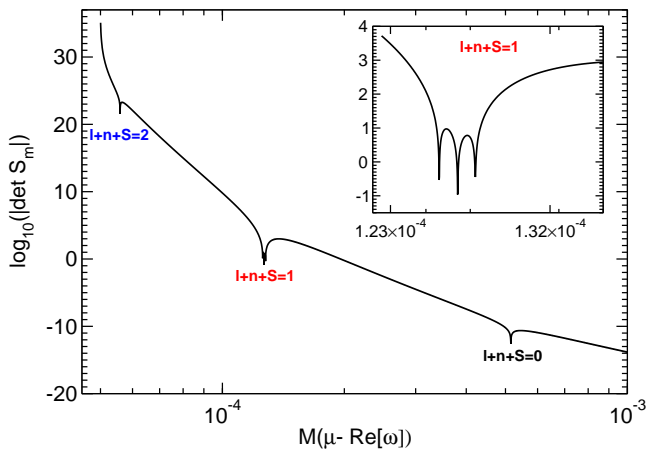}
\caption{(color online) Bound state modes obtained with a first-order Breit-Wigner method
  applied to the full system.  We show the determinant
  $|\det\mathbf{S}_m|$ as a function of the real part of the frequency
  for $M\mu=0.1$ and $\tilde{a}=0.1$. According to Eq.~\eqref{fit_wR},
  the real part of modes with the same $\ell+n+S$ is approximately
  degenerate for $M\mu\ll 1$.
\label{fig:BW}}
\end{center}
\end{figure}
\begin{figure*}[thb]
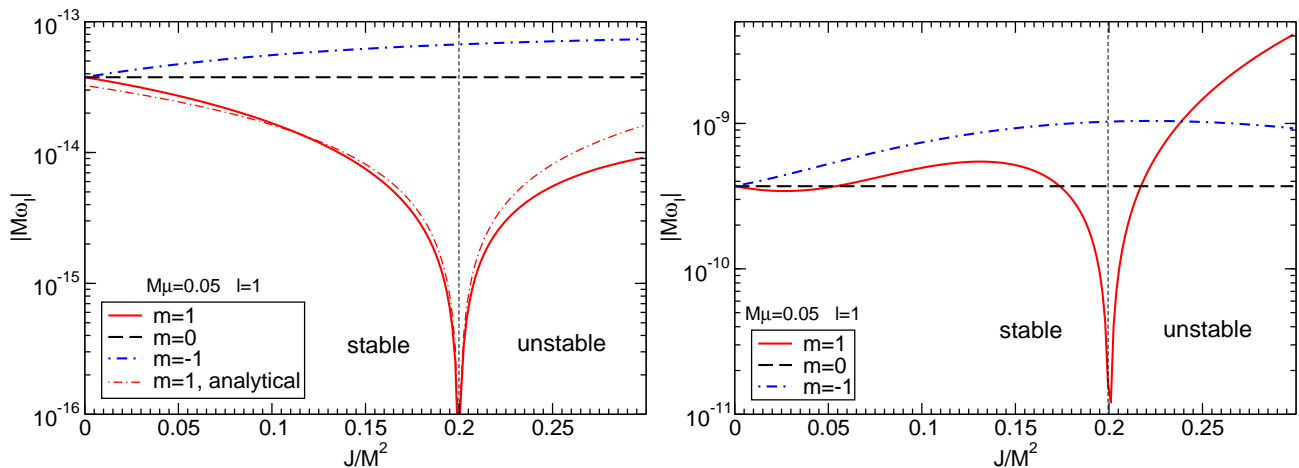

\begin{center}
\begin{tabular}{cc}
 \epsfig{file=fig5a.eps,width=8.5cm,angle=0,clip=true}&
\epsfig{file=fig5b.eps,width=8.5cm,angle=0,clip=true}
\end{tabular}
\caption{(color online) Absolute value of the imaginary part of the axial (left
  panel) and of the polar $S=-1$ (right panel) vector modes as a
  function of the BH rotation rate $\tilde a$ for $\ell=1$,
  $M\mu=0.05$ and different values of $m$, computed at first order. 
  For $m=1$, the modes cross
  the axis and become unstable when the superradiance
  condition~\eqref{superradiance} is met. In the left panel, the red
  dot-dashed line denotes the analytical result~\eqref{ana_l=1}.
  \label{fig:zeeman}}
\end{center}
\end{figure*}
%

%%%%%%%%%%%%%%%%%%%%%%%%%%%%%%%%%%%%%%%%%%%%%%%%%%%%%%%
\subsection{Proca modes to first order}
\label{res_num}
%%%%%%%%%%%%%%%%%%%%%%%%%%%%%%%%%%%%%%%%%%%%%%%%%%%%%%%%

Unlike the massless case, axial and polar modes for massive vector
perturbations are not isospectral. Furthermore there exist two classes
of polar modes, which can be distinguished by their
``polarization''~\cite{Rosa:2011my}. For small $M\mu$, Rosa and Dolan
found that the imaginary part of the bound states of a Schwarzschild
BH scales as
\begin{equation}
  {\omega_I}\sim {\mu}(M\mu)^{4\ell+5+2S}\,,\label{RDscaling}
\end{equation}
where $S=0$ for axial modes and $S=\pm1$ for the two classes of polar modes. The
monopole corresponds to $\ell=0$, $S=1$, in agreement with the rules for
addition of angular momenta~\cite{Rosa:2011my}. In the limit $\tilde a =0$ we
recover this scaling. We have also analyzed the $\tilde{a}$-dependence of the
modes for small $\tilde a$.
When $M\mu\ll 1$ we find the expected hydrogen-like behavior, $\omega_R\sim
\mu$, which remains valid also in the slowly rotating case. When $M\mu\lesssim
0.1$ the real part of the modes is roughly independent of $\tilde a$, and it is
very well approximated by the relation
\begin{equation}
 \omega_R^2=\mu^2\left[1-\left(\frac{M\mu}{\ell+n+S
+1}\right)^2\right]+{\cal O}\left(\mu^4\right)\,,\label{fit_wR}
\end{equation}
where $n\geq0$ is the overtone number (cf.~\cite{Dolan:2007mj}).  For the axial
case ($S=0$) this relation is validated by the analytical results presented in
Appendix~\ref{res_ana} [see in particular Eq.~\eqref{wRana}], where we solve the
axial equations in the limit $M\mu\ll 1$.  Equation \eqref{fit_wR} is also
supported by the nonrotating result given in Eq.~(49) of~\cite{Rosa:2011my}
(see also~\cite{Gal'tsov:1984nb}).

Equation~\eqref{fit_wR} predicts a degeneracy for modes with the same value of
$\ell+n+S$ when $M\mu\ll 1$. In the Breit-Wigner method, the mode frequencies
can be identified as minima of the real-valued function
$|\det{\mathbf{S}_m}|^2$.  The degeneracy predicted by Eq.~\eqref{fit_wR} is not
exact for $M\mu=0.1$, as illustrated in the inset of Fig.~\ref{fig:BW}, where we
display the minima of $|\det{\mathbf{S}_m}|^2$ when $\tilde{a}=0.1$.
The first minimum on the right corresponds to $\ell+n+S=0$, which can only be
achieved for the fundamental polar mode with $(\ell,n,S)=(1,0,-1)$. When
$\ell+n+S=1$ we have a three-fold degeneracy, corresponding to
$(\ell,n,S)=(1,0,0)$, $(2,0,-1)$, $(1,1,-1)$. The approximate nature of the
degeneracy is shown in the inset, where three distinct (albeit very close)
minima appear. For $\ell+n+S=2$ there is a five-fold degeneracy, which can be
resolved with high enough resolution. Note that according to Eq.~\eqref{fit} the
imaginary part of the modes is tiny when $\ell+n+S$ is large. This makes it
difficult to numerically resolve higher overtones and modes with large $\ell$.

\begin{figure*}[htb]
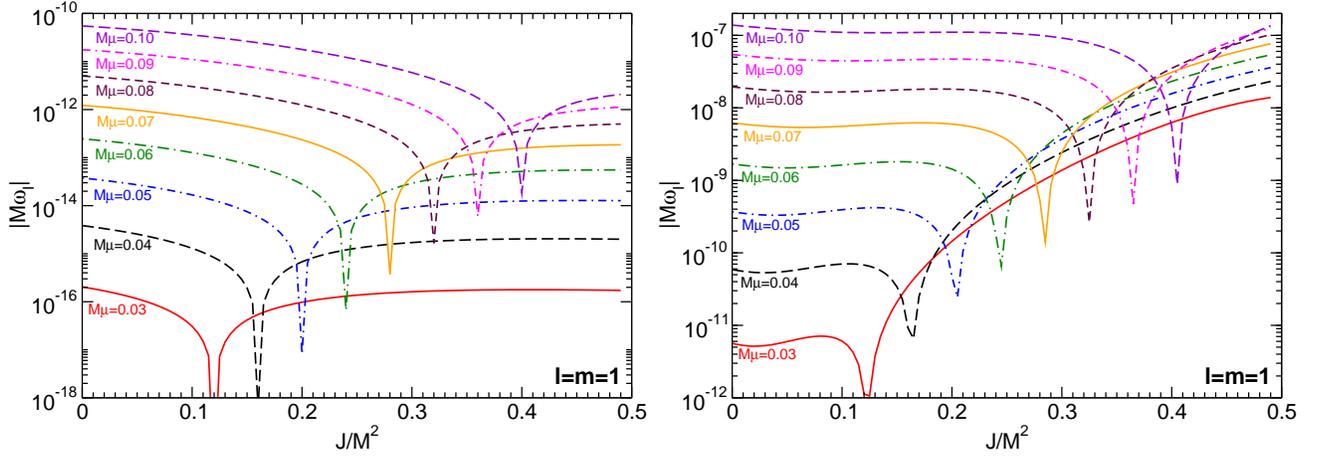

\begin{center}
\begin{tabular}{cc}
 \epsfig{file=fig6a.eps,width=8.5cm,angle=0,clip=true}&
 \epsfig{file=fig6b.eps,width=8.5cm,angle=0,clip=true}
\end{tabular}
\caption{(color online) Imaginary part of the axial (left panel) and of the polar
  $S=-1$ (right panel) vector modes as a function of the BH rotation
  rate $\tilde a=J/M^2$ for $\ell=m=1$ and several values of
  $\mu$. Although beyond the regime of validity of the first order
  approximation, the modes cross the axis and become unstable
  precisely when the superradiance condition~\eqref{superradiance} is
  met. As shown in Fig.~\ref{fig:wI_VS_J2nd}, the first-order results
  are also in qualitatively good agreement with those obtained at
  second order.
  \label{fig:crossover}}
\end{center}
\end{figure*}

The imaginary part of nonaxisymmetric modes shows the typical Zeeman-like
splitting for different values of $m$ when $\tilde{a}\neq 0$, as shown in
Fig.~\ref{fig:zeeman} for the axial modes and for the polar mode with $S=-1$.
For $m>0$ the imaginary part of the frequency decreases (in modulus) as $\tilde
a$ increases, and it has a zero crossing when
\begin{equation}
 \omega_R\sim\mu=m\Omega_H\sim m\frac{\tilde a}{4 M}+{\cal O}(\tilde a^3)\,,
\label{superradiance}
\end{equation}
which according to Eq.~\eqref{superradiance_cond} corresponds to the onset of
the superradiant regime.
Recall that a second-order calculation is needed to describe the superradiant
regime in a self-consistent way, because the latter is well beyond
the nominal regime of validity of the first-order approximation.
It is quite remarkable that even the first-order approximation predicts that the
instability should ``turn on'' at the right point, as shown in
Fig.~\ref{fig:crossover} for $\ell=m=1$ and several values of $\mu$.  The
quantitative accuracy of the first-order approximation is questionable in the
superradiant regime.
However, we will now show that second-order results indicate that we are
correctly capturing the order of magnitude of the instability even for
moderately large rotation.
%

%%%%%%%%%%%%%%%%%%%%%%%%%%%%%%%%%%%%%%%%%%%%%%%%%%%%%%%
\subsection{Proca modes to second order}
\label{sec:res_num2}
%%%%%%%%%%%%%%%%%%%%%%%%%%%%%%%%%%%%%%%%%%%%%%%%%%%%%%%%
%
\begin{figure*}[htb]
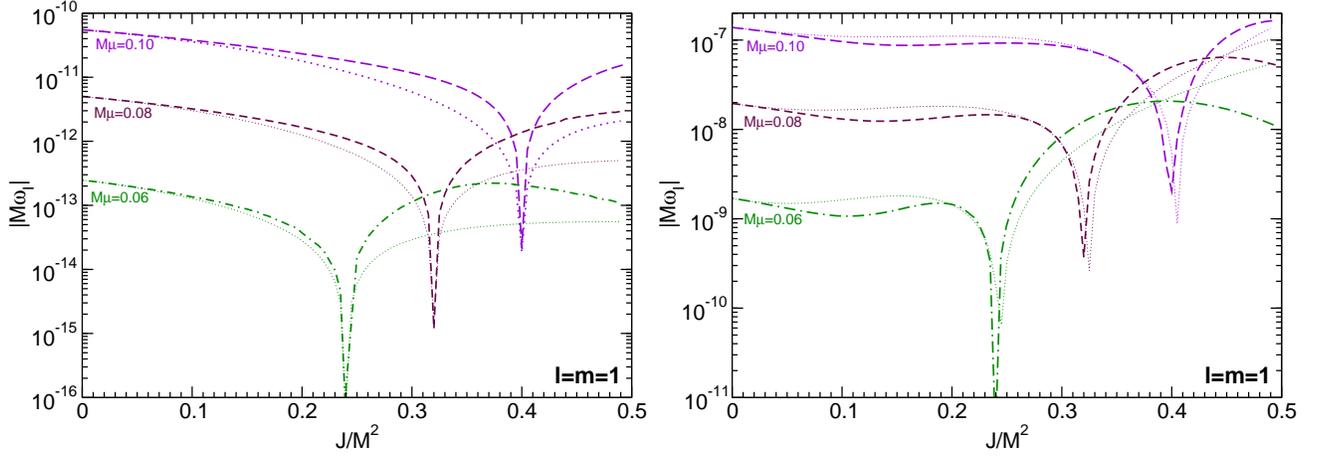

\begin{center}
\begin{tabular}{cc}
 \epsfig{file=fig7a.eps,width=8.5cm,angle=0,clip=true}&
 \epsfig{file=fig7b.eps,width=8.5cm,angle=0,clip=true}
\end{tabular}
\caption{(color online) Imaginary part of the axial $S=0$ vector modes (left panel)
  and for polar $S=-1$ vector modes (right panel) computed at second
  order in the rotation for $\ell=m=1$ and several values of
  $\mu$. The dotted thinner lines refer to the first order
  calculation. The modes cross the axis and become unstable when the
  superradiance condition is met.
  \label{fig:wI_VS_J2nd}}
\end{center}
\end{figure*}
As discussed in Sec.~\ref{sec:Proca2nd}, the perturbation equations
for massive vector fields at second order in rotation,
Eqs.~\eqref{systA} and~\eqref{systP}, are obtained from
Eqs.~\eqref{eqsec}, using the perturbation equations in the
nonrotating limit to eliminate the second derivatives. The explicit
form of the equations is available online~\cite{url}.

Solving the eigenvalue problem for these equations is numerically more
demanding than in the case of perturbations at first order in $\tilde
a$, especially when the imaginary part of the modes is tiny (as in the
axial case and in the small-mass limit).  For this reason we present
only a selection of results that allow us to validate the accuracy of
the first-order approximation, and (more in general) to estimate the
errors due the slow-rotation approximation.

The imaginary part of the fundamental $S=0$ axial and $S=-1$ polar
modes as a function of $\tilde{a}$ at first and second order are shown
in Fig.~\ref{fig:wI_VS_J2nd}.  The slow-rotation method correctly
predicts the onset of the Proca instability already at first order,
where (a priori) large deviations could be expected. Furthermore, the
value of $\tilde{a}$ which corresponds to the onset of the instability
is slightly smaller at second order, consistently with the fact that
the horizon location gets negative second-order contributions
[cf. Eq.~\eqref{cauchyhor}].  Fig.~\ref{fig:wI_VS_J2nd} actually
displays a remarkably good quantitative agreement between the first-
and second-order calculations. The main difference is that in the
superradiant region the instability predicted by the first-order
expansion is {\em weaker} (by a factor of a few) than the instability
computed at second order.  Another important point is that our
perturbative approach can consistently describe the superradiant
regime as long as $M\omega\ll1$ and $\tilde{a}\ll1$. Due to the
hydrogen-like spectrum of the bound states, this means that we are
limited to considering small values of $\mu M$. Again, this
expectation is validated by Fig.~\ref{fig:wI_VS_J2nd}.

At second order, all curves in Fig.~\ref{fig:wI_VS_J2nd} show a
maximum at large values of the spin. This maximum is an artifact of
the second-order approximation: a similar feature appears also in the
scalar case in which the exact, Teukolsky-based result is monotonic
(cf. Fig.~\ref{fig:scalar}). In what follows we have discarded
numerical data beyond such maxima, and we have taken this truncation
into account when estimating numerical errors.

The bottom line of this discussion is that the first order calculation
provides a reliable estimate of the order of magnitude of the
instability. The second-order calculation should be
\emph{quantitatively} reliable even in the unstable regime, at least
up to moderately large values of $\tilde{a}$.  Roughly speaking, the
reason why the first-order calculation captures the correct
qualitative properties of the instability can be traced back to the
fact that the superradiant instability condition $\omega<m\Omega_H$ is
a ``first order effect''. Because $\Omega_H$ is an odd function of
$\tilde{a}$, a first-order approximation of this relation can be
expected to be valid modulo third-order corrections.

%%%%%%%%%%%%%%%%%%%%%%%%%%%%%%%%%%%%%%%%%%%%%%%%%%%%%%%%%%%%%%%%%%%%%%%%%%%%%%%%
\subsection{The minimum instability growth scale}\label{sec:minimum_growth}
%%%%%%%%%%%%%%%%%%%%%%%%%%%%%%%%%%%%%%%%%%%%%%%%%%%%%%%%%%%%%%%%%%%%%%%%%%%%%%%%

The agreement between first- and second-order calculations can be used
to estimate the order of magnitude of the instability by
extrapolation. We will fit our data at the onset of the instability,
and then extrapolate to the superradiant region. This procedure should
give us at least a correct order of magnitude for the instability
timescale. We expect the extrapolation to work better in the axial
case, where bound-state frequencies have a monotonic behavior as
functions of $\tilde{a}$ (cf.~the left panel of
Fig.~\ref{fig:crossover}).
We estimate that close to the onset of the superradiant instability the
imaginary part of the modes should scale as follows:
\begin{equation}
M\omega_I\sim \gamma_{S\ell}\left(\tilde a m-2 r_+\mu \right) (M\mu)^{4\ell+5+2S}\,,
\label{fit}
\end{equation}
where $\gamma_{S\ell}$ is a coefficient that depends on $S$ and
$\ell$. For axial modes with $\ell=1$ our numerical data yield
$\gamma_{01}\approx0.09\pm0.03$, in good agreement with the analytical
formula in the limit $M\mu\ll 1$,
\begin{equation}
M\omega_I^{(\ell=1\,,\, {\rm axial})}=\frac{1}{12} 
\left(\tilde{a}m-2 r_+ \mu \right) (M\mu)^9\,,
\label{anl1}
\end{equation}
which is derived in Appendix~\ref{res_ana}. From the
equation above we get $\gamma_{01}=1/12\sim0.0833$.  Expression~\eqref{anl1}
is compared to the numerical results in Fig.~\ref{fig:zeeman}.

Here and in the following, numerical errors are estimated by comparing
the results at first and second order, and by taking the maximum
deviation between the fit and the data. We generate numerical data
with increasing accuracy until convergence is reached. In particular,
in the direct-integration method it is important to retain a
sufficiently large number of terms in the series expansion of the
boundary condition at infinity, Eq.~\eqref{BCinf}, in order to achieve
sufficient accuracy. This is expected because bound-state modes decay
exponentially, and therefore it is challenging to perform a numerical
integration at large distances.

We have applied the same method to compute unstable massive
\emph{scalar} modes~\cite{Detweiler:1980uk}, whose imaginary part
scales as in the case of axial vector modes with $S=0$. In this case
we find $\gamma_{01}^{\rm scalar}\approx0.03\pm0.01$, again in
reasonable agreement with analytical expectations (by setting $s=0$ in
Eq.~\eqref{anl1} we get $\gamma_{01}^{\rm scalar}=1/48\sim0.0208$).
Computing the modes in the limit $M\mu\ll 1$ is numerically
challenging because of the steep scaling of $\omega_I$ with $M\mu$
[cf. Eq.~\eqref{fit}] and this affects the precision of the
fit. Despite these numerical difficulties, we are able to recover the correct order of magnitude for the instability timescale
when $M\mu\ll 1$. In any case, the observational bounds that we will
address in the next section depend only mildly on the precise value of
$\gamma_{S\ell}$.  For the polar modes we find
$\gamma_{-11}\approx20\pm10$, while extracting the coefficient
$\gamma_{11}$ with sufficient precision is numerically challenging
even for the fundamental mode with $\ell=1$, since $\omega_I\sim
(M\mu)^{11}$.

Obviously our method becomes inaccurate when $\tilde{a}\to 1$, and the
fit~\eqref{fit} can provide at most an order of magnitude for the
instability timescale in the extremal limit. However, in the massive
scalar case Eq.~\eqref{fit} with $S=0$ is \emph{exact} in the
$M\mu\ll1$ limit, for any value of
$\tilde{a}$~\cite{Detweiler:1980uk}.
Furthermore, in the massive vector case, the good agreement between
the first- and second-order expansions suggests that the slow-rotation
approximation should be quite accurate even for moderately large spins,
$\tilde{a}\lesssim0.7$. Therefore it is reasonable to expect that
extrapolations of Eq.~\eqref{fit} from the slow-rotation limit should
at least provide the correct order of magnitude (and possibly a
reasonable quantitative estimate) of the instability timescale.

It is tempting to extrapolate our predictions to generic values of
$\tilde{a}$ and to discuss possible astrophysical implications of the
superradiant instability for massive vector fields around Kerr
BHs. Our conclusions rely on extrapolation, so they should be
confirmed by independent means. Let us assume that the imaginary part
of the modes is generically described by Eq.~\eqref{fit}, where we
left the coefficient $\gamma_{S\ell}$ undetermined, since its precise
value (whose order of magnitude approximately ranges between $0.1$ and
$10$, depending on $S$ and $\ell$) is not crucial for the following
discussion. From Eq.~\eqref{fit} we find that the instability growth
timescale has a minimum (i.e., $\omega_I$ has a maximum) for
\begin{equation}
\mu^{\rm min}=m\tilde{a}\frac{5+4\ell+2S}{4r_+(3+2\ell+S)}\,,
\end{equation}
and the corresponding value of the timescale $\tau=\omega_I^{-1}$
%$M\omega_I^{\rm min}$ 
is given by Eq.~\eqref{fit} evaluated at $\mu=\mu^{\rm min}$.  From
Eq.~\eqref{fit} we also see that (excluding the case $\tilde
am\simeq2r_+\mu$) this is a typical value of the instability
timescale.

If we apply the same procedure to scalar fields ($\ell=1$, $S=0$,
$\gamma_{01}^{-1}=48$), we find that the minimum instability
corresponds to $\tilde{a}=1$, $M\mu^{\rm min}=0.45$ and
$M\omega_I^{\rm max}=1.6\times 10^{-6}$. Notice that this simple
argument relies on an (a priori unjustified) extrapolation of
analytical expressions valid for $M\mu\ll 1$ to the regime $M\mu\sim
1$. In the scalar case this prediction overestimates the numerical
results by an order of magnitude: in fact,
Refs.~\cite{Cardoso:2005vk,Dolan:2007mj} found a minimum instability
$M\omega_I\sim1.5 \times10^{-7}$ at $M\mu\sim0.42$ for
$\tilde{a}\sim0.99$. This disagreement should not be surprising.
Actually, Eq.~\eqref{fit} performs remarkably well even for values of
$\tilde{a}$ close to extremality~\cite{Arvanitaki:2010sy}: it
overestimates the results in Table III in Ref.~\cite{Dolan:2007mj} by
only $3\%$ when $\tilde{a}=0.7$ and by less than $70\%$ when
$\tilde{a}=0.99$. In our view, this agreement is quite impressive.

Let us apply the same argument to vector fields. From Eq.~\eqref{fit}
we expect the stronger instability when $\ell=m=1$. For
$\tilde{a}=0.7$ this corresponds to
\begin{align}
&M\mu^{\rm min}\sim 0.187\,,\quad M\omega_I^{\rm max}\sim 5.8 
\gamma_{11}\times 10^{-10}\,,%\quad S=1\,,
\nn\\
&M\mu^{\rm min}\sim 0.184\,,\quad M\omega_I^{\rm max}\sim 1.7
\gamma_{01}\times 10^{-8}\,,%\quad S=0 \,,
\nn\\
&M\mu^{\rm min}\sim 0.179\,,\quad M\omega_I^{\rm max}\sim 5.1
{\gamma_{-11}}\times 10^{-7}\,,%\,\, S=-1\,.
\nn
%\label{sm1estimate}
\end{align}
for $S=1,0,-1$, respectively.  Our data at second order are compatible
with $\gamma_{01}\approx0.09$ and $\gamma_{-11}\approx20$. As we
discussed, we could not extract the coefficient $\gamma_{11}$ with
sufficient precision. Nevertheless it is not difficult to show that
the modes with $S=1$ are indeed those with the smallest (in modulus)
imaginary part, i.e. they are the least interesting for what concerns
the instability.
This confirms the expectation that polar perturbations with $S=-1$
should have the strongest instability in the rapidly rotating limit,
as conjectured in Ref.~\cite{Rosa:2011my}. In the Proca case the
strongest instability should occur on a timescale
\begin{equation}
\tau_{\rm vector}=\omega_I^{-1}\sim\frac{M(M\mu)^{-7}}{\gamma_{-11}
(\tilde{a}-2 \mu r_+)}\,.\label{tau_vector} 
\end{equation}
The timescale above must be compared with that for the massive scalar
case~\cite{Detweiler:1980uk} for $\ell=m=1$,
\begin{equation}
 \tau_{\rm scalar}\sim\frac{48M(M\mu)^{-9}}{\tilde{a}-2 \mu r_+}\,.
\end{equation}
Roughly speaking, the instability timescale against vector polar perturbations
is of order
%
%\begin{equation}
$\tau_{\rm vector}\sim 10^{-2}\gamma_{-11}^{-1} ({M}/{M_\odot})\,{\rm s}$.
%\end{equation}
%

In the next section we shall use the results obtained from our
numerics at second order to discuss some important astrophysical
consequences of the Proca instability. In this context it is crucial
to have a robust estimate of the instability timescale in the
$M\mu\ll1$ limit. In the axial case our numerical results are
supported by the analytical formula~\eqref{anl1}, which provides
strong support that the fit~\eqref{fit} represents the correct
behavior when $M\mu\ll 1$. Unfortunately in the polar case we do not
have the same analytical insight and, as discussed above, the
small-mass regime is challenging to investigate numerically.
In order to verify the reliability of Eq.~\eqref{fit} for polar
perturbations we have tried several choices of the fitting functions
in the most relevant case, $\ell=m=1$ and $S=-1$. It turns out that
Eq.~\eqref{fit} is a conservative choice in the polar case, as other
fits generically predict a {\em stronger} instability. Furthermore, by
comparing the results in Figs.~\ref{fig:crossover} and
\ref{fig:wI_VS_J2nd} for the axial and for the polar case, it is clear
that the polar case exhibits a more complex behavior, which is hard to
reproduce without some analytical insight. In fact our numerical data
are also very well reproduced by the following fitting function:
\begin{equation}
 M\omega_I\sim\left(\tilde a -\tilde a_{\rm SR}\right)
  \left[\eta_0(M\mu)^{\kappa_0}+\eta_1\tilde{a}
    (M\mu)^{\kappa_1}\right]\,,\label{fitII}
\end{equation}
%%%%
where
\begin{equation}
 \tilde a_{\rm SR}=\frac{-m+\sqrt{m^2+16 (M \mu )^2}}{2 \mu  M}\sim\frac{4M\mu}{m}+{\cal O}(\mu^3)\,,\label{aSR}
\end{equation}
%%%
corresponding to the superradiance threshold
[Eq.~\eqref{superradiance}] when $\omega_R\sim\mu$. The
fit~\eqref{fitII} has been obtained by imposing a second-order (in
$\tilde{a}$) functional form and $\omega_I=0$ when
$\tilde{a}=\tilde{a}_{\rm SR}$, consistently with our data. Finally,
the functional dependence on $\mu$ of the two remaining terms has been
obtained by fitting the data in the instability region and for
$0.03\lesssim M\mu\lesssim0.1$. We found $\eta_0\approx-6.5\pm2$,
$\eta_1\approx 2.1\pm1$, $\kappa_0\approx 6.0\pm0.1$ and
$\kappa_1\approx 5.0\pm0.3$. While the fit~\eqref{fit} is physically
more appealing~\cite{Rosa:2011my}, Eq.~\eqref{fitII} reproduces our
numerical data in the whole instability region within a factor of
two. 
%%%%
It would be interesting to better understand the behavior of the polar
instability in the limit $M\mu\ll 1$ using different approaches, such
as a time-evolution analysis or a full numerical evolution of the
Proca equation.
%%%%
In the following we shall discuss how the choice of either
Eq.~\eqref{fit} or Eq.~\eqref{fitII} affects the astrophysical
implications of our results.

%%%%%%%%%%%%%%%%%%%%%%%%%%%%%%%%%%%%%%%%%%%%%%%%%%%%%%%%%%%%%%%%%%%%%%%%%%%%%%%%
\section{Astrophysical implications of the Proca instability}
\label{sec:astro}
%%%%%%%%%%%%%%%%%%%%%%%%%%%%%%%%%%%%%%%%%%%%%%%%%%%%%%%%%%%%%%%%%%%%%%%%%%%%%%%%

Our numerical results in the previous section indicate that polar
perturbations with $S=-1$ have the shortest instability timescale, as
conjectured in~\cite{Rosa:2011my}. For fixed values of $\tilde{a}$ and
$\mu$, the instability timescale for polar perturbations is smaller by
two-three orders of magnitude than in the axial case.  While this
conclusion relies on an extrapolation of calculations that are valid
(strictly speaking) only in the slow-rotation limit, we have shown
that a similar extrapolation in the scalar case is in remarkable
quantitative agreement with numerical results that do {\it not} rely
on the slow rotation approximation. Furthermore, in the Proca case a
second-order calculation gives solid evidence that our extrapolation
is reliable, both qualitatively and quantitatively.  Therefore it is
reasonable to expect that extrapolations from the slow-rotation limit
should at least provide the correct order of magnitude (and possibly a
reliable quantitative estimate) of the polar instability
timescale. Here we explore the tantalizing astrophysical implications
of the Proca superradiant instability. Unless otherwise stated, we
shall focus on the most relevant modes, those with $\ell=m=1$, which
correspond to the stronger instability.

%%%%%%%
\subsection{Implications from the existence of supermassive black holes}
%%%%%%%
The most conservative assumption on the final state of the instability
is that the radiation will extract angular momentum from the BH,
leaving behind a Kerr metric with dimensionless spin parameter below
the superradiant threshold: in other words, the angular frequency of
the BH horizon must be such that $\Omega_H\lesssim \mu$.
By this argument, even a single reliable supermassive BH spin
measurement can impose a stringent constraint on the allowed mass
range of massive vector fields. Bounds on the mass of the vector field
follow from the requirement that astrophysical spinning BHs should be
stable, in the sense that the timescale~\eqref{tau_vector} should be
larger than some observational threshold.

\begin{figure}[thb]
\begin{center}
\epsfig{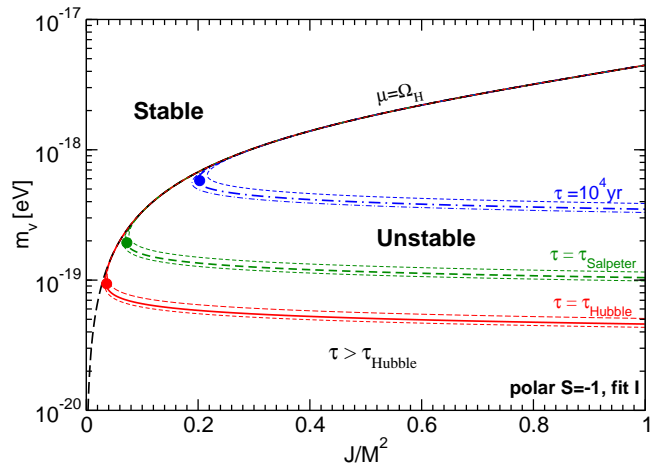}
\caption{(color online) Bounds on the photon mass $m_v=\hbar\mu$ obtained by
  extrapolating the instability timescale~\eqref{tau_vector} for
  $S=-1$ polar Proca modes of a Kerr BHs up to $\tilde a=1$. The
  region delimited by the curves corresponds to an instability
  timescale $\tau<\tau_{\rm Hubble}$ (continuous red line),
  $\tau<\tau_{\rm Salpeter}$ (dashed green line) and $\tau<10^{4}$~yr
  (dot-dashed blue line), respectively. We set
  $\gamma_{-11}\approx20\pm10$ in Eq.~\eqref{tau_vector} and consider
  a Kerr BH with $M=10^7 M_\odot$, but the results have a simple
  scaling with $\gamma_{-11}$ and with the BH mass
  (cf. Eqs.~\eqref{boundgen} and ~\eqref{bound}). Thin dashed lines
  bracket our estimated numerical errors.
  \label{fig:bound}}
\end{center}
\end{figure}

For isolated BHs we can consider the observational threshold to be the
age of the Universe, $\tau_{\rm Hubble}=1.38\times 10^{10}$~yr.  A
more conservative assumption is to include possible spin growth due to
mergers with other BHs and/or accretion. These effects are in
competition with superradiant angular momentum extraction.

\begin{figure*}[thb]
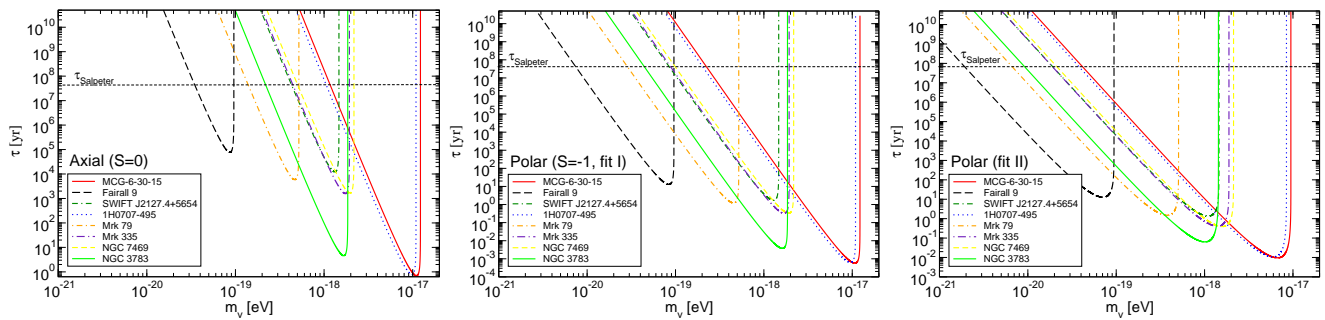

\begin{center}
\epsfig{file=fig9a.eps,width=5.75cm,angle=0,clip=true}
\epsfig{file=fig9b.eps,width=5.75cm,angle=0,clip=true}
\epsfig{file=fig9c.eps,width=5.75cm,angle=0,clip=true}
\caption{(color online) Instability timescale (in years) as a function of the vector
  mass $m_v$ (in eV) for the eight supermassive BHs listed in Table~2
  of~\cite{Brenneman:2011wz}. The left panel refers to the axial case
  ($S=0$, $\ell=m=1$); the middle panel to the polar case,
  Eq.~\eqref{fit} with $S=-1$, $\ell=m=1$ and $\gamma_{-11}\approx20$;
  the right panel to the polar case, but using the alternative fitting
  function given in Eq.~\eqref{fitII}. The horizontal line corresponds
  to $\tau=\tau_{\rm Salpeter}=4.5\times 10^7$~yr.
\label{fig:bound3}}
\end{center}
\end{figure*}

The most likely mechanism to produce fast-spinning
BHs\footnote{Supermassive BHs with moderate spin (say $\tilde a\sim
  0.7$, as in Fairall 9~\cite{Schmoll:2009gq}) could also be produced
  by comparable-mass BH mergers~\cite{Berti:2007fi}. The timescale for
  mergers depends on details of the ``final parsec problem'', which
  are poorly known~\cite{Merritt:2004gc}.  The {\em minimum} timescale
  necessary for BHs to merge is the gravitational-radiation timescale,
  which is of the same order as the Salpeter timescale (cf. Eq.~(10)
  in~\cite{Merritt:2004gc} or Eq.~(26) in~\cite{Sesana:2006ne}), but
  it is likely that the timescale for binary BHs to merge will be
  dominated by the dynamical friction timescale, which is typically of
  the order of a few Gyrs for satellites in a Milky Way-type halo
  ~\cite{Barausse:2012fy} and hence larger than the Salpeter
  timescale.} is prolonged accretion~\cite{Berti:2008af}. In this case
we can compare the superradiance timescale to the shortest timescale
over which accretion could spin up the BH. Thin-disk accretion can
increase the BH spin from $\tilde a=0$ to $\tilde a\approx1$ with a
corresponding mass increase by a factor $\sqrt{6}$
\cite{Bardeen:1970zz}. If we assume that mass growth occurs via
accretion at the Eddington limit, so that the BH mass grows
exponentially with $e$-folding time given by the Salpeter timescale
$\tau_{\rm Salpeter}=4.5\times 10^7$~yr, then the \emph{minimum}
timescale for the BH spin to grow from $\tilde a=0$ to $\tilde
a\approx1$ via thin-disk accretion is comparable to $\tau_{\rm
  Salpeter}$.  Note that accretion at the Eddington limit is a rather
conservative assumption, since more typically we would expect
accretion to be sub-Eddington. Furthermore fast-spinning BHs are hard
to produce in the presence of radiative effects~\cite{Thorne:1974ve}
or magnetohydrodynamics~\cite{Gammie:2003qi}, and ``chaotic''
accretion~\cite{King:2006uu} would make large spin parameters even
less likely.

For illustration, in Fig.~\ref{fig:bound} we consider a Kerr BH with
$M=10^7 M_\odot$, so that $m_v=10^{-17}(M\mu)$~eV. We assume
$\gamma_{-11}\approx20\pm10$ in Eq.~\eqref{tau_vector} (consistently
with our numerical results) and we plot: (i)~the superradiant
threshold (dashed black line), corresponding to the points where
$\mu\approx\Omega_H$ and the instability timescale (in the
perturbative treatment) becomes infinite; (ii)~the contours
corresponding to an instability timescale $\tau=\tau_{\rm Hubble}$
(continuous red line), $\tau=\tau_{\rm Salpeter}$ (dashed green line)
and $\tau=10^4$~yr (dash-dotted blue line), respectively. Each curve
is shown with the corresponding error bars.

In the region delimited by the continuous red line the instability
timescale is shorter than the Hubble time (of course, similar
considerations apply to the dashed green and dash-dotted blue lines).
For large $\tilde a$ the lower branches of the ``critical'' curves are
roughly horizontal. By solving $\omega_I^{-1}=\tau_c$ in the limit
$M\mu\ll1$ this regime can be well approximated by the formula
%%%%
\begin{equation}
 m_{v}^{*}=\hbar\mu^{*}= \frac{\hbar}{M}\left[\frac{M}{m\gamma_{S\ell}\tau_c}\right]^{1/(5+4\ell+2S)}\,,\label{boundgen}
\end{equation}
where $\tau_c$ is the threshold time. If $\tau_c=\tau_{\rm Hubble}$ and in the $S=-1$ polar case, we get
\begin{equation}
 m_{v}^{*}\sim\frac{7\times 10^{-20}}{
\gamma_{-11}^{1/7}}\left[\frac{10^7 M_\odot}{M}\right]^{6/7}{\rm eV}\,.\label{bound}
\end{equation}

Note that, although extracting the values of the fitting parameters
can be numerically challenging, our results depend very mildly on the
precise value of $\gamma_{-11}$.  From Eq.~\eqref{tau_vector} we see
that the instability timescale is inversely proportional to
$\gamma_{-11}$. Therefore the near-horizontal ``observability
threshold'' in Fig.~\ref{fig:bound3} that would correspond to choosing
$\tau = \tau_{\rm Salpeter}$ rather than $\tau = \tau_{\rm Hubble}$
can be obtained multiplying $\gamma_{-11}$ in Eq.~\eqref{bound} by the
ratio of the two timescales, $\tau_{\rm Salpeter}/\tau_{\rm
  Hubble}\simeq 3\times 10^{-3}$. Then the vector mass
bound~\eqref{bound} would increase by a modest factor of $2.3$. What
is crucial in obtaining strong astrophysical bounds is having a large
BH mass, because bounds scale like $\left(10^7
M_\odot/M\right)^{6/7}$.

Brenneman et al.~\cite{Brenneman:2011wz} have recently presented a
list of eight supermassive BH mass and spin estimates.  In
Fig.~\ref{fig:bound3} we show the superradiant instability timescale
$\tau$ (in years) as a function of $m_v$ (in eV) for these
supermassive BHs. The dashed horizontal line denotes $\tau_{\rm
  Salpeter}$, which we (conservatively) assume as the threshold for
the timescale to be observationally significant.  For each system we
compute instability timescales using the BH mass and spin estimates
given in Table~2 of~\cite{Brenneman:2011wz} (we select either the
average spin value quoted in the table, or the lower limit on the spin
when the observations imply $\tilde a>0.98$). The left panel refers to
axial modes with $S=0$ and $\ell=m=1$, for which we can compute the
instability timescale analytically: cf. Eq.~\eqref{anl1} and
Appendix~\ref{res_ana}.  The middle and right panels shows that
qualitatively similar results hold for polar modes with $S=-1$ and
$\ell=m=1$, which exhibit the strongest instability. In the polar
case, we computed the instability timescale using both Eq.~\eqref{fit}
(fit~I) and Eq.~\eqref{fitII} (fit~II). The results are qualitatively
similar and show that, independently of the fitting function, the
axial bounds on $m_v$ are typically one-two orders of magnitude less
stringent than the polar bounds. Fig.~\ref{fig:bound3} shows that {\em
  existing} measurements of supermassive BH spins rule out Proca field
masses in the whole range $10^{-20}$~eV$\lesssim m_v\lesssim
10^{-17}$~eV. 

%%%%%
Note that arguments based on the superradiant instability can only be used to set ``exclusion windows'' on the boson mass, rather then upper bounds~\cite{Arvanitaki:2009fg,Arvanitaki:2010sy}. However, observations of BHs in different mass ranges can exclude different mass windows. For example, stellar-size BHs, $M\simeq(5-20)M_\odot$, can set an exclusion region of $10^{-13}$~eV$\lesssim m_v\lesssim 10^{-9}$~eV. This mass range is relevant for so-called ``dark photons'' coupled to sterile neutrinos~\cite{Harnik:2012ni}. If the identified intermediate-mass BHs were confirmed to exist and if the largest known supermassive BHs with $M\simeq 2\times 10^{10}M_\odot$~\cite{2011Natur.480..215M,2012arXiv1203.1620M} were confirmed to have sufficiently large spin, combined observations of spinning BHs can potentially exclude the entire mass windows $10^{-21}$~eV$\lesssim m_v\lesssim 10^{-9}$~eV~\cite{Arvanitaki:2009fg,Arvanitaki:2010sy}. The latter is complementary to the bounds set by several other experiments and observations that exclude light bosonic particle with larger mass~\cite{PDG,Goldhaber:2008xy,Harnik:2012ni}.

%%%%%
Using current data on supermassive BHs~\cite{Brenneman:2011wz}, the best bound comes from Fairall
9~\cite{Schmoll:2009gq}, for which the polar instability implies a
conservative bound (including measurement errors) $m_v\lesssim
10^{-20}$~eV or $m_v\lesssim 10^{-21}$~eV, depending on the fitting
function~\cite{paperPRL}.  Although the instability for axial modes is weaker, in this
case our results are more precise and imply a bound as stringent as
$m_v\lesssim 4\times 10^{-20}$~eV if we consider $\tau<\tau_{\rm
  Salpeter}$, or $m_v\times 10^{-20}$~eV if we consider
$\tau<\tau_{\rm Hubble}$. Even under our very conservative assumptions
these results are of great significance, since the current best bound
on the photon mass is $m_\gamma<10^{-18}$~eV~\cite{PDG,Goldhaber:2008xy}.
Combining the exclusion window derived above, $10^{-20}\lesssim m_v\lesssim 10^{-17}$~eV, with existing bounds~\cite{PDG,Goldhaber:2008xy}, we find the most stringent upper bound on the photon mass, $m_\gamma\lesssim 10^{-20}$ eV.

We remind the reader that our results are also summarized in
Fig.~\ref{fig:ReggePlane}, where we show instability windows for axial
and polar modes together with supermassive BH mass and spin estimates
from Ref.~\cite{Brenneman:2011wz}.

%%%%%%%%%%%%%%%%%%%%%%%%%%%%%%%%%%%%%%%%%%%%%%%%%%%%%%%%%%%%%%%%%%%%%%%%%%%%%%
\subsection{Nonlinear effects, other couplings}
%%%%%%%%%%%%%%%%%%%%%%%%%%%%%%%%%%%%%%%%%%%%%%%%%%%%%%%%%%%%%%%%%%%%%%%%%%%%%%
An important ingredient that was not taken into account in our study
is the nonlinear evolution of the instability, that can modify the
background geometry. Photon self-interactions are very weak, being
suppressed by the mass of the electron. Therefore, it is quite likely
that the outcome of the instability will be a slow and gradual
drainage of the hole's rotational energy. However, exotic massive vector fields may show nonlinearities as those studied in Ref.~\cite{Kodama:2011zc,Yoshino:2012kn} for the scalar case.

Another important aspect that should be investigated is whether the coupling of
accreting matter to massive bosons can quench the instability. In principle
massive photons (unlike hidden $U(1)$ fields, for which the interaction with
matter is very small) can couple strongly to matter. However it is unlikely that
this will significantly affect the superradiant instability discussed here, for
two reasons. The first is that unstable modes are large-scale coherent modes
whose Compton wavelength is comparable to (or larger than) the size of the BH's
event horizon, and accretion disks are typically charge-neutral over these
lengthscales, so any possible coupling with ordinary neutral matter is
incoherent and most likely inefficient.
The second is that accretion disks are expected to be localized on the
equatorial plane, and therefore can affect at most some (but not all) of the unstable modes.  
The investigation of the superradiant instability in the
presence of matter requires further work, but the above considerations suggest
that vacuum estimates should be reliable. Spin estimates for slowly accreting
BHs (such as the BH at the center of our own galaxy) are the most reliable, in
that they should be less sensitive to details of the interaction of vector
fields with matter.

%%%%%%%%%%%%%%%%%%%%%%%%%%%%%%%%%%%%%%%%%%%%%%%%%%%%%%
\section{Conclusions and extensions}
\label{sec:conclusion}
%%%%%%%%%%%%%%%%%%%%%%%%%%%%%%%%%%%%%%%%%%%%%%%%%%%%%% 

BH perturbation theory is a powerful tool, with important applications
in astrophysics and high-energy physics, but its applicability is
limited when the perturbation equations cannot be separated. We have
discussed a general method to circumvent these difficulties in the
case of slowly rotating BH spacetimes. The method was originally
developed to study the modes of rotating stars but, as we showed, it applies to
\emph{any} slowly rotating BH background and to any kind of
perturbation. Within this general framework, we discussed two effects
induced by rotation: (i) the Zeeman-like splitting of the
eigenfrequencies, that breaks the degeneracy of modes with different
azimuthal number $m$, and (ii) a Laporte-like selection rule, in which
modes with multipolar index $\ell$ are coupled to those with
multipolar index $\ell\pm1$ and opposite parity and to those with
$\ell\pm2$ and same parity. We have extended Kojima's arguments to show
that these couplings do not affect the QNM frequencies of the
spacetime to first order in the rotation rate, and we verified this
claim by numerical integration of the perturbation equations.

As a first relevant application of this method, we studied massive
vector (Proca) perturbations on a slowly rotating Kerr background,
i.e. the only interesting class of perturbations of Kerr BHs in four
dimensions that does not seem to be separable. Using our approach we
reduced the equations to a set of coupled ODEs, that can be solved by
a straightforward extension of methods valid in the nonrotating case.
We have computed the corrections to the Proca eigenfrequencies of a
Kerr BH to first and to second order in the rotation rate. For the
first time we showed that the Proca bound-state modes become unstable
when the superradiant condition~\eqref{superradiance_cond} is met. We
proved this numerically for the polar modes, and both numerically and
analytically for the axial modes in the limit $M\mu\ll
1$. Unfortunately the method becomes intrinsically inaccurate for
large values of the rotation rate. This prevents us from accurately
computing the instability timescale close to extremality. By fitting
and extrapolating our data we confirmed the scaling with $m_v$
conjectured in Ref.~\cite{Rosa:2011my}, and we also estimated the
order of magnitude of the instability timescale. We expect our results
to be reliable up to $\tilde{a}\sim0.7$. Our estimates should be
confirmed by more accurate methods, but they provide strong evidence
that measurements of rotating supermassive BH spins set the most
stringent upper bounds on the photon mass
(cf. Fig.~\ref{fig:ReggePlane}).

Besides their astrophysical implications, our results are relevant
also for some classes of gravitational theories with higher-order
curvature terms in the action. For example, as proved by
Buchdahl~\cite{Buchdahl:1979ut}, a theory defined by ${\cal
  L}=\sqrt{-g}\left(R+\alpha R_{[ab]}R^{[ab]}\right)$ in the Palatini
approach is dynamically equivalent to the Einstein-Proca system when
the vector mass $\mu^2=3/|\alpha|$ (see also
Ref.~\cite{Vitagliano:2010pq}). The Proca instability of Kerr BHs in
this theory can put constraints on the parameter $\alpha$, which is
related to the nonmetricity of the connection.

While limited by the slow-rotation assumption, the power of the
present method consists in its generality. The slow-rotation
approximation allows us to study linear perturbations of any slowly
rotating spacetime even when the equations are nonseparable.  For
instance it can be used to study gravitational-electromagnetic
perturbations of Kerr-Newman BHs in four
dimensions~\cite{1992mtbh.book.....C}, QNMs of Myers-Perry
BHs~\cite{Durkee:2010qu} and BH greybody factors in higher-dimensional
rotating spacetimes~\cite{Cardoso:2012qm,Herdeiro:2011uu}.
Another interesting extension is the stability analysis of rotating BHs in
asymptotically anti-de Sitter spacetimes. This is relevant in the context of the
gauge/gravity duality, as the QNMs of anti-de Sitter BHs in five dimensions are
holographically related to specific correlation functions and transport
coefficients of the dual boundary theory~\cite{Berti:2009kk}.

We note that perturbations of rotating stars have been studied up to
second order in the Cowling approximation (see
e.g. Refs.~\cite{Sotani:2006at,Stavridis:2007xz,Passamonti:2007td}),
where metric perturbations are neglected and the equations have a
structure reminiscent of our general Eqs.~\eqref{epF1c} and
\eqref{epF2c}. 
Finally we remark that the techniques discussed in this paper are not
limited to general relativity. They are also useful to study
gravitational perturbations and the linear stability of slowly
rotating BH metrics in alternative theories of gravity, for example in
quadratic gravity~\cite{Yunes:2009hc,Pani:2011gy,Yagi:2012ya}.
We hope to report on these interesting extensions in the near future.

%%%%%%%%%%%%%%%%%%%%%%%%%%%%%%%%%%%%%%%%%%%%%%%%%%%%%%%%%%%%%%%%%%%%%%%%%%%%%%
\begin{acknowledgments}
  We wish to thank the Axiverse Project members (especially Hideo Kodama),
  Antonino Flachi, Stefano Liberati and Jo\~ao Rosa for valuable discussions, and the referees for their
  constructive comments. This work was supported by the DyBHo--256667 ERC
  Starting Grant, the NRHEP--295189 FP7--PEOPLE--2011--IRSES Grant, and by FCT -
  Portugal through PTDC projects FIS/098025/2008, FIS/098032/2008,
  CTE-ST/098034/2008, CERN/FP/123593/2011. A.I. was supported by JSPS
  Grant-in-Aid for Scientific Research Fund No. 22540299 and No. 22244030.
  E.B. was supported by NSF Grant No. PHY-0900735 and NSF CAREER Grant
  No. PHY-1055103. P.P. acknowledges financial support provided by the European
  Community through the Intra-European Marie Curie contract aStronGR-2011-298297
  and the kind hospitality of the Department of Physics, University of Rome
  ``Sapienza'' and of the International School for Advanced Studies (SISSA) in
  Trieste. V.C. thanks the Yukawa Institute for Theoretical Physics at Kyoto
  University for hospitality during the YITP-T-11-08 workshop on ``Recent
  advances in numerical and analytical methods for black hole dynamics''.
\end{acknowledgments}
%%%%%%%%%%%%%%%%%%%%%%%%%%%%%%%%%%%%%%%%%%%%%%%%%%%%%%%%%%%%%%%%%%%%%%%%%%%%%%
\appendix

 %%%%%%%%%%%%%%%%%%%%%%%%%%%%%%%%%%%%%%%%%%%%%%%%%%%%%% 
\section{Coefficients of the perturbation equations}
\label{app:coeffs}
 %%%%%%%%%%%%%%%%%%%%%%%%%%%%%%%%%%%%%%%%%%%%%%%%%%%%%%
In this Appendix we list the explicit form of the coefficients appearing in the
perturbation equations. We remark that the coefficients
$A_\ell,D_\ell,A_\ell^{(i)},B_\ell^{(i)},D_\ell^{(i)}$ incorporate terms at zeroth, first and
second order in the rotation parameter $\tilde a$. Therefore, there is no direct
correspondence between these coefficients and the quantities at fixed order in
$\tilde a$, such as ${\cal A}_{\ell}$, $\bar{\cal A}_{{\ell}}$, $\hat{{\cal
    A}}_\ell$, etc., which have been introduced in Sec.~\ref{sec:general}. In the following, a
prime denotes derivative with respect to the radial coordinate $r$.
%%%%%%%%%%%%%%%%%%%%%%%%%%%%%%%%%%%%%%%%%%%%%%%%%%%%%% 
\subsection{Coefficients of the scalar equation}\label{app:coeff_scalar}
%%%%%%%%%%%%%%%%%%%%%%%%%%%%%%%%%%%%%%%%%%%%%%%%%%%%%% 
%%%
The explicit form of the coefficients appearing in Eq.~\eqref{eq_expY} is 
%%%%
 \begin{eqnarray}
  A_{\ell} &=& -2 \left[M^2 r^2 \left(\left(4 M^2+2 M r \left(\Lambda-1+r^2 
\mu ^2\right)+r^4 \omega ^2\right.\right.\right.\nn\\
&&\left.\left.\left.-r^2 \left(\Lambda+r^2 \mu ^2\right)\right) 
\Psi_\ell+r (r-2M) \left(2 M \Psi_\ell' \right.\right.\right.\nn\\
&&\left.\left.\left.+r (r-2M) \Psi_\ell''\right)\right)\right]+8 \tilde{a} m M^4 r^3 \omega  \Psi_\ell +{\tilde{a}^2M^4} \left[\left(36 M^2 \right.\right.\nn\\
&&\left.\left.+2 M r \left(\Lambda-11+r^2 \mu
      ^2\right)-r^2 
\left(\Lambda-2+2 m^2+r^2 \mu ^2\right)\right.\right.\nn\\
&&\left.\left.+\frac{r^3\omega ^2}{r-2M} \left(8 M^2-2 M r+r^2
\right) \right) \Psi_\ell \right.\nn\\
&&\left.+r (r-2M) \left(10 M \Psi_\ell'-r (2 M+r) \Psi_\ell''\right)\right]\,,\\
%%%%
  D_{\ell} &=& 2 \tilde{a}^2M^4 \left[\left(4 M^2-\Lambda r^2+2 M r 
\left(\Lambda-1+r^2 \omega ^2\right)\right) \Psi_\ell\right.\nn\\
&&\left.+r (r-2M) \left(2 M 
\Psi_\ell'+r (r-2M) \Psi_\ell''\right)\right]  \,.
\end{eqnarray}
%%%%
\begin{widetext}
 %%%%%%%%%%%%%%%%%%%%%%%%%%%%%%%%%%%%%%%%%%%%%%%%%%%%%% 
\subsection{Coefficients of the Proca perturbation equations}\label{app:coeff}
%%%%%%%%%%%%%%%%%%%%%%%%%%%%%%%%%%%%%%%%%%%%%%%%%%%%%%%%%%%%%%%%%%%%%%%%%%%%%
In this Appendix we list the coefficients appearing in
Eqs.~\eqref{eq1i}--\eqref{eqL}. The coefficients read 
%%%%
\begin{eqnarray}
 A_\ell^{(0)}&=& \frac{{\tilde{a}}^2 M^2}{2 r^6} \left[\frac{{u}_{(1)}^\ell \left(-r^2
   \left(\Lambda +2 m^2+\mu ^2 r^2-2\right)+32 M^2+2 M r \left(\Lambda +\mu
   ^2 r^2-10\right)\right)}{2 M-r}+\right.\nn\\
   &&\left.\frac{r}{\ell
   (\ell+1) (r-2 M)^2} \left((2 M-r) \left(\ell
   (\ell+1) (2 M-r) \left(-12 M {u'}_{(1)}^\ell+r (2 M+r)
   {u''}_{(1)}^\ell+i r^2 \omega 
   {u'}_{(2)}^\ell\right)-2 i m^2 r^2 \omega 
   {u}_{(3)}^\ell\right)\right.\right.\nn\\
% \end{eqnarray}
% \begin{eqnarray}
   &&\left.\left.-i \Lambda r \omega 
   \left(-8 M^2-2 M r+r^2\right) {u}_{(2)}^\ell\right)\right]+\frac{2 {\tilde{a}} m M^2
   \left(\Lambda r \omega  {u}_{(1)}^\ell-i
   \left((2 M-r) {u'}_{(3)}^\ell-r^2 \omega ^2
   {u}_{(3)}^\ell+\Lambda
   {u}_{(2)}^\ell\right)\right)}{\Lambda r^4 (2
   M-r)}\nn\\
   &&-\frac{(2 M-r) \left(\Lambda +\mu ^2 r^2\right)
   {u}_{(1)}^\ell+r (r-2 M)^2 {u''}_{(1)}^\ell+i r \omega
    \left((2 M-r) \left({u}_{(3)}^\ell-r
   {u'}_{(2)}^\ell\right)+(r-4 M)
   {u}_{(2)}^\ell\right)}{r^3 (2 M-r)}\,,\\
%%%%%%%%%%%
 A_\ell^{(1)}&=& \frac{{\tilde{a}}^2 M^2}{2 \Lambda r^4 (2
   M-r)^3} \left[\Lambda
   {u}_{(2)}^\ell \left((2 M-r) \left({\ell^2} (2
   M+r)+\ell (2 M+r)+r \left(2 m^2+\mu ^2 r (2 M+r)\right)\right)+r^2
   \omega ^2 \left(8 M^2+2 M r+r^2\right)\right)\right.\nn\\
   &&\left.+2 m^2 r (r-2 M)^2
   {u'}_{(3)}^\ell+i \Lambda r \omega  (2 M-r)
   \left(8 M^2-2 M r+r^2\right) {u'}_{(1)}^\ell-i \ell
   (\ell+1) \omega  (2 M-r) \left(8 M^2-6 M r+3 r^2\right)
   {u}_{(1)}^\ell\right]\nn\\
   &&+\frac{2 {\tilde{a}} m M^2 \left(r^2 \omega  (r-2 M)
   {u'}_{(3)}^\ell+i \Lambda \left((2 M-r)
   {u}_{(1)}^\ell+r \left((r-2 M) {u'}_{(1)}^\ell+2 i r
   \omega  {u}_{(2)}^\ell\right)\right)\right)}{\ell
   (\ell+1) r^4 (r-2 M)^2}\nn\\
   &&+\frac{(2 M-r) \left((2 M-r)
   {u'}_{(3)}^\ell+i r^2 \omega 
   {u'}_{(1)}^\ell\right)+{u}_{(2)}^\ell \left((2 M-r)
   \left(\Lambda +\mu ^2 r^2\right)+r^3 \omega ^2\right)-i r \omega  (2 M-r)
   {u}_{(1)}^\ell}{r^2 (r-2 M)^2}\,,\\
%%%%%%%%%%%
 A_\ell^{(2)}&=&\frac{{\tilde{a}}^2 M^2  \left((2 M-r) \left(-2 m^2 r
   {u}_{(3)}^\ell-\Lambda r (r-2 M)
   {u'}_{(2)}^\ell-\Lambda (2 M-r)
   {u}_{(2)}^\ell\right)-i \Lambda r \omega 
   \left(8 M^2-2 M r+r^2\right) {u}_{(1)}^\ell\right)}{\ell
   (\ell+1) (r-2 M)^2}\nn\\
   &&+\frac{4 {\tilde{a}} m M^2 r  (r
   \omega  {u}_{(3)}^\ell+i \Lambda
   {u}_{(1)}^\ell)}{\Lambda (2 M-r)}+\frac{2 r^2
    \left((2 M-r) \left(r
   {u'}_{(2)}^\ell+{u}_{(2)}^\ell-{u_{(3)}}^\ell(r
   )\right)-i r^2 \omega  {u}_{(1)}^\ell\right)}{2 M-r}\,,\\
%%%%%%%%%%%
% \end{eqnarray}
% %%% aaa
% \begin{eqnarray}
%%%%%%%%%%%
 \tilde{A}_\ell^{(0)}&=&\frac{4 {\tilde{a}} M^2 {u}_{(4)}^\ell}{r^5}-\frac{4
   {\tilde{a}}^2 m M^3 \omega  {u}_{(4)}^\ell}{\Lambda  r^5} \,,\\
%%%%%%%%%%%
 \tilde{A}_\ell^{(1)}&=& -\frac{2 i {\tilde{a}}^2 m M^2 {u'}_{(4)}^\ell}{\Lambda  r^4}\,,\\
%%%%%%%%%%%
 \tilde{A}_\ell^{(2)}&=&0 \,,\\
%%%%%%%%%%%
 B_\ell^{(0)}&=& \frac{{\tilde{a}}^2 m M^2 \omega  {u}_{(4)}^\ell}{\ell
   (\ell+1) r^3 (2 M-r)}-\frac{2 {\tilde{a}} M^2 \left((r-2 M)
   {u'}_{(4)}^\ell+r^2 \omega ^2
   {u}_{(4)}^\ell\right)}{\Lambda r^4 (2 M-r)}\,,\\
%%%%%%%%%%%
 B_\ell^{(1)}&=& \frac{i {\tilde{a}}^2 m M^2 {u'}_{(4)}^\ell}{\ell
   (\ell+1) r^3 (2 M-r)}-\frac{2 i {\tilde{a}} M^2 \omega 
   {u'}_{(4)}^\ell}{\Lambda r^2 (2 M-r)}\,,\\
%%%%%%%%%%%
 B_\ell^{(2)}&=& \frac{2 i {\tilde{a}}^2 m M^2 r 
   {u}_{(4)}^\ell}{\Lambda (r-2 M)}-\frac{4 i
   {\tilde{a}} M^2 r^2 \omega  
   {u}_{(4)}^\ell}{\Lambda (r-2 M)}\,,\\
%%%%%%%%%%%
 \tilde{B}_\ell^{(0)}&=& \frac{4 {\tilde{a}}^2 M^3 (\Lambda
   {u}_{(1)}^\ell+i r \omega  {u}_{(3)}^\ell)}{\ell
   (\ell+1) r^6}\,,\\
%%%%%%%%%%%
 \tilde{B}_\ell^{(1)}&=& \frac{2 {\tilde{a}}^2 M^2 \left((r-2 M)
   {u'}_{(3)}^\ell-\Lambda
   {u}_{(2)}^\ell\right)}{\Lambda r^4 (2 M-r)} \,,\\
%%%%%%%%%%%
 \tilde{B}_\ell^{(2)}&=& 0 \,,
%%%%%%%%%%%
%%%%%%%%%%%
\end{eqnarray}
%%% 
\begin{eqnarray}
%%%%%%%%%%%
 D_\ell^{(0)}&=& \frac{{\tilde{a}}^2 M^2}{r^6 (2 M-r)} \left[-(2 M-r) (6 M-\Lambda r)
   {u}_{(1)}^\ell+r \left((2 M-r) \left(6 M
   {u'}_{(1)}^\ell-r (r-2 M) {u''}_{(1)}^\ell+i r \omega 
   \left({u}_{(3)}^\ell-r {u'}_{(2)}^\ell\right)\right)\right.\right.\nn\\
   &&\left.\left.-i r \omega  (10 M-r) {u}_{(2)}^\ell\right)\right]\,,\\
%%%%%%%%%%%
 D_\ell^{(1)}&=& \frac{{\tilde{a}}^2 M^2}{r^4 (r-2 M)^2} \left[{u}_{(2)}^\ell \left(-\ell
   (\ell+1) (2 M-r)-2 M r^2 \omega ^2\right)+(2 M-r) \left((r-2 M)
   {u'}_{(3)}^\ell-2 i M r \omega 
   {u'}_{(1)}^\ell\right)+2 i M \omega  (2 M-r)
   {u}_{(1)}^\ell\right]\,,\nn\\ \\
%%%%%%%%%%%
 D_\ell^{(2)}&=& \frac{2 {\tilde{a}}^2 M^2}{2 M-r}  \left[-(2 M-r) \left(r
   {u'}_{(2)}^\ell+{u}_{(2)}^\ell-{u_{(3)}}^\ell(r
   )\right)+2 i M r \omega  {u}_{(1)}^\ell\right]\,,
\end{eqnarray}
%%%
and
%%%%
\begin{eqnarray}
%%%%%%%%%%%
 \alpha_\ell&=& \frac{{\tilde{a}}^2 M^2}{2 \Lambda r^4 (r-2
   M)^2} \left[-48 M^3 {u'}_{(3)}^\ell+4 M^2 r
   {\ell^2} {u'}_{(2)}^\ell+4 \ell M^2 r
   {u'}_{(2)}^\ell+48 M^2 r {u'}_{(3)}^\ell-4 M r^3
   \omega ^2 {u}_{(3)}^\ell+i \Lambda r^2 \omega 
   (2 M+r) {u}_{(1)}^\ell\right.\nn\\
   &&\left.-4 M r^2 {\ell^2}
   {u'}_{(2)}^\ell-4 \ell M r^2 {u'}_{(2)}^\ell-12
   M r^2 {u'}_{(3)}^\ell-3 \Lambda (r-2 M)^2
   {u}_{(2)}^\ell+4 M r (r-2 M)^2 {u''}_{(3)}^\ell+r^3
   {\ell^2} {u'}_{(2)}^\ell+\ell r^3
   {u'}_{(2)}^\ell\right]\nn\\
   &&+\frac{2 {\tilde{a}} m M^2 (r \omega  {u}_{(3)}^\ell-i
   \Lambda {u}_{(1)}^\ell)}{\ell
   (\ell+1) r^3 (2 M-r)}+\frac{1}{\Lambda r^2 (2 M-r)}\left[4 M^2 {u'}_{(3)}^\ell-2 \mu ^2
   M r^2 {u}_{(3)}^\ell-2 M r {\ell^2}
   {u'}_{(2)}^\ell-2 \ell M r
   {u'}_{(2)}^\ell+\Lambda (2 M-r)
   {u}_{(2)}^\ell\right.\nn\\
   &&\left.-2 M r {u'}_{(3)}^\ell-r (r-2 M)^2
   {u''}_{(3)}^\ell+\mu ^2 r^3 {u}_{(3)}^\ell-r^3 \omega
   ^2 {u}_{(3)}^\ell+i \Lambda r^2 \omega 
   {u}_{(1)}^\ell+r^2 {\ell^2}
   {u'}_{(2)}^\ell+\ell r^2
   {u'}_{(2)}^\ell\right]\,,\\
%%%%%%%%%%%
 \beta_\ell&=& \frac{{\tilde{a}}^2 M^2 \left(-{u}_{(4)}^\ell \left(\ell
   (\ell+1) (r-2 M)^2-2 M r^3 \omega ^2\right)-2 M (r-2 M)^2 \left(r
   {u''}_{(4)}^\ell-3
   {u'}_{(4)}^\ell\right)\right)}{\Lambda r^4
   (r-2 M)^2}-\frac{2 {\tilde{a}} m M^2 \omega 
   {u}_{(4)}^\ell}{\Lambda r^2 (2
   M-r)}\nn\\
   &&+\frac{{u}_{(4)}^\ell \left((2 M-r) \left(\Lambda +\mu ^2
   r^2\right)+r^3 \omega ^2\right)+(2 M-r) \left(r (2 M-r)
   {u''}_{(4)}^\ell-2 M
   {u'}_{(4)}^\ell\right)}{\Lambda r^2 (2 M-r)}\,,\\
%%%%%%%%%%%
 \zeta_\ell&=& -\frac{{\tilde{a}}^2 m M^2 \left(2 \Lambda M r \omega 
   {u}_{(1)}^\ell-i \left(-2 M r^2 \omega ^2
   {u}_{(3)}^\ell+(r-2 M) \left(-8 M {u'}_{(3)}^\ell-r
   (r-2 M) {u''}_{(3)}^\ell+\Lambda r
   {u'}_{(2)}^\ell\right)+\Lambda (8 M-r)
   {u}_{(2)}^\ell\right)\right)}{\Lambda r^4 (2
   M-r)}\nn\\
   &&-\frac{6 {\tilde{a}} M^2 \left((2 M-r) {u}_{(1)}^\ell+r
   \left((r-2 M) {u'}_{(1)}^\ell+i r \omega 
   {u}_{(2)}^\ell\right)\right)}{r^4 (2 M-r)}\,,\\
%%%%%%%%%%%
 \eta_\ell&=& \frac{2 i {\tilde{a}} M^2 \omega  {u}_{(4)}^\ell}{2 M
   r^2-r^3}-\frac{i {\tilde{a}}^2 m M^2 \left(-2 \left(6 M^2-5 M
   r+r^2\right) {u'}_{(4)}^\ell+2 M \left(\Lambda +r^2 \omega
   ^2\right) {u}_{(4)}^\ell+r (r-2 M)^2
   {u''}_{(4)}^\ell\right)}{\Lambda r^4 (2 M-r)}\,,\\
%%%%%%%%%%%
 \rho_\ell&=& \frac{{\tilde{a}}^2 M^2}{\Lambda r^4 (2 M-r)} \left[12 M^2 {u'}_{(3)}^\ell-2 M r^2
   \omega ^2 {u}_{(3)}^\ell+2 i \Lambda M r
   \omega  {u}_{(1)}^\ell-2 M r {\ell^2}
   {u'}_{(2)}^\ell-2 \ell M r {u'}_{(2)}^\ell+3
   \Lambda (2 M-r) {u}_{(2)}^\ell\right.\nn\\
   &&\left.-10 M r
   {u'}_{(3)}^\ell-r (r-2 M)^2 {u''}_{(3)}^\ell+r^2
   {\ell^2} {u'}_{(2)}^\ell+\ell r^2
   {u'}_{(2)}^\ell+2 r^2
   {u'}_{(3)}^\ell\right]\,,\\
%%%%%%%%%%%
 \lambda_\ell&=& -\frac{4 {\tilde{a}}^2 M^3 {u}_{(4)}^\ell}{r^5}\,,\\
%%%%%%%%%%%
 \sigma_\ell&=& 0\,,\\
%%%%%%%%%%%
 \gamma_\ell&=& \frac{{\tilde{a}}^2 M^2}{\Lambda r^4 (2
   M-r)} \left[{u}_{(4)}^\ell \left(\ell
   (\ell+1) (2 M-r)+2 M r^2 \omega ^2\right)+(2 M-r) \left(r (2 M-r)
   {u''}_{(4)}^\ell-8 M
   {u'}_{(4)}^\ell\right)\right]\,.\\
%%%%%%%%%%%
\end{eqnarray}
\end{widetext}

%

%%%%%%%%%%%%%%%%%%%%%%%%%%%%%%%%%%%%%%%%%%%%%%%%%%%%%% 
\section{Proca equation for a slowly rotating Kerr BH}
\label{app:full}
%%%%%%%%%%%%%%%%%%%%%%%%%%%%%%%%%%%%%%%%%%%%%%%%%%%%%% 

In this Appendix we list the perturbation equations for a Proca field
on a slowly rotating Kerr background to first order in $\tilde{a}$. 
Equation~\eqref{eqset3} reads
%%%
\begin{eqnarray}
&&  \hat{\cal D}_2 u_{(4)}^{\ell}-\frac{4 \tilde aM^2 m\omega}{r^3}u_{(4)}^{\ell}\nn\\
&&=\frac{6 \tilde aM^2}{r^4} \left[(\ell+1) {\cal Q}_{\ell} \left(F u_{(1)}^{\ell-1}-\ii r \omega  u_{(2)}^{\ell-1}-F r {u'}_{(1)}^{\ell-1}\right)\right.\nn\\
&&\left.+\ell {\cal Q}_{\ell+1} \left(\ii r \omega  u_{(2)}^{\ell+1}-F u_{(1)}^{\ell+1}+F r {u'}_{(1)}^{\ell+1}\right)\right] \,,\label{axialRD_full}
\end{eqnarray}
%%%
where the operator $\hat{\cal D}_2$ is defined in Eq.~\eqref{D2}. The Lorenz
condition (Eq.~\eqref{eqset1} with $I=2$) becomes
\begin{align}
& \ii r \omega  u_{(1)}^{\ell}+F \left( u_{(2)}^{\ell}-  
 u_{(3)}^{\ell}+r  {u'}_{(2)}^{\ell}\right)\nn\\
&- \frac{2 \tilde aM^2 m}{r^2}\left(\ii  u_{(1)}^{\ell}+ \frac{r \omega}{\Lambda}  u_{(3)}^{\ell}\right)\nn\\
&=\frac{2 \ii \tilde aM^2  \omega}{r\Lambda} \left[(\ell+1) {\cal Q}_{\ell} u_{(4)}^{\ell-1}-\ell {\cal Q}_{\ell+1} u_{(4)}^{\ell+1}\right]\,. \label{lorenz_full}
\end{align}
%%%
Using Eq.~\eqref{lorenz_full}, Eqs.~\eqref{eqset1} with $I=0,1$ and
Eq.~\eqref{eqset2} read
\begin{align}
& \hat{\cal D}_2 u_{(3)}^\ell+\frac{2F\ell(\ell+1)}{r^2}u_{(2)}^\ell+\frac{2 \tilde aM^2 m
}{r^4}\left[r \omega  (3 u_{(2)}^\ell-2 u_{(3)}^\ell)\right.\nn\\
&\left.+3 \ii F \left(u_{(1)}^\ell-r {u'}_{(1)}^\ell\right)\right]=0\,,\label{pol1RD_full}\\
& \hat{\cal D}_2
u_{(2)}^\ell-\frac{2F}{r^2}\left(1-\frac{3M}{r}\right)\left[u_{(2)}^\ell-u_{(3)}^\ell\right]\nn\\
&-\frac{2 \tilde aM^2 m}{\ell(\ell+1) r^4}\left[\ell (\ell+1) (2 r \omega  u_{(2)}^\ell-3 \ii F u_{(1)}^\ell)-3 r \omega  F u_{(3)}^\ell\right]\nn\\
&=-\frac{6 \ii \tilde a M^2 F  \omega}{\ell(\ell+1) r^3}  \left[(\ell+1) {\cal Q}_{\ell} u_{(4)}^{\ell-1}-\ell {\cal Q}_{\ell+1} u_{(4)}^{\ell+1}\right]\,,\label{pol2RD_full}\\
& \hat{\cal D}_2 u_{(1)}^\ell-\frac{2 M}{r^2} \left(\ii \omega  u_{(2)}^\ell+F
    {u'}_{(1)}^\ell\right)\nn\\
&-\frac{2 \tilde aM^2 m}{\ell(\ell+1) r^4} \left[\ell(\ell+1) (2 r \omega  u_{(1)}^\ell-\ii u_{(2)}^\ell)+\ii r F {u'}_{(3)}^\ell\right]\nn\\
&=-\frac{2 \tilde aM^2 F}{\ell(\ell+1) r^4}  \left[(\ell+1) {\cal Q}_{\ell} \left(2 \ell u_{(4)}^{\ell-1}+r {u'}_{(4)}^{\ell-1}\right)\right.\nn\\
&\left.+\ell {\cal Q}_{\ell+1} \left(2 (\ell+1) u_{(4)}^{\ell+1}-r {u'}_{(4)}^{\ell+1}\right)\right]\,.\label{pol3RD_full}
\end{align}
The system~\eqref{axialRD_full}, \eqref{pol1RD_full}--\eqref{pol3RD_full} can be
greatly simplified by a systematic use of Eq.~\eqref{lorenz_full} to eliminate
$u_{(1)}^\ell$ and $u_{(1)}^{\ell\pm1}$. Solving Eq.~\eqref{lorenz_full} for
$u_{(1)}^\ell$ and substituting in Eq.~\eqref{pol1RD_full}--\eqref{pol2RD_full}
we get, to first order in $\tilde a$, the equations listed in the main text,
namely Eqs.\eqref{pol1RDb} and~\eqref{pol2RDb}.
%%%

In order to simplify the axial equation~\eqref{axialRD_full}, we consider
Eq.~\eqref{lorenz_full} with $\ell-1$ and $\ell+1$ and solve it for
$u_{(1)}^{\ell-1}$ and $u_{(1)}^{\ell+1}$, respectively. To first order in
$\tilde{a}$, and making also use of Eq.~\eqref{pol1RDb} and~\eqref{pol2RDb}, we
get Eq.~\eqref{axialRDb}.

%

%%%%%%%%%%%%%%%%%%%%%%%%%%%%%%%%%%%%%%%%%%%%%%%%%%%%%%%%%%%%%%%%%%%%
\section{Analytical results for the axial modes}
\label{res_ana}
%%%%%%%%%%%%%%%%%%%%%%%%%%%%%%%%%%%%%%%%%%%%%%%%%%%%%%%%%%%%%%%%%%%%

In this appendix we generalize Detweiler's calculation~\cite{Detweiler:1980uk}
of the unstable massive scalar modes of a Kerr BH to the case of a massive
vector field. We focus on the axial sector and, unlike Detweiler, we work in the
slow-rotation limit up to first order in $\tilde{a}$.

As discussed in the main text, the axial vector equation~\eqref{axialRD} and the
equation for scalar perturbations~\eqref{master} can be written in a unified
form, Eq.~\eqref{master}. Thus, in order to include both scalar and axial vector
perturbations, we shall solve Eq.~\eqref{master} in the Kerr case, where
$F=B=1-2M/r$ and $\varpi=2\tilde aM^2/r$.  By defining $R(r)=\Psi/r$ we get
\begin{eqnarray}
&& r^2F\frac{d}{dr}\left(r^2F\frac{dR}{dr}\right)+\left[r^4 
\omega ^2-4 \tilde{a} m r \omega\right.\nn\\
&&\left.-r^2F\left(\ell(\ell+1)+\mu^2 r^2-\frac{2 M s^2}{r}\right) \right]R=0\,.\label{det_full}
\end{eqnarray}
For $s=0$, the equation above is equivalent to
Eq.~(7) in Ref.~\cite{Detweiler:1980uk} at first order. For $s=\pm1$ and $\tilde{a}=0$
it is equivalent to Eq.~(51) in Ref.~\cite{Rosa:2011my}.

Following Starobinski's method of matching asymptotics~\cite{Starobinsky}, we first expand
Eq.~\eqref{det_full} for $r\gg M$:
\begin{equation}
 \frac{d^2(rR)}{dr^2}\left[\omega^2-\mu^2+\frac{2M\mu^2}{r}-\frac{\ell(\ell+1)}{r^2}\right]rR=0\,.
\end{equation}
%%%%
The solution of this equation with the correct boundary condition at infinity
reads
\begin{equation}
 R_\infty(x)\propto x^\ell e^{-x/2}U(\ell+1-\nu,2\ell+2,x)\,,
\end{equation}
where $U(p,q,x)$ is one of the confluent hypergeometric functions~\cite{abramowitz+stegun}, $x=2k_{\infty}r$, $k_\infty^2=\mu^2-\omega^2$ and
$\nu=M\mu^2/k_\infty$. The small distance ($x\ll1$) behavior
of the solution above reads
\begin{eqnarray}
 R_\infty(r)&\sim&(-1)^n\frac{(2\ell+1+n)!}{(2\ell+1)!}(2{k_\infty}r)^\ell\nn\\
&+&(-1)^{n+1}\delta\nu(2\ell)!(2{k_\infty}r)^{-\ell-1}\,,\label{RINFsmall}
\end{eqnarray}
where we have defined $\delta\nu=\nu-\ell-1-n$ and assumed $\delta\nu\sim x^{2\ell+1}\ll1$~\cite{Detweiler:1980uk}.

On the other hand, Eq.~\eqref{det_full} can be solved analytically also when
$r\ll{\rm max}(\ell/\omega,\ell/\mu)$. In this limit, by introducing a new
coordinate $z=(r-r_+)/r_+$, the equation reduces to
%%%
\begin{equation}
Z \frac{d}{dz}\left[Z\frac{dR}{dz}\right]+\left[P^2-\ell(\ell+1)Z+s^2z\right]R=0\,.
\label{eqHOR}
\end{equation}
We defined $Z=z(z+1)$ and
\begin{equation}
 P=-2Mk_H=-2M(\omega-m\Omega_H)\,,
\end{equation}
and we neglected ${\cal O}(\tilde{a}^2)$ terms in $P^2$. The general solution of
Eq.~\eqref{eqHOR} is a combination of hypergeometric functions. By imposing
ingoing waves at the horizon, i.e. $R\sim z^{iP}$ at $z\sim0$, we get the
general solution with the correct boundary condition:
\begin{eqnarray}
 R_H(r)&&\propto e^{-2 P \pi } (-1)^{2 i P} z^{i P} \times\nn\\
&& {}_2F_1\left[-\ell+\ii P+\sigma,1+\ell+\ii P+\sigma,1+2 i P,-z\right]\,,\nn
\end{eqnarray}
%%%
where ${}_2F_1(a,b,c,z)$ is the hypergeometric function~\cite{abramowitz+stegun} and $\sigma=\sqrt{s^2-P^2}$.
The large-distance limit of the equation above reads
\begin{eqnarray}
&& R_H(r)\sim \left[\frac{(2M)^{-\ell}  \Gamma[1+2 \ell] }{\Gamma\left[1
+\ell+\ii P-\sigma\right] \Gamma\left[1+\ell+\ii P+\sigma\right]}\right]r^\ell \nn\\
&&+\left[\frac{(2M)^{1+\ell}  \Gamma[-1-2 \ell] }{\Gamma\left[-\ell+\ii P
-\sigma\right] \Gamma\left[-\ell+\ii P+\sigma\right]}\right]r^{-\ell-1}\,.\label{RHlarge}
\end{eqnarray}
The key point of the method of matching asymptotics is that, when $|M\omega|\ll
\ell$ and $M\mu\ll\ell$, there exists an overlapping region where
Eqs.~\eqref{RHlarge} and~\eqref{RINFsmall} are both valid. By equating the
coefficients of $r^\ell$ and $r^{-\ell-1}$ one can fix the ratio of the overall
factors and the frequency of the mode. Finally, we get
%%%
\begin{eqnarray}
 \delta\nu&=&\frac{(4{k_\infty} M)^{2 \ell +1} \Gamma[-1-2 \ell ] \Gamma[2+n+2 \ell]}{\Gamma[1+n] \Gamma[1
+2 \ell ]^2 \Gamma[2+2 \ell ]}\times\nn\\
&&\frac{\Gamma\left[1+
i P-\sigma+\ell \right] \Gamma\left[1+\ii P+\sigma+\ell 
\right]}{\Gamma\left[i P-\sigma-\ell \right] \Gamma\left[i P+\sigma-\ell \right]}\,.\label{deltanu}
\end{eqnarray}
As in the case of massive scalar perturbations, the real and imaginary parts of
the mode are related to $\mu$ and $\delta\nu$ by~\cite{Detweiler:1980uk}
%%%
\begin{eqnarray}
 \mu^2-\omega_R^2&=&\mu^2\left(\frac{M\mu}{\ell+n+1}\right)^2\,,\label{wRana}\\
 \ii\omega_I&=&\frac{\delta\nu}{M}\left(\frac{M\mu}{\ell+n+1}\right)^3\,.
\end{eqnarray}
%%%
In the slow-rotation limit we can further simplify
Eq.~\eqref{deltanu}. To first order in $P$ we get
%%%
\begin{eqnarray}
 \delta\nu&&\sim\frac{i P(4{k_\infty} M)^{2 \ell+1}\Gamma[2+2 \ell+n] }{\Gamma[1+2 \ell]^2\Gamma[2+2 \ell]^2 
\Gamma[1+n]}\times\nn\\
&& \Gamma[1+\ell-s] 
\Gamma[1+\ell+s]\Gamma[1+\ell]^2\prod_{j=1}^s\left[\frac{j-1-s-\ell}{s-j-\ell}\right]\,,\nn\\ \label{deltanu2}
\end{eqnarray}
%%%
% \pp{to be removed: where we have used the property of the gamma function that
% $\Gamma[-q]=(-1)^{q-p}\Gamma[-p]\Gamma[p+1]/\Gamma[q+1]$ for any integer $p$
% and $q$.}
%
When $s=0$ we recover the result for the scalar
case~\cite{Detweiler:1980uk} to first order in $P$ (but see the
subsection below for a discussion).  In addition, our results
correctly reduce to those found in Ref.~\cite{Rosa:2011my} in the
nonrotating limit.  We can now evaluate the instability timescale. The
fundamental mode reads
\begin{eqnarray}
 M\omega_I^{(\ell=1)}&=&\frac{(s+1)!(s+1)}{48} \left(\tilde{a}m-2 r_+ \mu \right) (M\mu)^9\,,\nn\\\label{ana_l=1}\\
 M\omega_I^{(\ell=2)}&=&\frac{ \left[(2-s)! (2+s)!\right]^2}{885735} \left(\tilde{a}m-2 r_+ \mu \right)(M\mu)^{13}\,,\nn
\end{eqnarray}
for $\ell=1$ and $\ell=2$, respectively.
Note that $\omega_I\sim\mu^{4\ell+5}$ for any value of $s$, so the only
difference arises in the prefactor. We find
%%%
\begin{equation}
 \frac{\omega_I^{(s=\pm1)}}{\omega_I^{(s=0)}}=\frac{(1+\ell)^2}{\ell^2}\,,
\end{equation}
%%%%
which is independent of $\mu$, $\tilde a$ and $n$. The instability
timescale of the $\ell=1$ axial vector mode is $4$ times shorter than
that of the scalar mode. Equation~\eqref{ana_l=1} is compared to the
numerical results in Fig.~\ref{fig:zeeman}.
\vspace{0.5cm}
%%%%%
\subsubsection*{On the analytical result in the massive scalar case}\label{sec:Detweiler}
%%%%
Although this is not directly related to the Proca problem, we wish to
remark that when $s=0$ our equation~\eqref{deltanu2} differs from
Eq.~(28) in Ref.~\cite{Detweiler:1980uk} (expanded at first order) by
a factor $2$ for any $\ell$ and any $n$. We believe this is either due
to an inconsistent limit (explained below) or to a typo in
Ref.~\cite{Detweiler:1980uk} which propagates from Eqs.~(21)-(24) to
Eq.~(25), where a factor $1/2$ is missing.  Thus, Eq.~(29) in
Detweiler should read $\gamma=\mu\tilde{a}(\mu M)^8/48$ for
consistency with our Eq.~(\ref{ana_l=1}).

In order to understand the source of this discrepancy we refer the
reader to the work by Furuhashi and Nambu~\cite{Furuhashi:2004jk}, who
extended Detweiler's calculation to the case of Kerr-Newman BHs,
reproducing the results of Ref.~\cite{Detweiler:1980uk} in the
zero-charge limit. Ref.~\cite{Furuhashi:2004jk} presents a detailed
derivation of the instability timescale, facilitating intermediate
comparisons with our own calculation. We note that the results in
Ref.~\cite{Furuhashi:2004jk} seem to be partially flawed
%for two reasons: (i) their Eq.~(13) is not correct, because a
%different expansion of their Eq.~(12) should be used when $\ell$ is an
%integer: one should use Eq.~(15.3.12) rather then (15.3.6) in
%Ref.~\cite{abramowitz+stegun}; (ii) most importantly, 
because the limit defined in Eq.~(23) in Ref.~\cite{Furuhashi:2004jk},
$\Gamma(-n)/\Gamma(-m)=(-1)^{n-m}m!/n!$, is valid only when $n$ and $m$
are integers. However, in a rotating background the angular
eigenvalues are not (in general) integers. For example, in the scalar
case the separation constant $\lambda=\ell(\ell+1)+{\cal O}(\tilde a
M\omega)$ \cite{Detweiler:1980uk}. In the $M\omega\ll1$ limit, this
can be accounted for by considering $\ell\to\ell+\epsilon$, with
$\epsilon\ll1$. When $s=0$, the evaluation of Eq.~\eqref{deltanu}
above involves limits of the form
%%%
\begin{equation}
 \lim_{\epsilon\to0}\frac{\Gamma(-2\ell-2\epsilon-1)}{\Gamma(-\ell-\epsilon)}=\frac{(-1)^{\ell+1}}{2}\frac{\ell!}{(2\ell+1)!}\,,
\end{equation}
%%%
which differs by a factor $2$ from Eq.~(23) in Ref.~\cite{Furuhashi:2004jk} (see also the discussion in Ref.~\cite{Rosa:2012uz}).
%%%
We believe that this is the reason for our factor 2 discrepancy with
respect to Furuhashi and Nambu (and presumably also with respect to
Detweiler's calculation). Despite numerical difficulties in the
$M\mu\ll1$ limit, our numerical calculations for $\ell=m=1$ are in
better agreement with a prefactor of $1/48$, as computed in
Eq.~(\ref{ana_l=1}): see also Sec.~\ref{sec:minimum_growth} and Fig.~6
in Ref.~\cite{Dolan:2007mj}.
\bibliography{proca}

\begin{thebibliography}{95}%
\makeatletter
\providecommand \@ifxundefined [1]{%
 \@ifx{#1\undefined}
}%
\providecommand \@ifnum [1]{%
 \ifnum #1\expandafter \@firstoftwo
 \else \expandafter \@secondoftwo
 \fi
}%
\providecommand \@ifx [1]{%
 \ifx #1\expandafter \@firstoftwo
 \else \expandafter \@secondoftwo
 \fi
}%
\providecommand \natexlab [1]{#1}%
\providecommand \enquote  [1]{``#1''}%
\providecommand \bibnamefont  [1]{#1}%
\providecommand \bibfnamefont [1]{#1}%
\providecommand \citenamefont [1]{#1}%
\providecommand \href@noop [0]{\@secondoftwo}%
\providecommand \href [0]{\begingroup \@sanitize@url \@href}%
\providecommand \@href[1]{\@@startlink{#1}\@@href}%
\providecommand \@@href[1]{\endgroup#1\@@endlink}%
\providecommand \@sanitize@url [0]{\catcode `\\12\catcode `\$12\catcode
  `\&12\catcode `\#12\catcode `\^12\catcode `\_12\catcode `\%12\relax}%
\providecommand \@@startlink[1]{}%
\providecommand \@@endlink[0]{}%
\providecommand \url  [0]{\begingroup\@sanitize@url \@url }%
\providecommand \@url [1]{\endgroup\@href {#1}{\urlprefix }}%
\providecommand \urlprefix  [0]{URL }%
\providecommand \Eprint [0]{\href }%
\providecommand \doibase [0]{http://dx.doi.org/}%
\providecommand \selectlanguage [0]{\@gobble}%
\providecommand \bibinfo  [0]{\@secondoftwo}%
\providecommand \bibfield  [0]{\@secondoftwo}%
\providecommand \translation [1]{[#1]}%
\providecommand \BibitemOpen [0]{}%
\providecommand \bibitemStop [0]{}%
\providecommand \bibitemNoStop [0]{.\EOS\space}%
\providecommand \EOS [0]{\spacefactor3000\relax}%
\providecommand \BibitemShut  [1]{\csname bibitem#1\endcsname}%
\let\auto@bib@innerbib\@empty
%</preamble>
\bibitem [{\citenamefont {Nollert}(1999)}]{Nollert:1999ji}%
  \BibitemOpen
  \bibfield  {author} {\bibinfo {author} {\bibfnamefont {H.-P.}\ \bibnamefont
  {Nollert}},\ }\href {\doibase 10.1088/0264-9381/16/12/201} {\bibfield
  {journal} {\bibinfo  {journal} {Class.Quant.Grav.}\ }\textbf {\bibinfo
  {volume} {16}},\ \bibinfo {pages} {R159} (\bibinfo {year}
  {1999})}\BibitemShut {NoStop}%
%%CITATION = CQGRD,16,R159;%%
\bibitem [{\citenamefont {Kokkotas}\ and\ \citenamefont
  {Schmidt}(1999)}]{Kokkotas:1999bd}%
  \BibitemOpen
  \bibfield  {author} {\bibinfo {author} {\bibfnamefont {K.~D.}\ \bibnamefont
  {Kokkotas}}\ and\ \bibinfo {author} {\bibfnamefont {B.~G.}\ \bibnamefont
  {Schmidt}},\ }\href@noop {} {\bibfield  {journal} {\bibinfo  {journal}
  {Living Rev.Rel.}\ }\textbf {\bibinfo {volume} {2}},\ \bibinfo {pages} {2}
  (\bibinfo {year} {1999})},\ \Eprint {http://arxiv.org/abs/gr-qc/9909058}
  {arXiv:gr-qc/9909058 [gr-qc]} \BibitemShut {NoStop}%
\bibitem [{\citenamefont {Ferrari}\ and\ \citenamefont
  {Gualtieri}(2008)}]{Ferrari:2007dd}%
  \BibitemOpen
  \bibfield  {author} {\bibinfo {author} {\bibfnamefont {V.}~\bibnamefont
  {Ferrari}}\ and\ \bibinfo {author} {\bibfnamefont {L.}~\bibnamefont
  {Gualtieri}},\ }\href {\doibase 10.1007/s10714-007-0585-1} {\bibfield
  {journal} {\bibinfo  {journal} {Gen.Rel.Grav.}\ }\textbf {\bibinfo {volume}
  {40}},\ \bibinfo {pages} {945} (\bibinfo {year} {2008})},\ \Eprint
  {http://arxiv.org/abs/0709.0657} {arXiv:0709.0657 [gr-qc]} \BibitemShut
  {NoStop}%
\bibitem [{\citenamefont {Berti}\ \emph
  {et~al.}(2009{\natexlab{a}})\citenamefont {Berti}, \citenamefont {Cardoso},\
  and\ \citenamefont {Starinets}}]{Berti:2009kk}%
  \BibitemOpen
  \bibfield  {author} {\bibinfo {author} {\bibfnamefont {E.}~\bibnamefont
  {Berti}}, \bibinfo {author} {\bibfnamefont {V.}~\bibnamefont {Cardoso}}, \
  and\ \bibinfo {author} {\bibfnamefont {A.~O.}\ \bibnamefont {Starinets}},\
  }\href {\doibase 10.1088/0264-9381/26/16/163001} {\bibfield  {journal}
  {\bibinfo  {journal} {Class.Quant.Grav.}\ }\textbf {\bibinfo {volume} {26}},\
  \bibinfo {pages} {163001} (\bibinfo {year} {2009}{\natexlab{a}})},\ \Eprint
  {http://arxiv.org/abs/0905.2975} {arXiv:0905.2975 [gr-qc]} \BibitemShut
  {NoStop}%
%%CITATION = ARXIV:0905.2975;%%
\bibitem [{\citenamefont {Konoplya}\ and\ \citenamefont
  {Zhidenko}(2011)}]{Konoplya:2011qq}%
  \BibitemOpen
  \bibfield  {author} {\bibinfo {author} {\bibfnamefont {R.}~\bibnamefont
  {Konoplya}}\ and\ \bibinfo {author} {\bibfnamefont {A.}~\bibnamefont
  {Zhidenko}},\ }\href {\doibase 10.1103/RevModPhys.83.793} {\bibfield
  {journal} {\bibinfo  {journal} {Rev.Mod.Phys.}\ }\textbf {\bibinfo {volume}
  {83}},\ \bibinfo {pages} {793} (\bibinfo {year} {2011})},\ \Eprint
  {http://arxiv.org/abs/1102.4014} {arXiv:1102.4014 [gr-qc]} \BibitemShut
  {NoStop}%
%%CITATION = ARXIV:1102.4014;%%
\bibitem [{\citenamefont {Amaro-Seoane}\ \emph {et~al.}(2007)\citenamefont
  {Amaro-Seoane}, \citenamefont {Gair}, \citenamefont {Freitag}, \citenamefont
  {Coleman~Miller}, \citenamefont {Mandel} \emph
  {et~al.}}]{AmaroSeoane:2007aw}%
  \BibitemOpen
  \bibfield  {author} {\bibinfo {author} {\bibfnamefont {P.}~\bibnamefont
  {Amaro-Seoane}}, \bibinfo {author} {\bibfnamefont {J.~R.}\ \bibnamefont
  {Gair}}, \bibinfo {author} {\bibfnamefont {M.}~\bibnamefont {Freitag}},
  \bibinfo {author} {\bibfnamefont {M.}~\bibnamefont {Coleman~Miller}},
  \bibinfo {author} {\bibfnamefont {I.}~\bibnamefont {Mandel}},  \emph
  {et~al.},\ }\href {\doibase 10.1088/0264-9381/24/17/R01} {\bibfield
  {journal} {\bibinfo  {journal} {Class.Quant.Grav.}\ }\textbf {\bibinfo
  {volume} {24}},\ \bibinfo {pages} {R113} (\bibinfo {year} {2007})},\ \Eprint
  {http://arxiv.org/abs/astro-ph/0703495} {arXiv:astro-ph/0703495 [ASTRO-PH]}
  \BibitemShut {NoStop}%
\bibitem [{\citenamefont {Barack}(2009)}]{Barack:2009ux}%
  \BibitemOpen
  \bibfield  {author} {\bibinfo {author} {\bibfnamefont {L.}~\bibnamefont
  {Barack}},\ }\href {\doibase 10.1088/0264-9381/26/21/213001} {\bibfield
  {journal} {\bibinfo  {journal} {Class.Quant.Grav.}\ }\textbf {\bibinfo
  {volume} {26}},\ \bibinfo {pages} {213001} (\bibinfo {year} {2009})},\
  \Eprint {http://arxiv.org/abs/0908.1664} {arXiv:0908.1664 [gr-qc]}
  \BibitemShut {NoStop}%
%%CITATION = ARXIV:0908.1664;%%
\bibitem [{\citenamefont {Poisson}\ \emph {et~al.}(2011)\citenamefont
  {Poisson}, \citenamefont {Pound},\ and\ \citenamefont
  {Vega}}]{Poisson:2011nh}%
  \BibitemOpen
  \bibfield  {author} {\bibinfo {author} {\bibfnamefont {E.}~\bibnamefont
  {Poisson}}, \bibinfo {author} {\bibfnamefont {A.}~\bibnamefont {Pound}}, \
  and\ \bibinfo {author} {\bibfnamefont {I.}~\bibnamefont {Vega}},\ }\href@noop
  {} {\bibfield  {journal} {\bibinfo  {journal} {Living Rev.Rel.}\ }\textbf
  {\bibinfo {volume} {14}},\ \bibinfo {pages} {7} (\bibinfo {year} {2011})},\
  \Eprint {http://arxiv.org/abs/1102.0529} {arXiv:1102.0529 [gr-qc]}
  \BibitemShut {NoStop}%
%%CITATION = ARXIV:1102.0529;%%
\bibitem [{\citenamefont {Berti}\ \emph {et~al.}(2007)\citenamefont {Berti},
  \citenamefont {Cardoso}, \citenamefont {Gonzalez}, \citenamefont {Sperhake},
  \citenamefont {Hannam} \emph {et~al.}}]{Berti:2007fi}%
  \BibitemOpen
  \bibfield  {author} {\bibinfo {author} {\bibfnamefont {E.}~\bibnamefont
  {Berti}}, \bibinfo {author} {\bibfnamefont {V.}~\bibnamefont {Cardoso}},
  \bibinfo {author} {\bibfnamefont {J.~A.}\ \bibnamefont {Gonzalez}}, \bibinfo
  {author} {\bibfnamefont {U.}~\bibnamefont {Sperhake}}, \bibinfo {author}
  {\bibfnamefont {M.}~\bibnamefont {Hannam}},  \emph {et~al.},\ }\href
  {\doibase 10.1103/PhysRevD.76.064034} {\bibfield  {journal} {\bibinfo
  {journal} {Phys.Rev.}\ }\textbf {\bibinfo {volume} {D76}},\ \bibinfo {pages}
  {064034} (\bibinfo {year} {2007})},\ \Eprint
  {http://arxiv.org/abs/gr-qc/0703053} {arXiv:gr-qc/0703053 [GR-QC]}
  \BibitemShut {NoStop}%
%%CITATION = GR-QC/0703053;%%
\bibitem [{\citenamefont {Berti}\ \emph {et~al.}(2010)\citenamefont {Berti},
  \citenamefont {Cardoso}, \citenamefont {Hinderer}, \citenamefont {Lemos},
  \citenamefont {Pretorius} \emph {et~al.}}]{Berti:2010ce}%
  \BibitemOpen
  \bibfield  {author} {\bibinfo {author} {\bibfnamefont {E.}~\bibnamefont
  {Berti}}, \bibinfo {author} {\bibfnamefont {V.}~\bibnamefont {Cardoso}},
  \bibinfo {author} {\bibfnamefont {T.}~\bibnamefont {Hinderer}}, \bibinfo
  {author} {\bibfnamefont {M.}~\bibnamefont {Lemos}}, \bibinfo {author}
  {\bibfnamefont {F.}~\bibnamefont {Pretorius}},  \emph {et~al.},\ }\href
  {\doibase 10.1103/PhysRevD.81.104048} {\bibfield  {journal} {\bibinfo
  {journal} {Phys.Rev.}\ }\textbf {\bibinfo {volume} {D81}},\ \bibinfo {pages}
  {104048} (\bibinfo {year} {2010})},\ \Eprint {http://arxiv.org/abs/1003.0812}
  {arXiv:1003.0812 [gr-qc]} \BibitemShut {NoStop}%
%%CITATION = ARXIV:1003.0812;%%
\bibitem [{\citenamefont {Cardoso}\ \emph {et~al.}(2012)\citenamefont
  {Cardoso}, \citenamefont {Gualtieri}, \citenamefont {Herdeiro}, \citenamefont
  {Sperhake}, \citenamefont {Chesler} \emph {et~al.}}]{Cardoso:2012qm}%
  \BibitemOpen
  \bibfield  {author} {\bibinfo {author} {\bibfnamefont {V.}~\bibnamefont
  {Cardoso}}, \bibinfo {author} {\bibfnamefont {L.}~\bibnamefont {Gualtieri}},
  \bibinfo {author} {\bibfnamefont {C.}~\bibnamefont {Herdeiro}}, \bibinfo
  {author} {\bibfnamefont {U.}~\bibnamefont {Sperhake}}, \bibinfo {author}
  {\bibfnamefont {P.~M.}\ \bibnamefont {Chesler}},  \emph {et~al.},\
  }\href@noop {} {\  (\bibinfo {year} {2012})},\ \Eprint
  {http://arxiv.org/abs/1201.5118} {arXiv:1201.5118 [hep-th]} \BibitemShut
  {NoStop}%
\bibitem [{\citenamefont {Chandrasekhar}(1992)}]{1992mtbh.book.....C}%
  \BibitemOpen
  \bibfield  {author} {\bibinfo {author} {\bibfnamefont {S.}~\bibnamefont
  {Chandrasekhar}},\ }\href@noop {} {\emph {\bibinfo {title} {The mathematical
  theory of black holes}}}\ (\bibinfo  {publisher} {Oxford University Press},\
  \bibinfo {address} {New York},\ \bibinfo {year} {1992})\BibitemShut {NoStop}%
\bibitem [{\citenamefont {Teukolsky}(1973)}]{Teukolsky:1973ha}%
  \BibitemOpen
  \bibfield  {author} {\bibinfo {author} {\bibfnamefont {S.~A.}\ \bibnamefont
  {Teukolsky}},\ }\href {\doibase 10.1086/152444} {\bibfield  {journal}
  {\bibinfo  {journal} {Astrophys.J.}\ }\textbf {\bibinfo {volume} {185}},\
  \bibinfo {pages} {635} (\bibinfo {year} {1973})}\BibitemShut {NoStop}%
\bibitem [{\citenamefont {Durkee}\ and\ \citenamefont
  {Reall}(2011)}]{Durkee:2010qu}%
  \BibitemOpen
  \bibfield  {author} {\bibinfo {author} {\bibfnamefont {M.}~\bibnamefont
  {Durkee}}\ and\ \bibinfo {author} {\bibfnamefont {H.~S.}\ \bibnamefont
  {Reall}},\ }\href {\doibase 10.1088/0264-9381/28/3/035011} {\bibfield
  {journal} {\bibinfo  {journal} {Class.Quant.Grav.}\ }\textbf {\bibinfo
  {volume} {28}},\ \bibinfo {pages} {035011} (\bibinfo {year} {2011})},\
  \Eprint {http://arxiv.org/abs/1009.0015} {arXiv:1009.0015 [gr-qc]}
  \BibitemShut {NoStop}%
%%CITATION = ARXIV:1009.0015;%%
\bibitem [{\citenamefont {Torres~del Castillo}\ and\ \citenamefont
  {Silva-Ortigoza}(1990)}]{TorresdelCastillo:1990aw}%
  \BibitemOpen
  \bibfield  {author} {\bibinfo {author} {\bibfnamefont {G.}~\bibnamefont
  {Torres~del Castillo}}\ and\ \bibinfo {author} {\bibfnamefont
  {G.}~\bibnamefont {Silva-Ortigoza}},\ }\href {\doibase
  10.1103/PhysRevD.42.4082} {\bibfield  {journal} {\bibinfo  {journal}
  {Phys.Rev.}\ }\textbf {\bibinfo {volume} {D42}},\ \bibinfo {pages} {4082}
  (\bibinfo {year} {1990})}\BibitemShut {NoStop}%
%%CITATION = PHRVA,D42,4082;%%
\bibitem [{\citenamefont {Rosa}\ and\ \citenamefont
  {Dolan}(2012)}]{Rosa:2011my}%
  \BibitemOpen
  \bibfield  {author} {\bibinfo {author} {\bibfnamefont {J.~G.}\ \bibnamefont
  {Rosa}}\ and\ \bibinfo {author} {\bibfnamefont {S.~R.}\ \bibnamefont
  {Dolan}},\ }\href {\doibase 10.1103/PhysRevD.85.044043} {\bibfield  {journal}
  {\bibinfo  {journal} {Phys.Rev.}\ }\textbf {\bibinfo {volume} {D85}},\
  \bibinfo {pages} {044043} (\bibinfo {year} {2012})},\ \Eprint
  {http://arxiv.org/abs/1110.4494} {arXiv:1110.4494 [hep-th]} \BibitemShut
  {NoStop}%
%%CITATION = ARXIV:1110.4494;%%
\bibitem [{\citenamefont {Kojima}(1992)}]{Kojima:1992ie}%
  \BibitemOpen
  \bibfield  {author} {\bibinfo {author} {\bibfnamefont {Y.}~\bibnamefont
  {Kojima}},\ }\href {\doibase 10.1103/PhysRevD.46.4289} {\bibfield  {journal}
  {\bibinfo  {journal} {Phys.Rev.}\ }\textbf {\bibinfo {volume} {D46}},\
  \bibinfo {pages} {4289} (\bibinfo {year} {1992})}\BibitemShut {NoStop}%
%%CITATION = PHRVA,D46,4289;%%
\bibitem [{\citenamefont {{Kojima}}(1993{\natexlab{a}})}]{1993ApJ...414..247K}%
  \BibitemOpen
  \bibfield  {author} {\bibinfo {author} {\bibfnamefont {Y.}~\bibnamefont
  {{Kojima}}},\ }\href {\doibase 10.1086/173073} {\bibfield  {journal}
  {\bibinfo  {journal} {Astrophys.J.}\ }\textbf {\bibinfo {volume} {414}},\
  \bibinfo {pages} {247} (\bibinfo {year} {1993}{\natexlab{a}})}\BibitemShut
  {NoStop}%
\bibitem [{\citenamefont {{Kojima}}(1993{\natexlab{b}})}]{1993PThPh..90..977K}%
  \BibitemOpen
  \bibfield  {author} {\bibinfo {author} {\bibfnamefont {Y.}~\bibnamefont
  {{Kojima}}},\ }\href {\doibase 10.1143/PTP.90.977} {\bibfield  {journal}
  {\bibinfo  {journal} {Progress of Theoretical Physics}\ }\textbf {\bibinfo
  {volume} {90}},\ \bibinfo {pages} {977} (\bibinfo {year}
  {1993}{\natexlab{b}})}\BibitemShut {NoStop}%
\bibitem [{\citenamefont {Bruni}\ \emph {et~al.}(2003)\citenamefont {Bruni},
  \citenamefont {Gualtieri},\ and\ \citenamefont {Sopuerta}}]{Bruni:2002sma}%
  \BibitemOpen
  \bibfield  {author} {\bibinfo {author} {\bibfnamefont {M.}~\bibnamefont
  {Bruni}}, \bibinfo {author} {\bibfnamefont {L.}~\bibnamefont {Gualtieri}}, \
  and\ \bibinfo {author} {\bibfnamefont {C.~F.}\ \bibnamefont {Sopuerta}},\
  }\href {\doibase 10.1088/0264-9381/20/3/310} {\bibfield  {journal} {\bibinfo
  {journal} {Class.Quant.Grav.}\ }\textbf {\bibinfo {volume} {20}},\ \bibinfo
  {pages} {535} (\bibinfo {year} {2003})},\ \Eprint
  {http://arxiv.org/abs/gr-qc/0207105} {arXiv:gr-qc/0207105 [gr-qc]}
  \BibitemShut {NoStop}%
%%CITATION = GR-QC/0207105;%%
\bibitem [{\citenamefont {Chandrasekhar}\ and\ \citenamefont
  {Ferrari}(1991)}]{ChandraFerrari91}%
  \BibitemOpen
  \bibfield  {author} {\bibinfo {author} {\bibfnamefont {S.}~\bibnamefont
  {Chandrasekhar}}\ and\ \bibinfo {author} {\bibfnamefont {V.}~\bibnamefont
  {Ferrari}},\ }\href@noop {} {\bibfield  {journal} {\bibinfo  {journal}
  {Proc.Roy.Soc.Lond.}\ }\textbf {\bibinfo {volume} {A433}},\ \bibinfo {pages}
  {423} (\bibinfo {year} {1991})}\BibitemShut {NoStop}%
\bibitem [{\citenamefont {Yunes}\ and\ \citenamefont
  {Pretorius}(2009)}]{Yunes:2009hc}%
  \BibitemOpen
  \bibfield  {author} {\bibinfo {author} {\bibfnamefont {N.}~\bibnamefont
  {Yunes}}\ and\ \bibinfo {author} {\bibfnamefont {F.}~\bibnamefont
  {Pretorius}},\ }\href {\doibase 10.1103/PhysRevD.79.084043} {\bibfield
  {journal} {\bibinfo  {journal} {Phys.Rev.}\ }\textbf {\bibinfo {volume}
  {D79}},\ \bibinfo {pages} {084043} (\bibinfo {year} {2009})},\ \Eprint
  {http://arxiv.org/abs/0902.4669} {arXiv:0902.4669 [gr-qc]} \BibitemShut
  {NoStop}%
%%CITATION = ARXIV:0902.4669;%%
\bibitem [{\citenamefont {Konno}\ \emph {et~al.}(2009)\citenamefont {Konno},
  \citenamefont {Matsuyama},\ and\ \citenamefont {Tanda}}]{Konno:2009kg}%
  \BibitemOpen
  \bibfield  {author} {\bibinfo {author} {\bibfnamefont {K.}~\bibnamefont
  {Konno}}, \bibinfo {author} {\bibfnamefont {T.}~\bibnamefont {Matsuyama}}, \
  and\ \bibinfo {author} {\bibfnamefont {S.}~\bibnamefont {Tanda}},\ }\href
  {\doibase 10.1143/PTP.122.561} {\bibfield  {journal} {\bibinfo  {journal}
  {Prog.Theor.Phys.}\ }\textbf {\bibinfo {volume} {122}},\ \bibinfo {pages}
  {561} (\bibinfo {year} {2009})},\ \Eprint {http://arxiv.org/abs/0902.4767}
  {arXiv:0902.4767 [gr-qc]} \BibitemShut {NoStop}%
%%CITATION = ARXIV:0902.4767;%%
\bibitem [{\citenamefont {Pani}\ and\ \citenamefont
  {Cardoso}(2009)}]{Pani:2009wy}%
  \BibitemOpen
  \bibfield  {author} {\bibinfo {author} {\bibfnamefont {P.}~\bibnamefont
  {Pani}}\ and\ \bibinfo {author} {\bibfnamefont {V.}~\bibnamefont {Cardoso}},\
  }\href {\doibase 10.1103/PhysRevD.79.084031} {\bibfield  {journal} {\bibinfo
  {journal} {Phys.Rev.}\ }\textbf {\bibinfo {volume} {D79}},\ \bibinfo {pages}
  {084031} (\bibinfo {year} {2009})},\ \Eprint {http://arxiv.org/abs/0902.1569}
  {arXiv:0902.1569 [gr-qc]} \BibitemShut {NoStop}%
\bibitem [{\citenamefont {Pani}\ \emph {et~al.}(2011)\citenamefont {Pani},
  \citenamefont {Macedo}, \citenamefont {Crispino},\ and\ \citenamefont
  {Cardoso}}]{Pani:2011gy}%
  \BibitemOpen
  \bibfield  {author} {\bibinfo {author} {\bibfnamefont {P.}~\bibnamefont
  {Pani}}, \bibinfo {author} {\bibfnamefont {C.~F.}\ \bibnamefont {Macedo}},
  \bibinfo {author} {\bibfnamefont {L.~C.}\ \bibnamefont {Crispino}}, \ and\
  \bibinfo {author} {\bibfnamefont {V.}~\bibnamefont {Cardoso}},\ }\href
  {\doibase 10.1103/PhysRevD.84.087501} {\bibfield  {journal} {\bibinfo
  {journal} {Phys.Rev.}\ }\textbf {\bibinfo {volume} {D84}},\ \bibinfo {pages}
  {087501} (\bibinfo {year} {2011})},\ \Eprint {http://arxiv.org/abs/1109.3996}
  {arXiv:1109.3996 [gr-qc]} \BibitemShut {NoStop}%
%%CITATION = ARXIV:1109.3996;%%
\bibitem [{\citenamefont {Yagi}\ \emph {et~al.}(2012)\citenamefont {Yagi},
  \citenamefont {Yunes},\ and\ \citenamefont {Tanaka}}]{Yagi:2012ya}%
  \BibitemOpen
  \bibfield  {author} {\bibinfo {author} {\bibfnamefont {K.}~\bibnamefont
  {Yagi}}, \bibinfo {author} {\bibfnamefont {N.}~\bibnamefont {Yunes}}, \ and\
  \bibinfo {author} {\bibfnamefont {T.}~\bibnamefont {Tanaka}},\ }\href
  {\doibase 10.1103/PhysRevD.86.044037} {\bibfield  {journal} {\bibinfo
  {journal} {Phys.Rev.}\ }\textbf {\bibinfo {volume} {D86}},\ \bibinfo {pages}
  {044037} (\bibinfo {year} {2012})},\ \Eprint {http://arxiv.org/abs/1206.6130}
  {arXiv:1206.6130 [gr-qc]} \BibitemShut {NoStop}%
%%CITATION = ARXIV:1206.6130;%%
\bibitem [{\citenamefont {Andersson}(1998)}]{Andersson:1997xt}%
  \BibitemOpen
  \bibfield  {author} {\bibinfo {author} {\bibfnamefont {N.}~\bibnamefont
  {Andersson}},\ }\href {\doibase 10.1086/305919} {\bibfield  {journal}
  {\bibinfo  {journal} {Astrophys.J.}\ }\textbf {\bibinfo {volume} {502}},\
  \bibinfo {pages} {708} (\bibinfo {year} {1998})},\ \Eprint
  {http://arxiv.org/abs/gr-qc/9706075} {arXiv:gr-qc/9706075 [gr-qc]}
  \BibitemShut {NoStop}%
%%CITATION = GR-QC/9706075;%%
\bibitem [{\citenamefont {Kojima}(1998)}]{Kojima:1997vv}%
  \BibitemOpen
  \bibfield  {author} {\bibinfo {author} {\bibfnamefont {Y.}~\bibnamefont
  {Kojima}},\ }\href@noop {} {\bibfield  {journal} {\bibinfo  {journal}
  {Mon.Not.Roy.Astron.Soc.}\ }\textbf {\bibinfo {volume} {293}},\ \bibinfo
  {pages} {49} (\bibinfo {year} {1998})},\ \Eprint
  {http://arxiv.org/abs/gr-qc/9709003} {arXiv:gr-qc/9709003 [gr-qc]}
  \BibitemShut {NoStop}%
%%CITATION = GR-QC/9709003;%%
\bibitem [{\citenamefont {Lockitch}\ and\ \citenamefont
  {Friedman}(1999)}]{Lockitch:1998nq}%
  \BibitemOpen
  \bibfield  {author} {\bibinfo {author} {\bibfnamefont {K.~H.}\ \bibnamefont
  {Lockitch}}\ and\ \bibinfo {author} {\bibfnamefont {J.~L.}\ \bibnamefont
  {Friedman}},\ }\href {\doibase 10.1086/307580} {\bibfield  {journal}
  {\bibinfo  {journal} {Astrophys.J.}\ }\textbf {\bibinfo {volume} {521}},\
  \bibinfo {pages} {764} (\bibinfo {year} {1999})},\ \Eprint
  {http://arxiv.org/abs/gr-qc/9812019} {arXiv:gr-qc/9812019 [gr-qc]}
  \BibitemShut {NoStop}%
%%CITATION = GR-QC/9812019;%%
\bibitem [{\citenamefont {Lockitch}\ \emph {et~al.}(2001)\citenamefont
  {Lockitch}, \citenamefont {Andersson},\ and\ \citenamefont
  {Friedman}}]{Lockitch:2000aa}%
  \BibitemOpen
  \bibfield  {author} {\bibinfo {author} {\bibfnamefont {K.~H.}\ \bibnamefont
  {Lockitch}}, \bibinfo {author} {\bibfnamefont {N.}~\bibnamefont {Andersson}},
  \ and\ \bibinfo {author} {\bibfnamefont {J.~L.}\ \bibnamefont {Friedman}},\
  }\href {\doibase 10.1103/PhysRevD.63.024019} {\bibfield  {journal} {\bibinfo
  {journal} {Phys.Rev.}\ }\textbf {\bibinfo {volume} {D63}},\ \bibinfo {pages}
  {024019} (\bibinfo {year} {2001})},\ \Eprint
  {http://arxiv.org/abs/gr-qc/0008019} {arXiv:gr-qc/0008019 [gr-qc]}
  \BibitemShut {NoStop}%
%%CITATION = GR-QC/0008019;%%
\bibitem [{\citenamefont {Press}\ and\ \citenamefont
  {Teukolsky}(1972)}]{Press:1972zz}%
  \BibitemOpen
  \bibfield  {author} {\bibinfo {author} {\bibfnamefont {W.~H.}\ \bibnamefont
  {Press}}\ and\ \bibinfo {author} {\bibfnamefont {S.~A.}\ \bibnamefont
  {Teukolsky}},\ }\href {\doibase 10.1038/238211a0} {\bibfield  {journal}
  {\bibinfo  {journal} {Nature}\ }\textbf {\bibinfo {volume} {238}},\ \bibinfo
  {pages} {211} (\bibinfo {year} {1972})}\BibitemShut {NoStop}%
%%CITATION = NATUA,238,211;%%
\bibitem [{\citenamefont {Damour}\ \emph {et~al.}(1976)\citenamefont {Damour},
  \citenamefont {Deruelle},\ and\ \citenamefont {Ruffini}}]{Damour:1976kh}%
  \BibitemOpen
  \bibfield  {author} {\bibinfo {author} {\bibfnamefont {T.}~\bibnamefont
  {Damour}}, \bibinfo {author} {\bibfnamefont {N.}~\bibnamefont {Deruelle}}, \
  and\ \bibinfo {author} {\bibfnamefont {R.}~\bibnamefont {Ruffini}},\
  }\href@noop {} {\bibfield  {journal} {\bibinfo  {journal} {Lett.Nuovo Cim.}\
  }\textbf {\bibinfo {volume} {15}},\ \bibinfo {pages} {257} (\bibinfo {year}
  {1976})}\BibitemShut {NoStop}%
%%CITATION = NCLTA,15,257;%%
\bibitem [{\citenamefont {Cardoso}\ \emph {et~al.}(2004)\citenamefont
  {Cardoso}, \citenamefont {Dias}, \citenamefont {Lemos},\ and\ \citenamefont
  {Yoshida}}]{Cardoso:2004nk}%
  \BibitemOpen
  \bibfield  {author} {\bibinfo {author} {\bibfnamefont {V.}~\bibnamefont
  {Cardoso}}, \bibinfo {author} {\bibfnamefont {O.~J.}\ \bibnamefont {Dias}},
  \bibinfo {author} {\bibfnamefont {J.~P.}\ \bibnamefont {Lemos}}, \ and\
  \bibinfo {author} {\bibfnamefont {S.}~\bibnamefont {Yoshida}},\ }\href
  {\doibase 10.1103/PhysRevD.70.044039, 10.1103/PhysRevD.70.049903} {\bibfield
  {journal} {\bibinfo  {journal} {Phys.Rev.}\ }\textbf {\bibinfo {volume}
  {D70}},\ \bibinfo {pages} {044039} (\bibinfo {year} {2004})},\ \Eprint
  {http://arxiv.org/abs/hep-th/0404096} {arXiv:hep-th/0404096 [hep-th]}
  \BibitemShut {NoStop}%
%%CITATION = HEP-TH/0404096;%%
\bibitem [{\citenamefont {Cardoso}\ and\ \citenamefont
  {Yoshida}(2005)}]{Cardoso:2005vk}%
  \BibitemOpen
  \bibfield  {author} {\bibinfo {author} {\bibfnamefont {V.}~\bibnamefont
  {Cardoso}}\ and\ \bibinfo {author} {\bibfnamefont {S.}~\bibnamefont
  {Yoshida}},\ }\href {\doibase 10.1088/1126-6708/2005/07/009} {\bibfield
  {journal} {\bibinfo  {journal} {JHEP}\ }\textbf {\bibinfo {volume} {0507}},\
  \bibinfo {pages} {009} (\bibinfo {year} {2005})},\ \Eprint
  {http://arxiv.org/abs/hep-th/0502206} {arXiv:hep-th/0502206 [hep-th]}
  \BibitemShut {NoStop}%
%%CITATION = HEP-TH/0502206;%%
\bibitem [{\citenamefont {Dolan}(2007)}]{Dolan:2007mj}%
  \BibitemOpen
  \bibfield  {author} {\bibinfo {author} {\bibfnamefont {S.~R.}\ \bibnamefont
  {Dolan}},\ }\href {\doibase 10.1103/PhysRevD.76.084001} {\bibfield  {journal}
  {\bibinfo  {journal} {Phys.Rev.}\ }\textbf {\bibinfo {volume} {D76}},\
  \bibinfo {pages} {084001} (\bibinfo {year} {2007})},\ \Eprint
  {http://arxiv.org/abs/0705.2880} {arXiv:0705.2880 [gr-qc]} \BibitemShut
  {NoStop}%
%%CITATION = ARXIV:0705.2880;%%
\bibitem [{\citenamefont {Rosa}(2010)}]{Rosa:2009ei}%
  \BibitemOpen
  \bibfield  {author} {\bibinfo {author} {\bibfnamefont {J.}~\bibnamefont
  {Rosa}},\ }\href {\doibase 10.1007/JHEP06(2010)015} {\bibfield  {journal}
  {\bibinfo  {journal} {JHEP}\ }\textbf {\bibinfo {volume} {1006}},\ \bibinfo
  {pages} {015} (\bibinfo {year} {2010})},\ \Eprint
  {http://arxiv.org/abs/0912.1780} {arXiv:0912.1780 [hep-th]} \BibitemShut
  {NoStop}%
%%CITATION = ARXIV:0912.1780;%%
\bibitem [{\citenamefont {Cardoso}\ \emph {et~al.}(2011)\citenamefont
  {Cardoso}, \citenamefont {Chakrabarti}, \citenamefont {Pani}, \citenamefont
  {Berti},\ and\ \citenamefont {Gualtieri}}]{Cardoso:2011xi}%
  \BibitemOpen
  \bibfield  {author} {\bibinfo {author} {\bibfnamefont {V.}~\bibnamefont
  {Cardoso}}, \bibinfo {author} {\bibfnamefont {S.}~\bibnamefont
  {Chakrabarti}}, \bibinfo {author} {\bibfnamefont {P.}~\bibnamefont {Pani}},
  \bibinfo {author} {\bibfnamefont {E.}~\bibnamefont {Berti}}, \ and\ \bibinfo
  {author} {\bibfnamefont {L.}~\bibnamefont {Gualtieri}},\ }\href {\doibase
  10.1103/PhysRevLett.107.241101} {\bibfield  {journal} {\bibinfo  {journal}
  {Phys.Rev.Lett.}\ }\textbf {\bibinfo {volume} {107}},\ \bibinfo {pages}
  {241101} (\bibinfo {year} {2011})},\ \Eprint {http://arxiv.org/abs/1109.6021}
  {arXiv:1109.6021 [gr-qc]} \BibitemShut {NoStop}%
\bibitem [{\citenamefont {Zouros}\ and\ \citenamefont
  {Eardley}(1979)}]{Zouros:1979iw}%
  \BibitemOpen
  \bibfield  {author} {\bibinfo {author} {\bibfnamefont {T.}~\bibnamefont
  {Zouros}}\ and\ \bibinfo {author} {\bibfnamefont {D.}~\bibnamefont
  {Eardley}},\ }\href {\doibase 10.1016/0003-4916(79)90237-9} {\bibfield
  {journal} {\bibinfo  {journal} {Annals Phys.}\ }\textbf {\bibinfo {volume}
  {118}},\ \bibinfo {pages} {139} (\bibinfo {year} {1979})}\BibitemShut
  {NoStop}%
%%CITATION = APNYA,118,139;%%
\bibitem [{\citenamefont {{Hawking}}(1971)}]{1971MNRAS.152...75H}%
  \BibitemOpen
  \bibfield  {author} {\bibinfo {author} {\bibfnamefont {S.}~\bibnamefont
  {{Hawking}}},\ }\href@noop {} {\bibfield  {journal} {\bibinfo  {journal}
  {Mon.Not.Roy.Astron.Soc.}\ }\textbf {\bibinfo {volume} {152}},\ \bibinfo
  {pages} {75} (\bibinfo {year} {1971})}\BibitemShut {NoStop}%
\bibitem [{\citenamefont {{Zel'Dovich}}\ and\ \citenamefont
  {{Novikov}}(1966)}]{1966AZh....43..758Z}%
  \BibitemOpen
  \bibfield  {author} {\bibinfo {author} {\bibfnamefont {Y.~B.}\ \bibnamefont
  {{Zel'Dovich}}}\ and\ \bibinfo {author} {\bibfnamefont {I.~D.}\ \bibnamefont
  {{Novikov}}},\ }\href@noop {} {\bibfield  {journal} {\bibinfo  {journal}
  {Astron.Zh.}\ }\textbf {\bibinfo {volume} {43}},\ \bibinfo {pages} {758}
  (\bibinfo {year} {1966})}\BibitemShut {NoStop}%
\bibitem [{\citenamefont {{Carr}}\ and\ \citenamefont
  {{Hawking}}(1974)}]{1974MNRAS.168..399C}%
  \BibitemOpen
  \bibfield  {author} {\bibinfo {author} {\bibfnamefont {B.~J.}\ \bibnamefont
  {{Carr}}}\ and\ \bibinfo {author} {\bibfnamefont {S.~W.}\ \bibnamefont
  {{Hawking}}},\ }\href@noop {} {\bibfield  {journal} {\bibinfo  {journal}
  {Mon.Not.Roy.Astron.Soc.}\ }\textbf {\bibinfo {volume} {168}},\ \bibinfo
  {pages} {399} (\bibinfo {year} {1974})}\BibitemShut {NoStop}%
\bibitem [{\citenamefont {Arvanitaki}\ \emph {et~al.}(2010)\citenamefont
  {Arvanitaki}, \citenamefont {Dimopoulos}, \citenamefont {Dubovsky},
  \citenamefont {Kaloper},\ and\ \citenamefont
  {March-Russell}}]{Arvanitaki:2009fg}%
  \BibitemOpen
  \bibfield  {author} {\bibinfo {author} {\bibfnamefont {A.}~\bibnamefont
  {Arvanitaki}}, \bibinfo {author} {\bibfnamefont {S.}~\bibnamefont
  {Dimopoulos}}, \bibinfo {author} {\bibfnamefont {S.}~\bibnamefont
  {Dubovsky}}, \bibinfo {author} {\bibfnamefont {N.}~\bibnamefont {Kaloper}}, \
  and\ \bibinfo {author} {\bibfnamefont {J.}~\bibnamefont {March-Russell}},\
  }\href {\doibase 10.1103/PhysRevD.81.123530} {\bibfield  {journal} {\bibinfo
  {journal} {Phys.Rev.}\ }\textbf {\bibinfo {volume} {D81}},\ \bibinfo {pages}
  {123530} (\bibinfo {year} {2010})},\ \Eprint {http://arxiv.org/abs/0905.4720}
  {arXiv:0905.4720 [hep-th]} \BibitemShut {NoStop}%
%%CITATION = ARXIV:0905.4720;%%
\bibitem [{\citenamefont {Arvanitaki}\ and\ \citenamefont
  {Dubovsky}(2011)}]{Arvanitaki:2010sy}%
  \BibitemOpen
  \bibfield  {author} {\bibinfo {author} {\bibfnamefont {A.}~\bibnamefont
  {Arvanitaki}}\ and\ \bibinfo {author} {\bibfnamefont {S.}~\bibnamefont
  {Dubovsky}},\ }\href {\doibase 10.1103/PhysRevD.83.044026} {\bibfield
  {journal} {\bibinfo  {journal} {Phys.Rev.}\ }\textbf {\bibinfo {volume}
  {D83}},\ \bibinfo {pages} {044026} (\bibinfo {year} {2011})},\ \Eprint
  {http://arxiv.org/abs/1004.3558} {arXiv:1004.3558 [hep-th]} \BibitemShut
  {NoStop}%
%%CITATION = ARXIV:1004.3558;%%
\bibitem [{\citenamefont {Yunes}\ \emph {et~al.}(2012)\citenamefont {Yunes},
  \citenamefont {Pani},\ and\ \citenamefont {Cardoso}}]{Yunes:2011aa}%
  \BibitemOpen
  \bibfield  {author} {\bibinfo {author} {\bibfnamefont {N.}~\bibnamefont
  {Yunes}}, \bibinfo {author} {\bibfnamefont {P.}~\bibnamefont {Pani}}, \ and\
  \bibinfo {author} {\bibfnamefont {V.}~\bibnamefont {Cardoso}},\ }\href
  {\doibase 10.1103/PhysRevD.85.102003} {\bibfield  {journal} {\bibinfo
  {journal} {Phys.Rev.}\ }\textbf {\bibinfo {volume} {D85}},\ \bibinfo {pages}
  {102003} (\bibinfo {year} {2012})},\ \Eprint {http://arxiv.org/abs/1112.3351}
  {arXiv:1112.3351 [gr-qc]} \BibitemShut {NoStop}%
%%CITATION = ARXIV:1112.3351;%%
\bibitem [{\citenamefont {Alsing}\ \emph {et~al.}(2012)\citenamefont {Alsing},
  \citenamefont {Berti}, \citenamefont {Will},\ and\ \citenamefont
  {Zaglauer}}]{Alsing:2011er}%
  \BibitemOpen
  \bibfield  {author} {\bibinfo {author} {\bibfnamefont {J.}~\bibnamefont
  {Alsing}}, \bibinfo {author} {\bibfnamefont {E.}~\bibnamefont {Berti}},
  \bibinfo {author} {\bibfnamefont {C.~M.}\ \bibnamefont {Will}}, \ and\
  \bibinfo {author} {\bibfnamefont {H.}~\bibnamefont {Zaglauer}},\ }\href
  {\doibase 10.1103/PhysRevD.85.064041} {\bibfield  {journal} {\bibinfo
  {journal} {Phys.Rev.}\ }\textbf {\bibinfo {volume} {D85}},\ \bibinfo {pages}
  {064041} (\bibinfo {year} {2012})},\ \Eprint {http://arxiv.org/abs/1112.4903}
  {arXiv:1112.4903 [gr-qc]} \BibitemShut {NoStop}%
%%CITATION = ARXIV:1112.4903;%%
\bibitem [{\citenamefont {Kodama}\ and\ \citenamefont
  {Yoshino}(2011)}]{Kodama:2011zc}%
  \BibitemOpen
  \bibfield  {author} {\bibinfo {author} {\bibfnamefont {H.}~\bibnamefont
  {Kodama}}\ and\ \bibinfo {author} {\bibfnamefont {H.}~\bibnamefont
  {Yoshino}},\ }\href@noop {} {\  (\bibinfo {year} {2011})},\ \Eprint
  {http://arxiv.org/abs/1108.1365} {arXiv:1108.1365 [hep-th]} \BibitemShut
  {NoStop}%
%%CITATION = ARXIV:1108.1365;%%
\bibitem [{\citenamefont {Yoshino}\ and\ \citenamefont
  {Kodama}(2012)}]{Yoshino:2012kn}%
  \BibitemOpen
  \bibfield  {author} {\bibinfo {author} {\bibfnamefont {H.}~\bibnamefont
  {Yoshino}}\ and\ \bibinfo {author} {\bibfnamefont {H.}~\bibnamefont
  {Kodama}},\ }\href@noop {} {\bibfield  {journal} {\bibinfo  {journal}
  {Prog.Theor.Phys.}\ }\textbf {\bibinfo {volume} {128}},\ \bibinfo {pages}
  {153} (\bibinfo {year} {2012})},\ \Eprint {http://arxiv.org/abs/1203.5070}
  {arXiv:1203.5070 [gr-qc]} \BibitemShut {NoStop}%
%%CITATION = ARXIV:1203.5070;%%
\bibitem [{\citenamefont {Mocanu}\ and\ \citenamefont
  {Grumiller}(2012)}]{Mocanu:2012fd}%
  \BibitemOpen
  \bibfield  {author} {\bibinfo {author} {\bibfnamefont {G.}~\bibnamefont
  {Mocanu}}\ and\ \bibinfo {author} {\bibfnamefont {D.}~\bibnamefont
  {Grumiller}},\ }\href {\doibase 10.1103/PhysRevD.85.105022} {\bibfield
  {journal} {\bibinfo  {journal} {Phys.Rev.}\ }\textbf {\bibinfo {volume}
  {D85}},\ \bibinfo {pages} {105022} (\bibinfo {year} {2012})},\ \Eprint
  {http://arxiv.org/abs/1203.4681} {arXiv:1203.4681 [astro-ph.CO]} \BibitemShut
  {NoStop}%
%%CITATION = ARXIV:1203.4681;%%
\bibitem [{\citenamefont {Goodsell}\ \emph {et~al.}(2009)\citenamefont
  {Goodsell}, \citenamefont {Jaeckel}, \citenamefont {Redondo},\ and\
  \citenamefont {Ringwald}}]{Goodsell:2009xc}%
  \BibitemOpen
  \bibfield  {author} {\bibinfo {author} {\bibfnamefont {M.}~\bibnamefont
  {Goodsell}}, \bibinfo {author} {\bibfnamefont {J.}~\bibnamefont {Jaeckel}},
  \bibinfo {author} {\bibfnamefont {J.}~\bibnamefont {Redondo}}, \ and\
  \bibinfo {author} {\bibfnamefont {A.}~\bibnamefont {Ringwald}},\ }\href
  {\doibase 10.1088/1126-6708/2009/11/027} {\bibfield  {journal} {\bibinfo
  {journal} {JHEP}\ }\textbf {\bibinfo {volume} {0911}},\ \bibinfo {pages}
  {027} (\bibinfo {year} {2009})},\ \Eprint {http://arxiv.org/abs/0909.0515}
  {arXiv:0909.0515 [hep-ph]} \BibitemShut {NoStop}%
%%CITATION = ARXIV:0909.0515;%%
\bibitem [{\citenamefont {Jaeckel}\ and\ \citenamefont
  {Ringwald}(2010)}]{Jaeckel:2010ni}%
  \BibitemOpen
  \bibfield  {author} {\bibinfo {author} {\bibfnamefont {J.}~\bibnamefont
  {Jaeckel}}\ and\ \bibinfo {author} {\bibfnamefont {A.}~\bibnamefont
  {Ringwald}},\ }\href {\doibase 10.1146/annurev.nucl.012809.104433} {\bibfield
   {journal} {\bibinfo  {journal} {Ann.Rev.Nucl.Part.Sci.}\ }\textbf {\bibinfo
  {volume} {60}},\ \bibinfo {pages} {405} (\bibinfo {year} {2010})},\ \Eprint
  {http://arxiv.org/abs/1002.0329} {arXiv:1002.0329 [hep-ph]} \BibitemShut
  {NoStop}%
%%CITATION = ARXIV:1002.0329;%%
\bibitem [{\citenamefont {Camara}\ \emph {et~al.}(2011)\citenamefont {Camara},
  \citenamefont {Ibanez},\ and\ \citenamefont {Marchesano}}]{Camara:2011jg}%
  \BibitemOpen
  \bibfield  {author} {\bibinfo {author} {\bibfnamefont {P.~G.}\ \bibnamefont
  {Camara}}, \bibinfo {author} {\bibfnamefont {L.~E.}\ \bibnamefont {Ibanez}},
  \ and\ \bibinfo {author} {\bibfnamefont {F.}~\bibnamefont {Marchesano}},\
  }\href {\doibase 10.1007/JHEP09(2011)110} {\bibfield  {journal} {\bibinfo
  {journal} {JHEP}\ }\textbf {\bibinfo {volume} {1109}},\ \bibinfo {pages}
  {110} (\bibinfo {year} {2011})},\ \Eprint {http://arxiv.org/abs/1106.0060}
  {arXiv:1106.0060 [hep-th]} \BibitemShut {NoStop}%
%%CITATION = ARXIV:1106.0060;%%
\bibitem [{\citenamefont {Goldhaber}\ and\ \citenamefont
  {Nieto}(2010)}]{Goldhaber:2008xy}%
  \BibitemOpen
  \bibfield  {author} {\bibinfo {author} {\bibfnamefont {A.~S.}\ \bibnamefont
  {Goldhaber}}\ and\ \bibinfo {author} {\bibfnamefont {M.~M.}\ \bibnamefont
  {Nieto}},\ }\href {\doibase 10.1103/RevModPhys.82.939} {\bibfield  {journal}
  {\bibinfo  {journal} {Rev.Mod.Phys.}\ }\textbf {\bibinfo {volume} {82}},\
  \bibinfo {pages} {939} (\bibinfo {year} {2010})},\ \Eprint
  {http://arxiv.org/abs/0809.1003} {arXiv:0809.1003 [hep-ph]} \BibitemShut
  {NoStop}%
%%CITATION = ARXIV:0809.1003;%%
\bibitem [{\citenamefont {Gal'tsov}\ \emph {et~al.}(1984)\citenamefont
  {Gal'tsov}, \citenamefont {Pomerantseva},\ and\ \citenamefont
  {Chizhov}}]{Gal'tsov:1984nb}%
  \BibitemOpen
  \bibfield  {author} {\bibinfo {author} {\bibfnamefont {D.}~\bibnamefont
  {Gal'tsov}}, \bibinfo {author} {\bibfnamefont {G.}~\bibnamefont
  {Pomerantseva}}, \ and\ \bibinfo {author} {\bibfnamefont {G.}~\bibnamefont
  {Chizhov}},\ }\href {\doibase 10.1007/BF00893117, 10.1007/BF00893117}
  {\bibfield  {journal} {\bibinfo  {journal} {Sov.Phys.J.}\ }\textbf {\bibinfo
  {volume} {27}},\ \bibinfo {pages} {697} (\bibinfo {year} {1984})}\BibitemShut
  {NoStop}%
%%CITATION = SOPJA,27,697;%%
\bibitem [{\citenamefont {Herdeiro}\ \emph {et~al.}(2012)\citenamefont
  {Herdeiro}, \citenamefont {Sampaio},\ and\ \citenamefont
  {Wang}}]{Herdeiro:2011uu}%
  \BibitemOpen
  \bibfield  {author} {\bibinfo {author} {\bibfnamefont {C.}~\bibnamefont
  {Herdeiro}}, \bibinfo {author} {\bibfnamefont {M.~O.}\ \bibnamefont
  {Sampaio}}, \ and\ \bibinfo {author} {\bibfnamefont {M.}~\bibnamefont
  {Wang}},\ }\href {\doibase 10.1103/PhysRevD.85.024005} {\bibfield  {journal}
  {\bibinfo  {journal} {Phys.Rev.}\ }\textbf {\bibinfo {volume} {D85}},\
  \bibinfo {pages} {024005} (\bibinfo {year} {2012})},\ \Eprint
  {http://arxiv.org/abs/1110.2485} {arXiv:1110.2485 [gr-qc]} \BibitemShut
  {NoStop}%
%%CITATION = ARXIV:1110.2485;%%
\bibitem [{\citenamefont {Konoplya}(2006)}]{Konoplya:2005hr}%
  \BibitemOpen
  \bibfield  {author} {\bibinfo {author} {\bibfnamefont {R.}~\bibnamefont
  {Konoplya}},\ }\href {\doibase 10.1103/PhysRevD.73.024009} {\bibfield
  {journal} {\bibinfo  {journal} {Phys.Rev.}\ }\textbf {\bibinfo {volume}
  {D73}},\ \bibinfo {pages} {024009} (\bibinfo {year} {2006})},\ \Eprint
  {http://arxiv.org/abs/gr-qc/0509026} {arXiv:gr-qc/0509026 [gr-qc]}
  \BibitemShut {NoStop}%
%%CITATION = GR-QC/0509026;%%
\bibitem [{\citenamefont {Brenneman}\ \emph {et~al.}(2011)\citenamefont
  {Brenneman}, \citenamefont {Reynolds}, \citenamefont {Nowak}, \citenamefont
  {Reis}, \citenamefont {Trippe} \emph {et~al.}}]{Brenneman:2011wz}%
  \BibitemOpen
  \bibfield  {author} {\bibinfo {author} {\bibfnamefont {L.}~\bibnamefont
  {Brenneman}}, \bibinfo {author} {\bibfnamefont {C.}~\bibnamefont {Reynolds}},
  \bibinfo {author} {\bibfnamefont {M.}~\bibnamefont {Nowak}}, \bibinfo
  {author} {\bibfnamefont {R.}~\bibnamefont {Reis}}, \bibinfo {author}
  {\bibfnamefont {M.}~\bibnamefont {Trippe}},  \emph {et~al.},\ }\href
  {\doibase 10.1088/0004-637X/736/2/103} {\bibfield  {journal} {\bibinfo
  {journal} {Astrophys.J.}\ }\textbf {\bibinfo {volume} {736}},\ \bibinfo
  {pages} {103} (\bibinfo {year} {2011})},\ \Eprint
  {http://arxiv.org/abs/1104.1172} {arXiv:1104.1172 [astro-ph.HE]} \BibitemShut
  {NoStop}%
%%CITATION = ARXIV:1104.1172;%%
\bibitem [{\citenamefont {Schmoll}\ \emph {et~al.}(2009)\citenamefont
  {Schmoll}, \citenamefont {Miller}, \citenamefont {Volonteri}, \citenamefont
  {Cackett}, \citenamefont {Reynolds} \emph {et~al.}}]{Schmoll:2009gq}%
  \BibitemOpen
  \bibfield  {author} {\bibinfo {author} {\bibfnamefont {S.}~\bibnamefont
  {Schmoll}}, \bibinfo {author} {\bibfnamefont {J.}~\bibnamefont {Miller}},
  \bibinfo {author} {\bibfnamefont {M.}~\bibnamefont {Volonteri}}, \bibinfo
  {author} {\bibfnamefont {E.}~\bibnamefont {Cackett}}, \bibinfo {author}
  {\bibfnamefont {C.}~\bibnamefont {Reynolds}},  \emph {et~al.},\ }\href
  {\doibase 10.1088/0004-637X/703/2/2171} {\bibfield  {journal} {\bibinfo
  {journal} {Astrophys.J.}\ }\textbf {\bibinfo {volume} {703}},\ \bibinfo
  {pages} {2171} (\bibinfo {year} {2009})},\ \Eprint
  {http://arxiv.org/abs/0908.0013} {arXiv:0908.0013 [astro-ph.HE]} \BibitemShut
  {NoStop}%
%%CITATION = ARXIV:0908.0013;%%
\bibitem [{\citenamefont {Leaver}(1985)}]{Leaver:1985ax}%
  \BibitemOpen
  \bibfield  {author} {\bibinfo {author} {\bibfnamefont {E.}~\bibnamefont
  {Leaver}},\ }\href@noop {} {\bibfield  {journal} {\bibinfo  {journal}
  {Proc.Roy.Soc.Lond.}\ }\textbf {\bibinfo {volume} {A402}},\ \bibinfo {pages}
  {285} (\bibinfo {year} {1985})}\BibitemShut {NoStop}%
%%CITATION = PRSLA,A402,285;%%
\bibitem [{\citenamefont {Pani}\ \emph {et~al.}(2012)\citenamefont {Pani},
  \citenamefont {Cardoso}, \citenamefont {Gualtieri}, \citenamefont {Berti},\
  and\ \citenamefont {Ishibashi}}]{paperPRL}%
  \BibitemOpen
  \bibfield  {author} {\bibinfo {author} {\bibfnamefont {P.}~\bibnamefont
  {Pani}}, \bibinfo {author} {\bibfnamefont {V.}~\bibnamefont {Cardoso}},
  \bibinfo {author} {\bibfnamefont {L.}~\bibnamefont {Gualtieri}}, \bibinfo
  {author} {\bibfnamefont {E.}~\bibnamefont {Berti}}, \ and\ \bibinfo {author}
  {\bibfnamefont {A.}~\bibnamefont {Ishibashi}},\ }\href@noop {} {\  (\bibinfo
  {year} {2012})},\ \bibinfo {note} {to appear in Physical Review
  Letters.}\BibitemShut {Stop}%
\bibitem [{\citenamefont {Detweiler}(1980)}]{Detweiler:1980uk}%
  \BibitemOpen
  \bibfield  {author} {\bibinfo {author} {\bibfnamefont {S.~L.}\ \bibnamefont
  {Detweiler}},\ }\href {\doibase 10.1103/PhysRevD.22.2323} {\bibfield
  {journal} {\bibinfo  {journal} {Phys. Rev.}\ }\textbf {\bibinfo {volume}
  {D22}},\ \bibinfo {pages} {2323} (\bibinfo {year} {1980})}\BibitemShut
  {NoStop}%
%%CITATION = PHRVA,D22,2323;%%
\bibitem [{url()}]{url}%
  \BibitemOpen
  \href@noop {} {\ }\bibinfo {note} {\noindent
  \url{http://blackholes.ist.utl.pt/?page=Files},\\
  \url{http://www.phy.olemiss.edu/~berti/qnms.html}}\BibitemShut {NoStop}%
\bibitem [{\citenamefont {Hartle}(1967)}]{Hartle:1967he}%
  \BibitemOpen
  \bibfield  {author} {\bibinfo {author} {\bibfnamefont {J.~B.}\ \bibnamefont
  {Hartle}},\ }\href@noop {} {\bibfield  {journal} {\bibinfo  {journal}
  {Astrophys.J.}\ }\textbf {\bibinfo {volume} {150}},\ \bibinfo {pages} {1005}
  (\bibinfo {year} {1967})}\BibitemShut {NoStop}%
%%CITATION = ASJOA,150,1005;%%
\bibitem [{\citenamefont {Ferrari}\ \emph {et~al.}(2007)\citenamefont
  {Ferrari}, \citenamefont {Gualtieri},\ and\ \citenamefont
  {Marassi}}]{Ferrari:2007rc}%
  \BibitemOpen
  \bibfield  {author} {\bibinfo {author} {\bibfnamefont {V.}~\bibnamefont
  {Ferrari}}, \bibinfo {author} {\bibfnamefont {L.}~\bibnamefont {Gualtieri}},
  \ and\ \bibinfo {author} {\bibfnamefont {S.}~\bibnamefont {Marassi}},\ }\href
  {\doibase 10.1103/PhysRevD.76.104033} {\bibfield  {journal} {\bibinfo
  {journal} {Phys.Rev.}\ }\textbf {\bibinfo {volume} {D76}},\ \bibinfo {pages}
  {104033} (\bibinfo {year} {2007})},\ \Eprint {http://arxiv.org/abs/0709.2925}
  {arXiv:0709.2925 [gr-qc]} \BibitemShut {NoStop}%
%%CITATION = ARXIV:0709.2925;%%
\bibitem [{\citenamefont {Berti}\ \emph {et~al.}(2006)\citenamefont {Berti},
  \citenamefont {Cardoso},\ and\ \citenamefont {Casals}}]{Berti:2005gp}%
  \BibitemOpen
  \bibfield  {author} {\bibinfo {author} {\bibfnamefont {E.}~\bibnamefont
  {Berti}}, \bibinfo {author} {\bibfnamefont {V.}~\bibnamefont {Cardoso}}, \
  and\ \bibinfo {author} {\bibfnamefont {M.}~\bibnamefont {Casals}},\ }\href
  {\doibase 10.1103/PhysRevD.73.024013, 10.1103/PhysRevD.73.109902} {\bibfield
  {journal} {\bibinfo  {journal} {Phys.Rev.}\ }\textbf {\bibinfo {volume}
  {D73}},\ \bibinfo {pages} {024013} (\bibinfo {year} {2006})},\ \Eprint
  {http://arxiv.org/abs/gr-qc/0511111} {arXiv:gr-qc/0511111 [gr-qc]}
  \BibitemShut {NoStop}%
%%CITATION = GR-QC/0511111;%%
\bibitem [{\citenamefont {Gerlach}\ and\ \citenamefont
  {Sengupta}(1980)}]{Gerlach:1980tx}%
  \BibitemOpen
  \bibfield  {author} {\bibinfo {author} {\bibfnamefont {U.}~\bibnamefont
  {Gerlach}}\ and\ \bibinfo {author} {\bibfnamefont {U.}~\bibnamefont
  {Sengupta}},\ }\href {\doibase 10.1103/PhysRevD.22.1300} {\bibfield
  {journal} {\bibinfo  {journal} {Phys.Rev.}\ }\textbf {\bibinfo {volume}
  {D22}},\ \bibinfo {pages} {1300} (\bibinfo {year} {1980})}\BibitemShut
  {NoStop}%
%%CITATION = PHRVA,D22,1300;%%
\bibitem [{\citenamefont {Teukolsky}\ and\ \citenamefont
  {Press}(1974)}]{Teukolsky:1974yv}%
  \BibitemOpen
  \bibfield  {author} {\bibinfo {author} {\bibfnamefont {S.}~\bibnamefont
  {Teukolsky}}\ and\ \bibinfo {author} {\bibfnamefont {W.}~\bibnamefont
  {Press}},\ }\href {\doibase 10.1086/153180} {\bibfield  {journal} {\bibinfo
  {journal} {Astrophys.J.}\ }\textbf {\bibinfo {volume} {193}},\ \bibinfo
  {pages} {443} (\bibinfo {year} {1974})}\BibitemShut {NoStop}%
%%CITATION = ASJOA,193,443;%%
\bibitem [{\citenamefont {Berti}\ \emph {et~al.}(2005)\citenamefont {Berti},
  \citenamefont {White}, \citenamefont {Maniopoulou},\ and\ \citenamefont
  {Bruni}}]{Berti:2004ny}%
  \BibitemOpen
  \bibfield  {author} {\bibinfo {author} {\bibfnamefont {E.}~\bibnamefont
  {Berti}}, \bibinfo {author} {\bibfnamefont {F.}~\bibnamefont {White}},
  \bibinfo {author} {\bibfnamefont {A.}~\bibnamefont {Maniopoulou}}, \ and\
  \bibinfo {author} {\bibfnamefont {M.}~\bibnamefont {Bruni}},\ }\href
  {\doibase 10.1111/j.1365-2966.2005.08812.x/abs/} {\bibfield  {journal}
  {\bibinfo  {journal} {Mon.Not.Roy.Astron.Soc.}\ }\textbf {\bibinfo {volume}
  {358}},\ \bibinfo {pages} {923} (\bibinfo {year} {2005})},\ \Eprint
  {http://arxiv.org/abs/gr-qc/0405146} {arXiv:gr-qc/0405146 [gr-qc]}
  \BibitemShut {NoStop}%
\bibitem [{\citenamefont {Lockitch}\ and\ \citenamefont
  {Andersson}(2004)}]{Lockitch:2001hq}%
  \BibitemOpen
  \bibfield  {author} {\bibinfo {author} {\bibfnamefont {K.~H.}\ \bibnamefont
  {Lockitch}}\ and\ \bibinfo {author} {\bibfnamefont {N.}~\bibnamefont
  {Andersson}},\ }\href {\doibase 10.1088/0264-9381/21/19/012} {\bibfield
  {journal} {\bibinfo  {journal} {Class.Quant.Grav.}\ }\textbf {\bibinfo
  {volume} {21}},\ \bibinfo {pages} {4661} (\bibinfo {year} {2004})},\ \Eprint
  {http://arxiv.org/abs/gr-qc/0106088} {arXiv:gr-qc/0106088 [gr-qc]}
  \BibitemShut {NoStop}%
%%CITATION = GR-QC/0106088;%%
\bibitem [{\citenamefont {Leaver}(1990)}]{Leaver:1990zz}%
  \BibitemOpen
  \bibfield  {author} {\bibinfo {author} {\bibfnamefont {E.~W.}\ \bibnamefont
  {Leaver}},\ }\href {\doibase 10.1103/PhysRevD.41.2986} {\bibfield  {journal}
  {\bibinfo  {journal} {Phys.Rev.}\ }\textbf {\bibinfo {volume} {D41}},\
  \bibinfo {pages} {2986} (\bibinfo {year} {1990})}\BibitemShut {NoStop}%
%%CITATION = PHRVA,D41,2986;%%
\bibitem [{\citenamefont {{Simmendinger}}\ \emph {et~al.}(1999)\citenamefont
  {{Simmendinger}}, \citenamefont {{Wunderlin}},\ and\ \citenamefont
  {{Pelster}}}]{1999PhRvE..59.5344S}%
  \BibitemOpen
  \bibfield  {author} {\bibinfo {author} {\bibfnamefont {C.}~\bibnamefont
  {{Simmendinger}}}, \bibinfo {author} {\bibfnamefont {A.}~\bibnamefont
  {{Wunderlin}}}, \ and\ \bibinfo {author} {\bibfnamefont {A.}~\bibnamefont
  {{Pelster}}},\ }\href {\doibase 10.1103/PhysRevE.59.5344} {\bibfield
  {journal} {\bibinfo  {journal} {\pre}\ }\textbf {\bibinfo {volume} {59}},\
  \bibinfo {pages} {5344} (\bibinfo {year} {1999})},\ \Eprint
  {http://arxiv.org/abs/arXiv:math-ph/0106011} {arXiv:math-ph/0106011}
  \BibitemShut {NoStop}%
\bibitem [{\citenamefont {{Thorne}}(1969)}]{1969ApJ...158....1T}%
  \BibitemOpen
  \bibfield  {author} {\bibinfo {author} {\bibfnamefont {K.~S.}\ \bibnamefont
  {{Thorne}}},\ }\href {\doibase 10.1086/150168} {\bibfield  {journal}
  {\bibinfo  {journal} {\apj}\ }\textbf {\bibinfo {volume} {158}},\ \bibinfo
  {pages} {1} (\bibinfo {year} {1969})}\BibitemShut {NoStop}%
\bibitem [{\citenamefont {Chandrasekhar}\ and\ \citenamefont
  {Ferrari}(1992)}]{Chandrasekhar:1992ey}%
  \BibitemOpen
  \bibfield  {author} {\bibinfo {author} {\bibfnamefont {S.}~\bibnamefont
  {Chandrasekhar}}\ and\ \bibinfo {author} {\bibfnamefont {V.}~\bibnamefont
  {Ferrari}},\ }\href@noop {} {\bibfield  {journal} {\bibinfo  {journal}
  {Proc.Roy.Soc.Lond.}\ }\textbf {\bibinfo {volume} {A437}},\ \bibinfo {pages}
  {133} (\bibinfo {year} {1992})}\BibitemShut {NoStop}%
%%CITATION = PRSLA,A437,133;%%
\bibitem [{\citenamefont {Berti}\ \emph
  {et~al.}(2009{\natexlab{b}})\citenamefont {Berti}, \citenamefont {Cardoso},\
  and\ \citenamefont {Pani}}]{Berti:2009wx}%
  \BibitemOpen
  \bibfield  {author} {\bibinfo {author} {\bibfnamefont {E.}~\bibnamefont
  {Berti}}, \bibinfo {author} {\bibfnamefont {V.}~\bibnamefont {Cardoso}}, \
  and\ \bibinfo {author} {\bibfnamefont {P.}~\bibnamefont {Pani}},\ }\href
  {\doibase 10.1103/PhysRevD.79.101501} {\bibfield  {journal} {\bibinfo
  {journal} {Phys.Rev.}\ }\textbf {\bibinfo {volume} {D79}},\ \bibinfo {pages}
  {101501} (\bibinfo {year} {2009}{\natexlab{b}})},\ \Eprint
  {http://arxiv.org/abs/0903.5311} {arXiv:0903.5311 [gr-qc]} \BibitemShut
  {NoStop}%
%%CITATION = ARXIV:0903.5311;%%
\bibitem [{\citenamefont {Ruffini}(1973)}]{Ruffini}%
  \BibitemOpen
  \bibfield  {author} {\bibinfo {author} {\bibfnamefont {R.}~\bibnamefont
  {Ruffini}},\ }\href@noop {} {\emph {\bibinfo {title} {Black Holes: les Astres
  Occlus}}}\ (\bibinfo  {publisher} {Gordon and Breach Science Publishers},\
  \bibinfo {year} {1973})\BibitemShut {NoStop}%
\bibitem [{\citenamefont {Merritt}\ and\ \citenamefont
  {Milosavljevic}(2005)}]{Merritt:2004gc}%
  \BibitemOpen
  \bibfield  {author} {\bibinfo {author} {\bibfnamefont {D.}~\bibnamefont
  {Merritt}}\ and\ \bibinfo {author} {\bibfnamefont {M.}~\bibnamefont
  {Milosavljevic}},\ }\href@noop {} {\bibfield  {journal} {\bibinfo  {journal}
  {Living Rev.Rel.}\ }\textbf {\bibinfo {volume} {8}},\ \bibinfo {pages} {8}
  (\bibinfo {year} {2005})},\ \Eprint {http://arxiv.org/abs/astro-ph/0410364}
  {arXiv:astro-ph/0410364 [astro-ph]} \BibitemShut {NoStop}%
%%CITATION = ASTRO-PH/0410364;%%
\bibitem [{\citenamefont {Sesana}\ \emph {et~al.}(2007)\citenamefont {Sesana},
  \citenamefont {Haardt},\ and\ \citenamefont {Madau}}]{Sesana:2006ne}%
  \BibitemOpen
  \bibfield  {author} {\bibinfo {author} {\bibfnamefont {A.}~\bibnamefont
  {Sesana}}, \bibinfo {author} {\bibfnamefont {F.}~\bibnamefont {Haardt}}, \
  and\ \bibinfo {author} {\bibfnamefont {P.}~\bibnamefont {Madau}},\ }\href
  {\doibase 10.1086/513016} {\bibfield  {journal} {\bibinfo  {journal}
  {Astrophys.J.}\ }\textbf {\bibinfo {volume} {660}},\ \bibinfo {pages} {546}
  (\bibinfo {year} {2007})},\ \Eprint {http://arxiv.org/abs/astro-ph/0612265}
  {arXiv:astro-ph/0612265 [astro-ph]} \BibitemShut {NoStop}%
%%CITATION = ASTRO-PH/0612265;%%
\bibitem [{\citenamefont {Barausse}(2012)}]{Barausse:2012fy}%
  \BibitemOpen
  \bibfield  {author} {\bibinfo {author} {\bibfnamefont {E.}~\bibnamefont
  {Barausse}},\ }\href {\doibase 10.1111/j.1365-2966.2012.21057.x} {\bibfield
  {journal} {\bibinfo  {journal} {Mon.Not.Roy.Astron.Soc.}\ }\textbf {\bibinfo
  {volume} {423}},\ \bibinfo {pages} {2533} (\bibinfo {year} {2012})},\ \Eprint
  {http://arxiv.org/abs/1201.5888} {arXiv:1201.5888 [astro-ph.CO]} \BibitemShut
  {NoStop}%
%%CITATION = ARXIV:1201.5888;%%
\bibitem [{\citenamefont {Berti}\ and\ \citenamefont
  {Volonteri}(2008)}]{Berti:2008af}%
  \BibitemOpen
  \bibfield  {author} {\bibinfo {author} {\bibfnamefont {E.}~\bibnamefont
  {Berti}}\ and\ \bibinfo {author} {\bibfnamefont {M.}~\bibnamefont
  {Volonteri}},\ }\href {\doibase 10.1086/590379} {\bibfield  {journal}
  {\bibinfo  {journal} {Astrophys.J.}\ }\textbf {\bibinfo {volume} {684}},\
  \bibinfo {pages} {822} (\bibinfo {year} {2008})},\ \Eprint
  {http://arxiv.org/abs/0802.0025} {arXiv:0802.0025 [astro-ph]} \BibitemShut
  {NoStop}%
%%CITATION = ARXIV:0802.0025;%%
\bibitem [{\citenamefont {Bardeen}(1970)}]{Bardeen:1970zz}%
  \BibitemOpen
  \bibfield  {author} {\bibinfo {author} {\bibfnamefont {J.~M.}\ \bibnamefont
  {Bardeen}},\ }\href@noop {} {\bibfield  {journal} {\bibinfo  {journal}
  {Nature}\ }\textbf {\bibinfo {volume} {226}},\ \bibinfo {pages} {64}
  (\bibinfo {year} {1970})}\BibitemShut {NoStop}%
%%CITATION = NATUA,226,64;%%
\bibitem [{\citenamefont {Thorne}(1974)}]{Thorne:1974ve}%
  \BibitemOpen
  \bibfield  {author} {\bibinfo {author} {\bibfnamefont {K.~S.}\ \bibnamefont
  {Thorne}},\ }\href {\doibase 10.1086/152991} {\bibfield  {journal} {\bibinfo
  {journal} {Astrophys.J.}\ }\textbf {\bibinfo {volume} {191}},\ \bibinfo
  {pages} {507} (\bibinfo {year} {1974})}\BibitemShut {NoStop}%
%%CITATION = ASJOA,191,507;%%
\bibitem [{\citenamefont {Gammie}\ \emph {et~al.}(2004)\citenamefont {Gammie},
  \citenamefont {Shapiro},\ and\ \citenamefont {McKinney}}]{Gammie:2003qi}%
  \BibitemOpen
  \bibfield  {author} {\bibinfo {author} {\bibfnamefont {C.~F.}\ \bibnamefont
  {Gammie}}, \bibinfo {author} {\bibfnamefont {S.~L.}\ \bibnamefont {Shapiro}},
  \ and\ \bibinfo {author} {\bibfnamefont {J.~C.}\ \bibnamefont {McKinney}},\
  }\href {\doibase 10.1086/380996} {\bibfield  {journal} {\bibinfo  {journal}
  {Astrophys.J.}\ }\textbf {\bibinfo {volume} {602}},\ \bibinfo {pages} {312}
  (\bibinfo {year} {2004})},\ \Eprint {http://arxiv.org/abs/astro-ph/0310886}
  {arXiv:astro-ph/0310886 [astro-ph]} \BibitemShut {NoStop}%
%%CITATION = ASTRO-PH/0310886;%%
\bibitem [{\citenamefont {King}\ and\ \citenamefont
  {Pringle}(2006)}]{King:2006uu}%
  \BibitemOpen
  \bibfield  {author} {\bibinfo {author} {\bibfnamefont {A.~R.}\ \bibnamefont
  {King}}\ and\ \bibinfo {author} {\bibfnamefont {J.}~\bibnamefont {Pringle}},\
  }\href {\doibase 10.1111/j.1745-3933.2006.00250.x} {\bibfield  {journal}
  {\bibinfo  {journal} {Mon.Not.Roy.Astron.Soc.Lett.}\ }\textbf {\bibinfo
  {volume} {373}},\ \bibinfo {pages} {L93} (\bibinfo {year} {2006})},\ \Eprint
  {http://arxiv.org/abs/astro-ph/0609598} {arXiv:astro-ph/0609598 [astro-ph]}
  \BibitemShut {NoStop}%
%%CITATION = ASTRO-PH/0609598;%%
\bibitem [{\citenamefont {Harnik}\ \emph {et~al.}(2012)\citenamefont {Harnik},
  \citenamefont {Kopp},\ and\ \citenamefont {Machado}}]{Harnik:2012ni}%
  \BibitemOpen
  \bibfield  {author} {\bibinfo {author} {\bibfnamefont {R.}~\bibnamefont
  {Harnik}}, \bibinfo {author} {\bibfnamefont {J.}~\bibnamefont {Kopp}}, \ and\
  \bibinfo {author} {\bibfnamefont {P.~A.}\ \bibnamefont {Machado}},\ }\href
  {\doibase 10.1088/1475-7516/2012/07/026} {\bibfield  {journal} {\bibinfo
  {journal} {JCAP}\ }\textbf {\bibinfo {volume} {1207}},\ \bibinfo {pages}
  {026} (\bibinfo {year} {2012})},\ \Eprint {http://arxiv.org/abs/1202.6073}
  {arXiv:1202.6073 [hep-ph]} \BibitemShut {NoStop}%
%%CITATION = ARXIV:1202.6073;%%
\bibitem [{\citenamefont {{McConnell}}\ \emph {et~al.}(2011)\citenamefont
  {{McConnell}}, \citenamefont {{Ma}}, \citenamefont {{Gebhardt}},
  \citenamefont {{Wright}}, \citenamefont {{Murphy}}, \citenamefont {{Lauer}},
  \citenamefont {{Graham}},\ and\ \citenamefont
  {{Richstone}}}]{2011Natur.480..215M}%
  \BibitemOpen
  \bibfield  {author} {\bibinfo {author} {\bibfnamefont {N.~J.}\ \bibnamefont
  {{McConnell}}}, \bibinfo {author} {\bibfnamefont {C.-P.}\ \bibnamefont
  {{Ma}}}, \bibinfo {author} {\bibfnamefont {K.}~\bibnamefont {{Gebhardt}}},
  \bibinfo {author} {\bibfnamefont {S.~A.}\ \bibnamefont {{Wright}}}, \bibinfo
  {author} {\bibfnamefont {J.~D.}\ \bibnamefont {{Murphy}}}, \bibinfo {author}
  {\bibfnamefont {T.~R.}\ \bibnamefont {{Lauer}}}, \bibinfo {author}
  {\bibfnamefont {J.~R.}\ \bibnamefont {{Graham}}}, \ and\ \bibinfo {author}
  {\bibfnamefont {D.~O.}\ \bibnamefont {{Richstone}}},\ }\href {\doibase
  10.1038/nature10636} {\bibfield  {journal} {\bibinfo  {journal} {\nat}\
  }\textbf {\bibinfo {volume} {480}},\ \bibinfo {pages} {215} (\bibinfo {year}
  {2011})},\ \Eprint {http://arxiv.org/abs/1112.1078} {arXiv:1112.1078
  [astro-ph.CO]} \BibitemShut {NoStop}%
\bibitem [{\citenamefont {{McConnell}}\ \emph {et~al.}(2012)\citenamefont
  {{McConnell}}, \citenamefont {{Ma}}, \citenamefont {{Murphy}}, \citenamefont
  {{Gebhardt}}, \citenamefont {{Lauer}}, \citenamefont {{Graham}},
  \citenamefont {{Wright}},\ and\ \citenamefont
  {{Richstone}}}]{2012arXiv1203.1620M}%
  \BibitemOpen
  \bibfield  {author} {\bibinfo {author} {\bibfnamefont {N.~J.}\ \bibnamefont
  {{McConnell}}}, \bibinfo {author} {\bibfnamefont {C.-P.}\ \bibnamefont
  {{Ma}}}, \bibinfo {author} {\bibfnamefont {J.~D.}\ \bibnamefont {{Murphy}}},
  \bibinfo {author} {\bibfnamefont {K.}~\bibnamefont {{Gebhardt}}}, \bibinfo
  {author} {\bibfnamefont {T.~R.}\ \bibnamefont {{Lauer}}}, \bibinfo {author}
  {\bibfnamefont {J.~R.}\ \bibnamefont {{Graham}}}, \bibinfo {author}
  {\bibfnamefont {S.~A.}\ \bibnamefont {{Wright}}}, \ and\ \bibinfo {author}
  {\bibfnamefont {D.~O.}\ \bibnamefont {{Richstone}}},\ }\href@noop {}
  {\bibfield  {journal} {\bibinfo  {journal} {ArXiv e-prints}\ } (\bibinfo
  {year} {2012})},\ \Eprint {http://arxiv.org/abs/1203.1620} {arXiv:1203.1620
  [astro-ph.CO]} \BibitemShut {NoStop}%
\bibitem [{\citenamefont {Beringer}\ \emph {et~al.}(2012)\citenamefont
  {Beringer} \emph {et~al.}}]{PDG}%
  \BibitemOpen
  \bibfield  {author} {\bibinfo {author} {\bibfnamefont {J.}~\bibnamefont
  {Beringer}} \emph {et~al.} (\bibinfo {collaboration} {Particle Data Group}),\
  }\href {\doibase 10.1103/PhysRevD.86.010001} {\bibfield  {journal} {\bibinfo
  {journal} {Phys.Rev.}\ }\textbf {\bibinfo {volume} {D86}},\ \bibinfo {pages}
  {010001} (\bibinfo {year} {2012})}\BibitemShut {NoStop}%
%%CITATION = PHRVA,D86,010001;%%
\bibitem [{\citenamefont {Buchdahl}(1979)}]{Buchdahl:1979ut}%
  \BibitemOpen
  \bibfield  {author} {\bibinfo {author} {\bibfnamefont {H.~A.}\ \bibnamefont
  {Buchdahl}},\ }\href {\doibase 10.1088/0305-4470/12/8/018} {\bibfield
  {journal} {\bibinfo  {journal} {J.Phys.A}\ }\textbf {\bibinfo {volume}
  {A12}},\ \bibinfo {pages} {1235} (\bibinfo {year} {1979})}\BibitemShut
  {NoStop}%
%%CITATION = JPAGB,A12,1235;%%
\bibitem [{\citenamefont {Vitagliano}\ \emph {et~al.}(2010)\citenamefont
  {Vitagliano}, \citenamefont {Sotiriou},\ and\ \citenamefont
  {Liberati}}]{Vitagliano:2010pq}%
  \BibitemOpen
  \bibfield  {author} {\bibinfo {author} {\bibfnamefont {V.}~\bibnamefont
  {Vitagliano}}, \bibinfo {author} {\bibfnamefont {T.~P.}\ \bibnamefont
  {Sotiriou}}, \ and\ \bibinfo {author} {\bibfnamefont {S.}~\bibnamefont
  {Liberati}},\ }\href {\doibase 10.1103/PhysRevD.82.084007} {\bibfield
  {journal} {\bibinfo  {journal} {Phys.Rev.}\ }\textbf {\bibinfo {volume}
  {D82}},\ \bibinfo {pages} {084007} (\bibinfo {year} {2010})},\ \Eprint
  {http://arxiv.org/abs/1007.3937} {arXiv:1007.3937 [gr-qc]} \BibitemShut
  {NoStop}%
%%CITATION = ARXIV:1007.3937;%%
\bibitem [{\citenamefont {Sotani}\ \emph {et~al.}(2007)\citenamefont {Sotani},
  \citenamefont {Kokkotas},\ and\ \citenamefont {Stergioulas}}]{Sotani:2006at}%
  \BibitemOpen
  \bibfield  {author} {\bibinfo {author} {\bibfnamefont {H.}~\bibnamefont
  {Sotani}}, \bibinfo {author} {\bibfnamefont {K.}~\bibnamefont {Kokkotas}}, \
  and\ \bibinfo {author} {\bibfnamefont {N.}~\bibnamefont {Stergioulas}},\
  }\href {\doibase 10.1111/j.1365-2966.2006.11304.x} {\bibfield  {journal}
  {\bibinfo  {journal} {Mon.Not.Roy.Astron.Soc.}\ }\textbf {\bibinfo {volume}
  {375}},\ \bibinfo {pages} {261} (\bibinfo {year} {2007})},\ \Eprint
  {http://arxiv.org/abs/astro-ph/0608626} {arXiv:astro-ph/0608626 [astro-ph]}
  \BibitemShut {NoStop}%
%%CITATION = ASTRO-PH/0608626;%%
\bibitem [{\citenamefont {Stavridis}\ \emph {et~al.}(2007)\citenamefont
  {Stavridis}, \citenamefont {Passamonti},\ and\ \citenamefont
  {Kokkotas}}]{Stavridis:2007xz}%
  \BibitemOpen
  \bibfield  {author} {\bibinfo {author} {\bibfnamefont {A.}~\bibnamefont
  {Stavridis}}, \bibinfo {author} {\bibfnamefont {A.}~\bibnamefont
  {Passamonti}}, \ and\ \bibinfo {author} {\bibfnamefont {K.}~\bibnamefont
  {Kokkotas}},\ }\href {\doibase 10.1103/PhysRevD.75.064019} {\bibfield
  {journal} {\bibinfo  {journal} {Phys.Rev.}\ }\textbf {\bibinfo {volume}
  {D75}},\ \bibinfo {pages} {064019} (\bibinfo {year} {2007})},\ \Eprint
  {http://arxiv.org/abs/gr-qc/0701122} {arXiv:gr-qc/0701122 [gr-qc]}
  \BibitemShut {NoStop}%
%%CITATION = GR-QC/0701122;%%
\bibitem [{\citenamefont {Passamonti}\ \emph {et~al.}(2008)\citenamefont
  {Passamonti}, \citenamefont {Stavridis},\ and\ \citenamefont
  {Kokkotas}}]{Passamonti:2007td}%
  \BibitemOpen
  \bibfield  {author} {\bibinfo {author} {\bibfnamefont {A.}~\bibnamefont
  {Passamonti}}, \bibinfo {author} {\bibfnamefont {A.}~\bibnamefont
  {Stavridis}}, \ and\ \bibinfo {author} {\bibfnamefont {K.}~\bibnamefont
  {Kokkotas}},\ }\href {\doibase 10.1103/PhysRevD.77.024029} {\bibfield
  {journal} {\bibinfo  {journal} {Phys.Rev.}\ }\textbf {\bibinfo {volume}
  {D77}},\ \bibinfo {pages} {024029} (\bibinfo {year} {2008})},\ \Eprint
  {http://arxiv.org/abs/0706.0991} {arXiv:0706.0991 [gr-qc]} \BibitemShut
  {NoStop}%
%%CITATION = ARXIV:0706.0991;%%
\bibitem [{\citenamefont {Starobinsky}(1973)}]{Starobinsky}%
  \BibitemOpen
  \bibfield  {author} {\bibinfo {author} {\bibfnamefont {A.~A.}\ \bibnamefont
  {Starobinsky}},\ }\href@noop {} {\bibfield  {journal} {\bibinfo  {journal}
  {Zh. Eksp. Teor. Fiz}\ }\textbf {\bibinfo {volume} {64}},\ \bibinfo {pages}
  {48} (\bibinfo {year} {1973})},\ \bibinfo {note} {sov. Phys. JETP 37, 28
  (1973)}\BibitemShut {NoStop}%
\bibitem [{\citenamefont {Abramowitz}\ and\ \citenamefont
  {Stegun}(1964)}]{abramowitz+stegun}%
  \BibitemOpen
  \bibfield  {author} {\bibinfo {author} {\bibfnamefont {M.}~\bibnamefont
  {Abramowitz}}\ and\ \bibinfo {author} {\bibfnamefont {I.~A.}\ \bibnamefont
  {Stegun}},\ }\href@noop {} {\emph {\bibinfo {title} {Handbook of Mathematical
  Functions with Formulas, Graphs, and Mathematical Tables}}},\ \bibinfo
  {edition} {ninth dover printing, tenth gpo printing}\ ed.\ (\bibinfo
  {publisher} {Dover},\ \bibinfo {address} {New York},\ \bibinfo {year}
  {1964})\BibitemShut {NoStop}%
\bibitem [{\citenamefont {Furuhashi}\ and\ \citenamefont
  {Nambu}(2004)}]{Furuhashi:2004jk}%
  \BibitemOpen
  \bibfield  {author} {\bibinfo {author} {\bibfnamefont {H.}~\bibnamefont
  {Furuhashi}}\ and\ \bibinfo {author} {\bibfnamefont {Y.}~\bibnamefont
  {Nambu}},\ }\href {\doibase 10.1143/PTP.112.983} {\bibfield  {journal}
  {\bibinfo  {journal} {Prog.Theor.Phys.}\ }\textbf {\bibinfo {volume} {112}},\
  \bibinfo {pages} {983} (\bibinfo {year} {2004})},\ \Eprint
  {http://arxiv.org/abs/gr-qc/0402037} {arXiv:gr-qc/0402037 [gr-qc]}
  \BibitemShut {NoStop}%
%%CITATION = GR-QC/0402037;%%
\bibitem [{\citenamefont {Rosa}(2012)}]{Rosa:2012uz}%
  \BibitemOpen
  \bibfield  {author} {\bibinfo {author} {\bibfnamefont {J.~G.}\ \bibnamefont
  {Rosa}},\ }\href@noop {} {\  (\bibinfo {year} {2012})},\ \Eprint
  {http://arxiv.org/abs/1209.4211} {arXiv:1209.4211 [hep-th]} \BibitemShut
  {NoStop}%
%%CITATION = ARXIV:1209.4211;%%
\end{thebibliography}%
\end{document}